MANOEL BENEDITO NIRDO DA SILVA CAMPOS

# CONTRIBUIÇÃO DO ECOTURISMO PARA O USO SUSTENTÁVEL DOS RECURSOS HÍDRICOS DO MUNICÍPIO DE RONDONÓPOLIS-MT

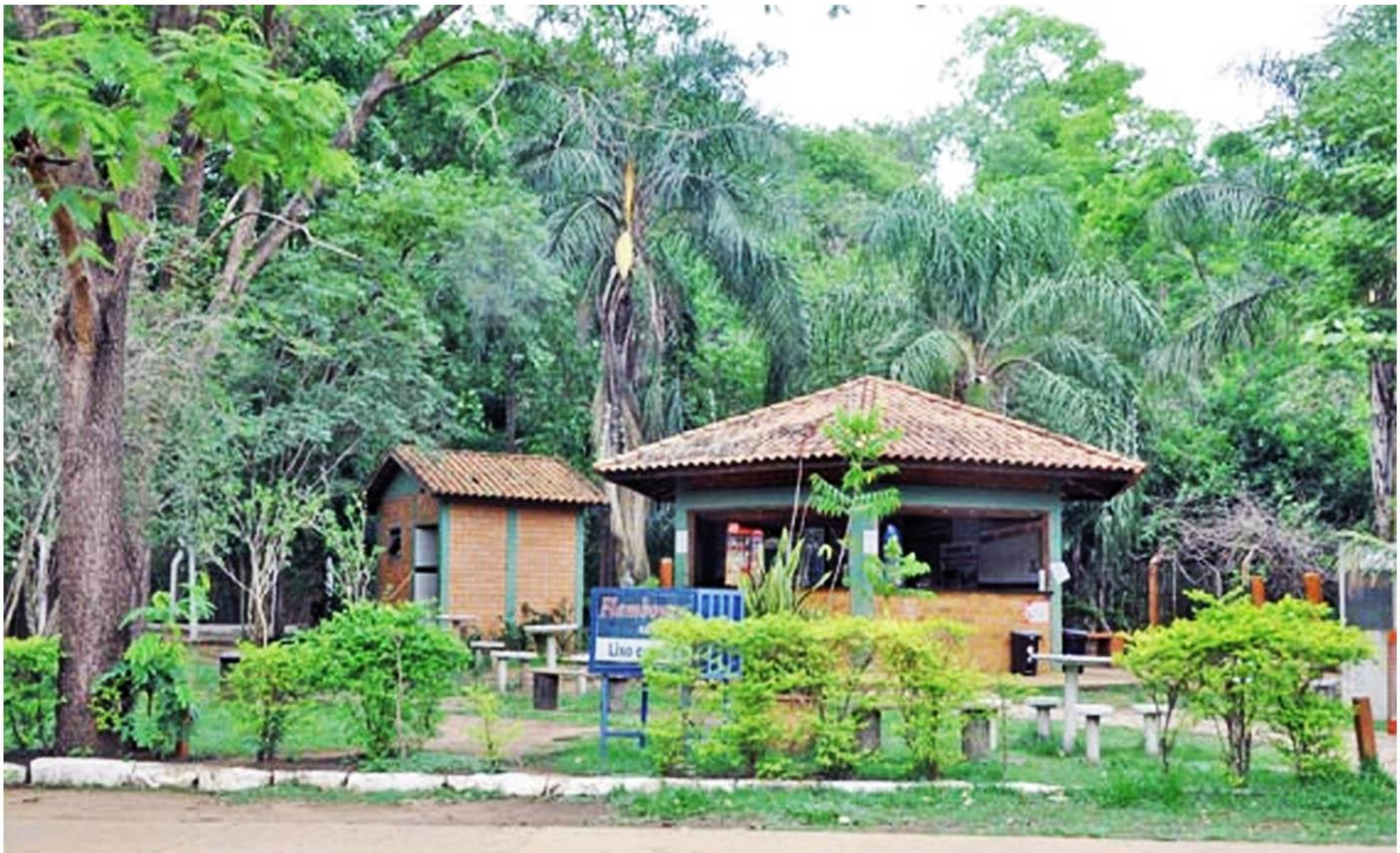

MANOEL BENEDITO NIRDO DA SILVA CAMPOS

# CONTRIBUIÇÃO DO ECOTURISMO PARA O USO SUSTENTÁVEL DOS RECURSOS HÍDRICOS DO MUNICÍPIO DE RONDONÓPOLIS-MT

Dedico este livro à minha esposa Carmen, aos meus filhos, Emanuel, Matheus, Emanuella, Andrew, Alessandro e aos meus pais Manoel e Antônia (*in memoriam*).

# AGRADECIMENTOS

Primeiramente agradeço a Deus, sabedoria superior do universo, da qual sou criatura e filho. Na sua palavra sempre busco fé e inspiração para passar pelos desertos da vida.

Ao Campus Universitário de Rondonópolis da Universidade Federal de Mato Grosso (UFMT), por intermédio do Departamento de Geografia e ao Programa de Pós-Graduação em Geografia, pelo acolhimento e pela possibilidade de aprimoramento acadêmico e profissional.

À minha amiga, professora Dra. Antônia Marília Medeiros Nardes, por ter aceitado o desafio de me orientar, e com toda dedicação, mostrou-me o caminho em que devia conduzir a pesquisa, pelos seus ensinamentos, pela compreensão e, sobretudo, pela confiança e liberdade de trabalho.

Ao coordenador do curso, professor Dr. Jeater W. Maciel Correa dos Santos, pela amizade e apoio.

Aos professores das disciplinas cursadas no mestrado pelas contribuições, pela dedicação e ensinamentos.

Aos meus amigos Adinael, Alex Teodoro, Camila, e ao professor Dr. Alexandre Gusmão pelo apoio e cooperação.

Enfim, a todos que direta e indiretamente contribuíram para a realização e sucesso desta obra, resta-me apenas dizer: MUITO OBRIGADO!

## RESUMO

O Município de Rondonópolis possui atrativos turísticos reconhecidos pela grande diversidade das cachoeiras e prainhas situadas no entorno da malha urbana, pois atraem turistas de várias localidades. Objetivando compreender como o ecoturismo pode contribuir para a conservação dos recursos hídricos nas áreas de lazer, bem como suas possibilidades de desenvolvimento das atividades turísticas nesses locais. Para alcançar o objetivo proposto, realizou-se uma análise acerca da geração dos benefícios socioambientais no qual se optou por uma abordagem quali-quantitativa procurando-se estabelecer a frequência da atitude dos usuários, através de questionários, levantamento fotográfico e Escala Likerts de cinco pontos, buscando dados que traçassem o perfil dos usuários que auxiliassem na compreensão do uso relacionados com as atividades ecoturísticas. O estudo justifica-se pela inexistência de referenciais teóricos metodológicos da temática em questão. A pesquisa foi desenvolvida em três etapas metodológicas: pesquisa bibliográfica, identificação das áreas de lazer e elaboração dos mapas temáticos para representação do potencial turístico. Os procedimentos que incluíram o uso de várias técnicas subsidiadas em ferramentas de sensoriamento remoto e geoprocessamento que permitiram a análise e espacialização das atividades de ecoturismo das principais áreas de lazer. A espacialização das cachoeiras e seu entorno, observam-se os caracteres biofísicos como: a vegetação endêmica, as cachoeiras, as quedas d'água, os afloramentos rochosos, os rios, as prainhas e os espraiados. Os resultados apresentaram uma correta percepção dos pesquisados sobre as inter-relações existentes entre as práticas ecoturísticas e o uso sustentável dos recursos hídricos. Conclui-se ainda que, um longo caminho deve ser realizado de forma a impedir que os benefícios econômicos do ecoturismo acabem gerando um quadro de exploração de recursos naturais de forma inadequada, ocasionando problemas ambientais em especial, aos recursos hídricos nas localidades.

PALAVRAS-CHAVE: Potencial Turístico. Recursos hídricos. Ecoturismo. Rondonópolis.

## ABSTRACT

The Municipality of Rondonópolis possesses several touristic attractions such as a great diversity of waterfalls and little beaches located in the surroundings of the urban area, which attract tourists from various locations. Aiming to understand how ecotourism can contribute to the conservation of water resources in the leisure areas, as well as their potential development of touristic activities in those places. To achieve the proposed objective, there was an analysis of the generation of social and environmental benefits in which it has opted for a qualitative and quantitative approach seeking to establish the utilization frequency of the users through questionnaires, photographic survey and Likerts scale of five points, seeking data to draw the profile of users that would help to understand the use related to ecotourism activities. The study is justified by the inexistence of methodological and theoretical framework of the theme in question. The research was conducted in three methodological steps: bibliographical research, identification of leisure areas and thematic maps elaboration for representation of touristic potential. The procedures included the use of various techniques subsidized in remote sensing and geoprocessing tools that allowed the analysis and spatial distribution of tourism activities of the main leisure areas. The spatial distribution of the waterfalls and its surroundings, we observe the biophysical characters such as: the endemic vegetation, the *cachoeiras*, the waterfalls, the rocky outcrops, rivers, little beaches and *espraiados*. The results showed a correct perception of respondents on existing inter-relationships between ecotourism practices and the sustainable use of water resources. In conclusion though, a long way must be performed in order to prevent the economic benefits of ecotourism generate an inappropriate exploitation of natural resources, causing environmental problems, particularly to water resources in the surroundings.

KEYWORDS: Touristic Potential. Water resources. Ecotourism. Rondonópolis.

# LISTA DE FIGURAS

## LISTA DE FOTOS

## LISTA DE MAPAS

## LISTA DE GRÁFICOS

# LISTA DE TABELAS

## LISTA DE QUADROS

## LISTA DE SIGLAS

ANPTUR – Associação Nacional Pesquisa em Turismo
ATR – Agências de Turismo Receptivo
CNTUR – Conselho Nacional de Turismo
SEDTUR – Secretaria de Estado de Desenvolvimento do Turismo
DEM – Modelo Digital de Elevação
DNIT – Departamento Nacional de Infraestrutura de Transportes
EMBRATUR – Empresa Brasileira de Turismo
IATA – International Air Transport Association
IBGE – Instituto Brasileiro de Geografia e Estatística
INCRA – Instituto Nacional de Colonização e Reforma Agrária
INTERNET – Sistema Global de Rede de Computadores Interligados
IP – Instituições Públicas
ML – Moradores das Localidades
MMA – Ministério do Meio Ambiente
OMT – Organização Mundial de Turismo
ONU – Organização das Nações Unidas
PCNs – Parâmetros Curriculares Nacionais
PNT – Plano Nacional do Turismo
PNUMA – Programas das Noções Unidas para o Meio Ambiente
PIB – Produto Interno Bruto
SEBRAE/MT – Serviço Brasileiro de Apoio às Micro e Pequenas Empresas de Mato Grosso
SEMMA – Secretaria Municipal de Meio Ambiente
SIGs – Sistemas de Informações Geográficas
UF – Usuários Frequentadores

**SUMÁRIO**

# 1 INTRODUÇÃO

Este estudo trata sobre as potencialidades nas áreas de lazer no Município de Rondonópolis, trazendo alguns elementos e consideram a viabilidade da exploração do turismo sustentável, com o aproveitamento dos recursos hídricos, fauna e flora, que vislumbram em possíveis das atividades turísticas.

A conciliação entre viabilidade econômica e conservação dos recursos hídricos com sustentabilidade constitui um desafio na atualidade. Poucas são as perspectivas, pois a acumulação de bens e riquezas é um dos maiores causadores de impactos ambientais negativos de ordem econômica e social da localidade (RODRIGUES e AMARANTE-JUNIOR, 2009).

Dentro desse contexto, o desenvolvimento de atividades turísticas que fomentem a economia local, desponta como possibilidade de transformação do espaço em que é praticada, é significativo no campo econômico pela criação de empregos e a abertura de novos empreendimentos aos municípios (PORTUGUEZ et al., 2012).

De acordo com Souza (2011, p. 21), "[...] o turismo é um impulsionador do desenvolvimento social e econômico pela geração de emprego e renda, mas deve ocorrer de forma organizada para que não beneficie apenas uma minoria". É uma atividade que pode ainda contribuir com o desenvolvimento do município, quando as atividades são planejadas.

Etimologicamente, tem-se que a palavra turismo é de procedência latina, significa ação e movimento, em um sentido mais amplo a ida e volta de algum lugar (DIAS e AGUIAR, 2002).

O turismo é hoje considerado uma das atividades que mais cresce, sendo, portanto, uma das maiores indústrias mundiais. Está associado a muitos dos principais setores da economia mundial. Esse fenômeno, esta estreitamente entrelaçada no tecido da vida – econômica, sociocultural e ambiental, e depende dos níveis primário, secundário e terciário de produção e serviços (FENNELL, 2002, p. 15).

Nesta mesma linha de compreensão, o turismo é definido por Fennell (2002, p.17) "como um sistema inter-relacionado que inclui os turistas e os serviços associados fornecidos e utilizados para auxiliar a movimentação do turista, tal como é concebido pela OMT".

A Organização Mundial do Turismo (OMT) organizou em 1991, juntamente com o governo do Canadá, a conferência internacional sobre Estatística de Viagens e Turismo (conhecida como Conferência de Ottawa). Nessa conferência foi aceito como critério e aprovado, o conceito do lado da demanda, e não da oferta de bens/serviços particulares. A definição de turismo tem por base a definição do tipo de consumidor cuja atividade constitui o turismo, e não o tipo de produto consumido (PAKMAN, 2014).

Ainda de acordo com o autor houve, em 1999 uma atualização da definição de 1991, adotando uma posição em relação a um dos pontos mais polêmicos do que pode ou não ser considerado turismo. Assim, conceitua-se turismo:

> O turismo compreende as atividades realizadas pelas pessoas durante suas viagens e estadias em lugares diferentes de seu entorno habitual, por um período de tempo consecutivo inferior a um ano, tendo em vista lazer, negócios ou outros motivos **não relacionados ao exercício de atividade no lugar visitado** (PAKMAN, 2014, p. 13, grifo nosso).

Dessa forma, este conceito conclui o entendimento internacional do turismo e de sua "definição oficial" no século XX.

Nesse contexto, com o objetivo de fortalecer os países nos fundamentos metodológicos e operacionais das estatísticas de turismo foram publicadas pelas Nações Unidas recomendações que apresentam um enfoque diferente dos convencionais e, finalmente reforçam o lazer entendido como prazer principal na motivação da saída de pessoas do entorno habitual. Assim, o turismo conquista uma nova concepção conceitual:

> O turismo é um fenômeno social, cultural e econômico, que envolve o movimento de pessoas para lugares fora do seu local de residência habitual, geralmente por prazer adaptado do conceito da (ONU/OMT, publicado pela IRTS - 2008).

A atividade turística, nas últimas décadas, obteve crescimento bastante significativo, favorecido por fenômenos culturais, econômicos e sociais.

> O desenvolvimento tecnológico dos transportes, o maior tempo livre e as melhores condições das pessoas, aliados às necessidades de evasão, de fuga dos grandes centros (como forma de recuperação do equilíbrio físico e espiritual de seus moradores), alteram o setor turístico. Como resultado obteve-se o acréscimo no número, de pessoas que viajam e o desenvolvimento da infraestrutura e dos equipamentos turísticos (ANSARAH, 1999, p. 17).

O cenário de movimentação do fluxo de turista é apresentado no estudo realizado pelo Plano Nacional de Turismo (PNT) 2013-2016 quanto à movimentação do transporte aéreo internacional projetado pela International Air Transport Association (IATA) para os principais países da Europa e da África:

> A International Air Transport association (IATA) apresentou também projeção para o ano de 2016 referentes ao transporte aéreo (doméstico e internacional). Estima-se que para o mercado internacional 1,45 bilhões de passageiros sejam transportados, com um incremento de 331 milhões em relação ao ano de 2011, tendo como os cinco principais países em movimentação os Estados Unidos (223,1 milhões), o Reino Unido (200,8 milhões), a Alemanha (172,9 milhões, a Espanha (124,6 milhões) e a França (123,1 milhões). Entretanto, entre os países que mais crescerão, figuram Uzbequistão (11,1%), Sudão (9,2%), Uruguai (9,0%), Azerbaijão (8,9%), Ucrânia (8,8%), Camboja (8,7%), Chile (8,5%), Panamá (8,5%) e Rússia (8,4%) (PNT, 2012, p. 29-30).

Especificamente no Brasil, o turismo é uma atividade que se destacou a partir de 1964, com a mudança de regime político e a concepção de uma atividade econômica extremamente lucrativa. Em 1966, a partir do Decreto Lei nº 55 foi implantado o Conselho Nacional de Turismo (CNTUR) e a Empresa Brasileira de Turismo (EMBRATUR), com objetivo de intensificar o turismo no Brasil (FENNELL, 2002).

De acordo com dados do IBGE (2012, p. 25), em 2009, a renda gerada no Brasil pelas atividades do turismo foi de 103,7 bilhões de reais contra 90,5 bilhões de reais em 2008. Isto significa um crescimento real na economia de 14,5% em média.

Ainda segundo os mesmo dados do IBGE, nas atividades do turismo, a remuneração por ocupação cresceu 15,2%, em termos nominais, de 2008 para 2009. Neste mesmo período, a média por ocupação para todas as atividades econômicas do País cresceu 8,0%. Um dado revelador nas atividades recreativas, culturais e desportivas foram 194 mil ocupações a mais e nos serviços de alimentação o acréscimo de 134 mil.

No Estado de Mato Grosso, o turismo ecológico é impulsionado pelo seu grande potencial e pela sua progressiva descoberta das inúmeras belezas naturais, bem como pela singularidade e diversidade de sua cultura, carros chefes das atividades ecoturistas (LÚCIO, 2008).

A atividade turística está ligada a aspectos econômicos, de modo que o investimento por parte poder público ou privado em infraestrutura para esta prática geram, mesmo sob a concepção de conservação ambiental, importante instrumento de crescimento para comunidades que possuem áreas de belezas naturais.

As ações apontadas do turismo, como segmento econômico, são, no entanto "predador do espaço e gerador de agravos ambientais e danos irreversível" Coriolano (*apud* PINTO 2008, p. 21), além de ser uma atividade com amplitude de ação, Begnini e Silva (2003) considera o turismo uma atividade que ao ser executada se envolve e interfere na área social e, também no campo cultural, econômico, ecológico e político da comunidade em que é executada.

Há ainda a necessidade de um estudo prévio, verificando se a ocupação ocorre do espaço geográfico de forma ordenada, para destacar as potencialidades, e também seguindo a critérios conservacionistas baseados na legislação vigente e critérios técnicos, principalmente nas áreas mais frágeis, pois alterações como a extração da vegetação, seriam causadores de grandes prejuízos ao ambiente (SOUZA, 2004).

Para Marinho e Bruhns (2003, p.54), a atividade turística não se constitui somente como uma atividade econômica, mas também como: "[...] prática social complexa e multifacetada, implica essencialmente o deslocamento de pessoas e a relação dessas pessoas entre si, com a comunidade e com o lugar visitado".

Nesta esteira, compreende-se que a atividade turística não somente se utiliza dos recursos disponíveis na natureza, como também tem importante papel com relação à transformação do meio rural e urbano, seja de forma direta ou indireta (COSTA et al., 2003).

O Brasil apresenta-se como um dos países com maior potencialidade para práticas de atividades ligadas ao ecoturismo.

> A nova ordem econômica mundial sinaliza para a conciliação da utilização racional dos recursos ambientais abrindo novos espaços para um campo de atividade que se torna cada dia mais importante dentro do enfoque de desenvolvimento sustentável: o termo ecológico, ou mais popularmente, ecoturismo (BARBOSA, 2003, p. 32).

O ecoturismo é conceituado como "o apoio à conservação ambiental, com o uso dito sustentável dos recursos" Pires (*apud* RODRIGUES 2003, p.31). Esta forma de turismo consiste na utilização de quatro componentes: turista, prestador de serviço, governo e comunidade.

A definição adotada pelo Ministério do Turismo (2010) e da Sociedade Internacional de Ecoturismo (TIES) apresenta uma conceituação semelhante "ecoturismo é uma viagem responsável às áreas naturais, visando preservar o meio ambiente e promover o bem-estar da população local". Entre as diversas interpretações e definições estabelecidas esta continua sendo referência no país.

A OMT e o Programa das Nações Unidas para o Meio Ambiente (PNUMA) referem-se ao Ecoturismo como um segmento do turismo, enquanto princípios que se almejam para o turismo sustentável são aplicáveis e devem servir de premissa para todos os tipos de turismo em quaisquer destinos (BRASIL, 2008).

O termo ecoturismo teve origem na década de 1960, quando foi utilizado com o objetivo de demonstrar a relação entre os turistas, o meio ambiente e a cultura. Segundo Viana e Nascimento (2009, p. 3) "muito já foi escrito sobre ecoturismo, pouco é o consenso sobre o seu significado [...], praticadas por uma variedade ainda maior de tipos de turistas".

Sob esse enfoque, o ecoturismo assenta-se em um tripé: interpretação, conservação e sustentabilidade.

O princípio da sustentabilidade é definido como algo que vai além da dimensão ecológica, pois também a melhoria das condições econômicas, sociais das populações locais e a satisfação dos turistas.

Compreende-se que:

> 1. Sustentabilidade ecológica, entendida como a proteção da natureza e da diversidade biológica; portanto, o desenvolvimento turístico deve respeitar a "capacidade de suporte" dos ecossistemas, limitarem o consumo dos recursos naturais, e provocar o mínimo de danos aos sistemas de sustentação da vida;
> 2. Sustentabilidade social, fundamentada no estabelecimento de um processo de desenvolvimento que conduza a um padrão estável de crescimento, com uma distribuição mais equitativa de renda, redução das atuais diferenças sociais e a garantia dos direitos de cidadania;
> 3. Sustentabilidade cultural implica a necessidade de se buscar soluções de âmbito local, utilizando-se as potencialidades das culturas específicas, considerando a identidade cultural e o modo de vida local, assim como a participação da população local nos processos decisórios e na formulação e gestão de programas e planos de desenvolvimento turístico;

> 4. Sustentabilidade econômica, que assegure o crescimento econômico para as gerações atuais e, ao mesmo tempo, o manejo responsável dos recursos naturais, que deverão satisfazer as necessidades das gerações futuras;
> 5. Sustentabilidade espacial baseia-se na distribuição geográfica mais equilibrada dos assentamentos turísticos para evitar a superconcentração de pessoas, de equipamentos e de infraestrutura turísticas e, conseqüentemente, diminuir a destruição de ecossistemas frágeis e a deterioração da qualidade da experiência do turista (SACHS, 1993 *apud* SILVEIRA, 1997, p.90-91).

Os princípios de sustentabilidade podem propiciar modalidades para o desenvolvimento de turismo alternativo, ecoturismo, turismo verde, turismo leve, turismo responsável e turismo rural.

Rondonópolis oferece atrativos como paisagens de beleza cênica, naturais e culturais, destacando-se as cachoeiras e as prainhas. Assim, o ecoturismo caracteriza-se como uma importante atividade para o município "propiciando oportunidades para os visitantes vivenciarem manifestações da natureza e da cultura aprendendo sobre a importância da conservação da biodiversidade [...] e benefícios para as comunidades residentes" (MARULO, 2012, p.16).

Os atrativos naturais como ribeirões, córregos, morros, serras, formações rochosas, trilhas, cachoeiras, cascatas, quedas d'água que enriquecem as belezas naturais do município, são consideradas potenciais para a prática de vários segmentos do turismo como o ecoturismo e outros (BARBOSA, 2005).

De acordo com Ely (1998, p.89), esses componentes configuram a síntese da estruturação paisagística "[...] sustentam as formas que se desenvolvem no relevo, a qual proporciona o entendimento da composição da paisagem rondonopolitana". Essas características do relevo com suas formas são importantes para as atividades turísticas.

As atividades turísticas representam impactos sobre o ambiente, efetivados sem uma preocupação com a conservação das potencialidades naturais. O Município de Rondonópolis apresenta uma paisagem que foi ocupada e vem sendo utilizada de maneira desordenada, contribuindo para processos erosivos, assoreamento dos rios e para a diminuição da capacidade produtiva dos corpos hídricos. No **Quadro1,** identifica-se o uso e a ocupação da terra por unidade geomorfológica.

Com a presença de afloramentos rochosos, região de encosta e topos de morros formam um ambiente de grande riqueza, o que constitui um cenário propício ao ecoturismo. Na concepção de Santos "Contempla uma diversidade de características ambientais relacionadas com a estética e beleza cênica da paisagem, que se configuram em atrativos especiais aproveitados no desenvolvimento do turismo ecológico local" (SANTOS et al., 2007, p.01).

| COMPARTIMENTAÇÃO GEOMORFOLÓGICA | | | | | | | |
|---|---|---|---|---|---|---|---|
| Critérios | Morfoestrutura | Compartimentos Morfoesculturais | Unidades de padrões de formas semelhantes | Formas de Acumulação e Degradação | Arcabouço Geológico | Embasamento Pedológico | Ocupação |
| Compartimentos, Formas associadas, Arcabouço Geológico / Pedológico e Padrões Gerais de Ocupação do solo | Bacia Sedimentar do Paraná | 1 - Planalto Ocidental | A - Superfície de padrões de formas semelhantes. | "Dales" | Cobertura Detrito-Lateriticos | Latosolo vermelho-escuro álico | Agricultura |
| | | | B - Domínio de formas tabulares. | Escarpas "Dales" | Formações Furnas, Ponta Grossa e Bauru | Latosolo vermelho-escuro álico, areias quartzosas álicos e hidromóficos | Agricultura |
| | | | C - Domínio de formas suavemente convexizadas. | Escarpas | Formação Aquidauana | Latosolo vermelho-escurso álico, podzólico vermelho-amarelo entrófico, areias quartzosa áticos e litílicos álicos | Pecuária e Agricultura |
| | | | D - Depressão do Ribeirão Ponte de Pedra. | Escarpas e Estruturas Residuais | Formação Furnas | Podzólico vermelho-amarelo distrófico, litólicos distróficos e areias quartzosas áticos | Vegetação |
| | | 2 - Planalto Oriental | A - Domínio de formas de baixa convexidade. | Escarpas | Formação Ponta Grossa | Latosolo vermelho-escuro álico, podzólico vermelho-amarelo entrófico e areias quartzosas álicos | Pecuária e Ocupação Urbana |
| | | | B - Domínio de formas de média convexidade. | Escarpas, Morros Testemunho e Estruturas Residuais | Formação Ponta Grossa, Aquidauana | Podzólico vermelho-amarelo entrófico e cambisolos | Pecuária |
| | | | C - Domínio de formas de alta convexidade. | Escarpas, Morros Testemunhos e Cristais Isoclimais | Formação Aquidauana | Podzólico vermelho-amarelo entrófico | Pecuária |
| | | | D - Forma Residual de Serra Formosa. | ... | Formação Aquidauana | Solos litólicos e latosolo vandro-escuro álico | Vegetação |
| | | 3 - Planície do Rio Vermelho | A - Planície de inundação do Rio Vermelho/Tadarimana. | Terraços e Meandros Abandonados | Formação Ponta Grossa | Solos hidromórficos gleizados | Pecuária, Ocupação Urbana e Vegetação |
| | | | B - Planície de inundação do Rio Vermelho/Rio São Lourenço | Leques Aluviais, Terras e Meandros Abandonados | Depósitos Detriticos e Aluviões Atuais | Solos hidromórficos gleizados | Vegetação |

**Quadro 1 - Compartimentação Geomorfológica da Paisagem de Rondonópolis**

Fonte: Ely (1998, p. 66)

Org: CAMPOS, M. B. N. S. (2015)

As diversas categorias da compartimentação geomorfológica de Rondonópolis caracterizam seus vários atributos paisagísticos. O relevo articulado com as potencialidades dos recursos naturais define o desenvolvimento de modalidades turísticas.

Em função desta biodiversidade e riquezas naturais, o ecoturismo também colabora como uma atividade econômica extremamente promissora. Sendo assim, surge uma necessidade de buscar e apresentar ao mercado um produto diferenciado, incrementando as opções de atividades para os turistas.

A escolha do Município de Rondonópolis para aplicação desta pesquisa se deu pelo fato de possuir atrativos turísticos reconhecidos, e pela grande diversidade das cachoeiras e prainhas situadas no entorno da malha urbana (NARDES, 1997), o que atrai turistas de várias localidades.

## 1.1 PROBLEMA DE PESQUISA

As atividades turísticas nas localidades objeto de estudo atraem nos finais de semana grande fluxo de visitantes gerando impactos negativos ao ambiente.

A visita sem orientação das trilhas de acesso pode aumentar o desmatamento, a poluição, a compactação do solo, o escoamento superficial e a erosão (COSTA et al., 2008). Desse modo, o desenvolvimento do turismo em áreas naturais pode resultar numa série de impactos negativos.

Diante desses fatos, da dificuldade da viabilidade econômica e dos problemas em torno de questões ambientais e sociais, para a conservação e o desenvolvimento de atividades ecoturísticas, torna-se fundamental o estudo das possibilidades de compreensão e uma interação entre atividades turísticas e a conservação do atrativo natural, conforme o Ministério do Turismo (2010), o que podemos considerar um dos principais atrativos turísticos do Município de Rondonópolis. Assim, observa-se a necessidade e o interesse em conhecer as áreas de lazer em cada uma das localidades.

Percebe-se então, a importância da ocupação do espaço geográfico de forma ordenada, seguindo critérios conservacionistas baseados na legislação vigente e em critérios técnicos, principalmente nas áreas mais frágeis, uma vez que alterações como a extração da vegetação, podem causar grandes prejuízos a esse ambiente. O desenvolvimento de atividades turísticas é entendido como pressuposto da percepção de satisfação e auxilia juntamente com outras variáveis, na criação da imagem do atrativo ou destino turístico (MONDO e FIATES, 2015).

No intuito de analisar os atrativos turísticos do Município de Rondonópolis por meio do diagnóstico socioambiental, foram encaminhados questionários para moradores próximos das localidades das áreas de lazer, alguns frequentadores, instituições públicas e empresas envolvidas com o turismo local, bem como foram realizadas visitas técnicas *in loco*, buscando caracterizar as potencialidades turísticas.

## 1.2 PERGUNTA DA PESQUISA

Para alcançar os objetivos propostos o presente estudo parte do seguinte questionamento: O Município de Rondonópolis situado na porção sudeste mato-grossense apresenta recursos hídricos que podem desenvolver práticas de atividades do ecoturismo?

## 1.3 JUSTIFICATIVA

A atividade turística de modo consciente e planejado pode gerar alternativas de desenvolvimento local e regional. O turismo ecológico surge como uma opção de qualidade de vida para as comunidades (PORTUGUEZ et al., 2012). Áreas de lazer com potencialidade turística e valor histórico, cultural e econômico, como por exemplo, cachoeiras, prainhas ao longo do rio vermelho, podem também serem utilizadas para atividades turísticas.

De acordo com o Ministério do Turismo (2010), o uso dos recursos naturais relacionados com atividades ecoturísticas de forma sustentável, pode auxiliar as localidades receptoras na expansão de suas áreas de lazer.

Do ponto de vista acadêmico, espera-se que o estudo contribua para melhorar as discussões relacionadas ao ecoturismo, há que se ressaltar ainda que turismo é um dos temas transversais no ensino fundamental da disciplina Geografia nos Parâmetros Curriculares Nacionais – PCNs (SOUZA, 2008), no entanto, se verifica uma escassez de estudos relacionados às potencialidades e atividades turísticas.

Debater o turismo é uma iniciativa positiva, o ecoturismo precisa encontrar alternativa para estabelecer uma relação harmoniosa com os atributos físicos, mas, sobretudo, faz-se necessária uma relação harmoniosa com as comunidades receptoras (CANDIOTTO, 2011).

## 1.4 OBJETIVOS DO ESTUDO

### 1.4.1 Objetivo Geral

Compreender como o ecoturismo pode contribuir para a conservação dos recursos hídricos no Município de Rondonópolis-MT.

### 1.4.2 Objetivos Específicos

a) Caracterizar os atributos turísticos que contribuem com o ecoturismo no Município de Rondonópolis;

b) Identificar os tipos de lazer nas áreas selecionadas, estabelecendo um perfil socioambiental dos moradores das localidades, dos usuários frequentadores, das instituições públicas e agências de turismo receptivo;

c) Analisar os diferentes espaços turísticos por meio dos parâmetros socioambientais.

Para alcançar os objetivos definidos, de acordo com a metodologia elaborada para o estudo, a pesquisa está dividida em cinco capítulos: na primeira parte enfatiza-se a temática considerando sua visão introdutória, bem como sua importância na investigação, trata-se da apresentação de conceitos e significados importantes, os quais permitem a redefinição do objeto e do papel da Geografia em relação turismo e ecoturismo.

O capítulo segundo traz diferentes abordagens da construção teórica, bem como diversos recortes espaciais buscando a interdisciplinaridade com as diversas ciências que abordam a temática em questão.

O capítulo terceiro apresenta tanto os procedimentos metodológicos, como a estrutura conceitual para abordagem sistêmica e multidisciplinar aplicada ao estudo do ecoturismo e dos recursos hídricos do Município de Rondonópolis.

O capítulo quarto tece uma discussão das potencialidades ecoturísticas existente no município de Rondonópolis cujos eixos são resultantes da conjunção de quatro principais elementos: cachoeiras, infraestrutura, localização e práticas desenvolvidas. A sinergia entre esses elementos proporciona condições favoráveis para o desenvolvimento do ecoturismo.

Os capítulos que são apresentados nesta dissertação contemplam estudos significativos da temática teórica e empírica. Este estudo é um convite essencial na atual conjuntura para a tomada de decisões, alerta a respeito da importância do ecoturismo no município e sinaliza quais preocupações devem direcionar seu processo de implementação.

# 2 FUNDAMENTAÇÃO TEÓRICA

## 2.1 TURISMO: Histórico e Perspectivas

As primeiras informações históricas sobre povos descrevem o gosto de viajar, iniciação científica e a caracterização de diversos elementos que ordenam o espaço.

> Os romanos podem ser considerados os primeiros a viajar por prazer. Diversas pesquisas científicas (análise de azulejos, placas, vasos e mapas) revelaram que o povo romano ia à praia e a centros de rejuvenescimento e tratamento do corpo, buscando sempre divertimento e relaxamento (SOUTO MAIOR, 1990, p.138).

Ainda segundo Souto Maior (1990) o marco histórico deste processo começou com o turismo na Grécia no século VIII a.C., período no qual as pessoas viajavam para ver os jogos olímpicos. O grego Heródoto foi o primeiro historiador da humanidade que viajou por várias regiões do seu país, visitou até mesmo o Egito, relatou atributos naturais e culturais, com intuito de ampliar seu conhecimento.

No Antigo Testamento há relação entre os povos mesopotâmicos, no Egito e na China as viagens de prazer, de aventura ou de descanso se davam entre os reis e os faraós.

> A partir das grandes civilizações clássicas, como a Grécia e Roma, que as viagens foram gradativamente tomando maior relevância. Os romanos exerceram um papel fundamental nas viagens de lazer, prazer e comércio, muitas estradas foram construídas pelo Império Romano, possibilitando e determinando que os cidadãos viajassem entre o século II a.C. e o século II d.C. (SILVA e KEMP, 2008, p.02).

Neste sentido, aponta-se a importância do povo romano no desenvolvimento do turismo, sendo estes os primeiros a viajar por prazer, conhecimento, cultura e diversão. Entre os séculos VII e IX houve um processo de expansão da atividade turística, relacionado com as festas populares, como a da primavera, da colheita ou as festas religiosas (BADARÓ, 2010).

Este autor explica que no princípio a busca por alimentos, ou seja, a própria sobrevivência foi um dos primeiros motivos geradores do turismo, com o passar do tempo, outros elementos foram sendo considerados, como motivos econômicos, sociais, políticos, culturais e esportivos, na Grécia Antiga, porém, já ocorria o turismo cultural e o esportivo a partir dos jogos olímpicos.

Grande parte do que se tem no mundo moderno deve-se a inspiração da Grécia antiga, é dela que se recebe, imita, transforma, parodia ou recicla muitos hábitos, técnicas e expressões artísticas atuais. Conhecida como a casa, a Grécia acaba por ser o lar original de toda a civilização ocidental.

As relações que se estabelecem com outros tempos e espaços, além de valorizar o patrimônio sociocultural e respeitar a diversidade, devem ser reconhecidos como um direito de todos os povos e indivíduos.

> [...] o turismo cultural desempenha o mesmo papel em mostrar a valorização de patrimônios, respeito à diversidade no propósito de deslocar o indivíduo á encontros artísticos, científicos, de formação e informação através do tempo e do espaço; portanto a história dá as informações e mostram os fatos enquanto o turismo se ocupa em transladar o viajante a descobrir tudo que há de bom ou ruim na História, fazendo-o conhecer costumes e peculiaridades de épocas e civilizações distantes; e como e quantos são influenciáveis em nossos dias atuais (GUERRA et al., 2008, p.03).

A evolução dos meios de transporte e o surgimento de outras formas de deslocamento como as viagens ultramarinas foram os fatores, que propiciaram o aumento das viagens turísticas na segunda metade do século XV e todo o século XVI, promovendo desenvolvimento comercial entre os povos e originando uma produção capitalista (BADARÓ, 2010).

Na interpretação de Fourastié (1979), o ser humano evoluiu a partir do contato com os demais povos mesopotâmicos, em alguns casos longe de sua localidade. Os primeiros relatos históricos apontam para a origem do turismo somente no século XIX, ainda que as antigas civilizações tivessem diferentes formas de ações turísticas. O desenvolvimento do turismo trouxe consigo não apenas o desenvolvimento da produção e comércio, mas gerou também dinâmicas espaciais, instrumentos legais, infraestrutura técnica, equipamentos e serviços. A exemplo:

> - Legislação social que determina o tempo de trabalho, o repouso semanal e férias anuais remuneradas;
> - O crescimento demográfico, provocando maiores concentrações urbano e a busca do repouso e recuperação física e mental, levando o homem ao campo;
> - O progresso nos transportes em geral;
> - O incentivo criado por alguns governos objetivando a geração de maiores fluxos turísticos;
> - O desenvolvimento de equipamentos e a criação de novos serviços na oferta turística (ROSE, 2002, p.6).

O turismo está entre as quatro atividades econômicas mais importantes, responsável por aproximadamente 10% do Produto Interno Bruto (PIB) mundial. Diante dessa magnitude econômica, a OMT prevê um crescimento entre 4% a 5% ao ano, podendo atingir cifras de 6,7% no ano de 2020, enquanto a estimativa para as demais atividades da economia apresentam um índice de crescimento inferior a 3% no mesmo período (BARTELÓ, 2000 *apud* OLIVEIRA, 2008).

Atualmente, o turismo é uma atividade socioambiental que pode representar importante alternativa para melhoria da qualidade de vida dos sujeitos nela envolvidos conforme Sousa (2011, p.17): "[...] para que isso ocorra é preciso que sejam implantadas ações planejadas com o intuito de trazer benefícios sociais, econômicos e culturais, através do respeito ao patrimônio natural das localidades turísticas".

As belezas naturais e o patrimônio cultural histórico simbolizam abundante potencial turístico para Mato Grosso, Estado que é contemplado com atributos naturais, os quais podem ser ofertados em forma de produtos turísticos em muitos municípios mato-grossenses.

## 2.2 ECOTURISMO

O ecoturismo, praticado em locais com atrativo ecológico, deve ser praticado de maneira sustentável procurando articular turismo com ambiente, condições culturais e recursos naturais.

Essa compreensão é reafirmada por Schneider (2006), o autor coloca que conservar o meio ambiente e desenvolver atividades de turismo não podem ser elementos separados no ecoturismo, ao contrário estão fortemente interligados. Isto porque não existe ecoturismo quando ocorrem prejuízos para os recursos naturais ou um processo de degradação ambiental, pois seu objetivo não pode ser apenas o lucro econômico da comunidade em que está inserida, mas esta atividade deve estar alicerçada no princípio da sustentabilidade.

Neste sentido, Hetzer (*apud* FENNELL, 2002) identificou quatro aspectos que devem ser respeitados para que haja uma convivência harmônica entre o turista e o meio ambiente: (1) impacto ambiental mínimo; (2) impacto mínimo às culturas anfitriãs; (3) máximos benefícios econômicos para as comunidades do país anfitrião; e (4) satisfação recreacional máxima para os turistas participantes.

De acordo com Brasil (2008), o termo ecoturismo foi introduzido no país somente no final da década de 1980, quando houve a valorização do meio ambiente e o reconhecimento de que seria necessário usar racionalmente os recursos naturais.

Nesse mesmo ano, também foram autorizados os primeiros estudos para inclusão do ecoturismo como iniciativa para direcionar e ordenar o segmento das atividades turísticas.

> Ainda na mesma década, foram autorizados os primeiros cursos de guia de turismo especializados, porém, foi na década seguinte, com a Conferência das Nações Unidas para o Meio Ambiente – ECO 92, realizado em 1992 no Rio de Janeiro, que esse tipo de turismo ganhou visibilidade e impulsionou um mercado com tendência de franco crescimento, propondo diretrizes e tratadas com aplicação de âmbito mundial, a partir da aceitação ou consignação de cada nação (MINISTÉRIO DO TURISMO, 2010, p.14).

No ano de 1994, o Ministério do Meio Ambiente trouxe como Política Nacional de Ecoturismo o "turismo ecológico", o qual passou a ser considerado um segmento da atividade turística que, segundo Brasil (2008, p.31): "[...] faz uso, de forma sustentável, do patrimônio natural e cultural, incentiva sua conservação e busca a formação de uma consciência ambientalista por meio da interpretação do ambiente, promovendo o bem-estar das populações".

Importante salientar, ainda conforme Brasil (2008), que o território brasileiro possui uma rica biodiversidade, com diferentes biomas como a Amazônia, Mata Atlântica, Campos Sulinos, Caatinga, Cerrado, Pantanal, Zona Costeira e Marítima, bem como, os milhares de ecossistemas, têm qualidades que permitem o desenvolvimento do ecoturismo, porém necessita de desenvolver mecanismos eficazes para a conservação e proteção dos mesmos.

Autores como Rodrigues e Amarante-Júnior (2009) enfatizam que o ecoturismo deve constituir uma maneira turística sustentável, não apenas o uso dos recursos naturais para atrair turistas em uma localidade, mas ações de conservação do patrimônio natural e cultural. Para eles ecoturismo se objetiva nos seguintes:

> - Vincula-se a um tipo de uso que minimiza os impactos negativos no meio ambiente e na população local.
> - Aumenta a consciência e a compreensão em relação aos sistemas naturais e culturais da área e o conseqüente envolvimento dos visitantes nas questões que afetam esses sistemas.
> - Contribui à conservação e à gestão de áreas legalmente protegidas e outras naturais.
> - Maximiza a participação prévia e em longo prazo da população local nas decisões que determinam o tipo de quantidade de turismo a ser implantado.

> - Direciona os benefícios econômicos e outros tipos de benefícios à população local, que pode assim complementar a renda decorrente das práticas tradicionais em vez de eliminar ou substituí-las (agricultura, a pesca, os sistemas sociais).
> - Oferece oportunidades especiais para a população local e os funcionários do turismo na natureza de utilizarem e visitarem as áreas naturais e aprenderem mais sobre aquelas maravilha que os outros visitantes vêm conhecer (WALLACE e PIERCE, 1996 *apud* FENNELL, 2002, p.49).

Estes princípios trazem entendimento de que o ecoturismo não é somente a relação homem-turista/recursos ambientais, mas um processo de sensibilização de conservar os recursos ambientais nos paradigmas sociais, culturais, econômicos e ambientais de forma sustentável.

Ainda sobre os princípios do ecoturismo, segundo Tourism Concern (1992) dez podem ser mencionados: 1. Usar de forma sustentável os recursos; 2. Reduzir o desperdício dos recursos disponíveis no meio ambiente; 3. Perpetuar a diversidade; 4. Integrar turismo/planejamento; 5. Apoio à economia do local em que é desenvolvido o ecoturismo; 6. Envolvimento da comunidade local; 7. Consultar investidores e o público para conciliar toda a comunidade local; 8. Treinar a equipe que irá integrar o turismo sustentável; 9. Elaborar um marketing consciente e 10. Realizar pesquisas sobre os problemas e benefícios que possam advir na comunidade pelo ecoturismo.

De acordo com os apontamentos de Silva, Santos e Benevides (2007), problemas causados pelo ecoturismo são ainda mais evidentes em países subdesenvolvidos, como o Brasil, em que o uso inadequado dos recursos naturais em nome do crescimento turístico e econômico, é causador de problemas ambientais, especialmente, os hídricos.

As potencialidades nas áreas de lazer no Município de Rondonópolis trazem elementos que apontam para a viabilidade da exploração do turismo sustentável, com o aproveitamento dos recursos hídricos, fauna e flora, o que estimula a atividade turística.

Sobre este tema, tem-se:

> A atividade turística, entendida não somente como atividade econômica, mas como prática social complexa e multifacetada, implica essencialmente o deslocamento de pessoas e a relação dessas pessoas entre si, com a comunidade e com o lugar visitado (MARINHO; BRUHNS, 2003, p.54).

O ecoturismo, por apresentar sua base de desenvolvimento na sustentabilidade, deve considerar a gestão dos recursos naturais, como abordado a seguir:

> Os princípios e os critérios para o desenvolvimento do segmento devem considerar a gestão socioambiental dos recursos naturais, para que os impactos positivos do Ecoturismo sejam maximizados, e os negativos sejam minimizados

> na esfera ambiental, social e econômica, em especial aos que estão relacionados aos sítios turísticos naturais no Brasil, e àqueles relacionados às Unidades de Conservação que permitem a visitação pública (MINISTÉRIO DO TURISMO, 2010, p.16).

Sintetizam com bastante propriedade essas consequências, o estabelecimento de parcerias e formação de redes para a inserção e a participação da comunidade local são pontos fundamentais:

> É de fundamental importância a articulação interinstitucional para que se possam estabelecer parcerias entre todos os atores sociais do turismo - governo, empresários, organizações não governamentais, instituições de ensino e pesquisa, e associações comunitárias, - para que sejam realizadas ações, como planejamento participativo, ordenamento, monitoramento, implantação de infraestrutura e qualificação profissional para a efetiva inserção de produtos no mercado turístico (MINISTÉRIO DO TURISMO, 2010, p.67).

A **Figura 01** aborda os aspectos da demanda turística, referentes aos processos de articulações interinstitucionais entre diversos atores e setores dos destinos de ecoturismo (MINISTÉRIO DO TURISMO, 2010).

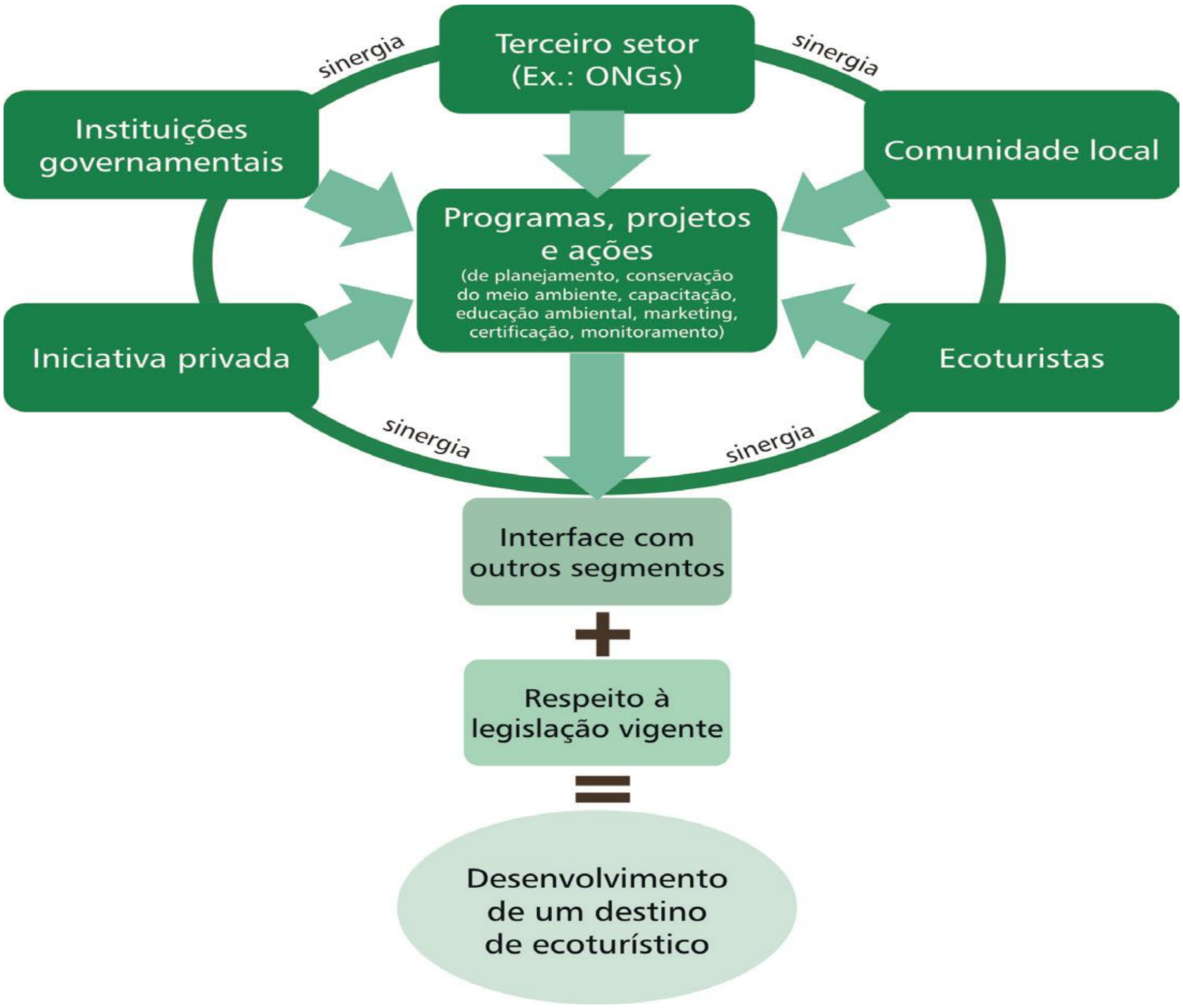

**Figura 1 - Inter-relações entre os diversos atores e setores dos destinos de Ecoturismo**
Fonte: Ministério do Turismo (2010)

Partindo dessa premissa, faz-se importante ter conhecimento dos destinos do ecoturismo e os atores sociais, como se deve agir e interagir com as atividades turísticas sustentáveis nas localidades.

Na implantação de um projeto para as atividades turísticas, para Milaré (2004), somente a partir da conservação dos recursos naturais, é possível que os atores envolvidos tenham qualidade de vida e possam proteger a fauna e flora

De acordo com Oliveira e Ricco (2013), o conhecimento das potencialidades turísticas e suas limitações pelas comunidades é um processo demorado, porém representa a essência de um desenvolvimento sustentável.

## 2.3 POSSIBILIDADES

Como foi anteriormente comentado neste estudo, o ecoturismo traz diversas possibilidades, inclusive para as comunidades mais distantes dos grandes centros, em que as oportunidades de trabalho são restritas, viabilizando o crescimento econômico e social como aponta (BUDOWSKI, 2001 *apud* DIAS, 2003).

Silva, Santos e Benevides (2007) reiteram que o ecoturismo surge como um aporte econômico e social. No entanto, é preciso fazer uma análise detalhada das possibilidades e limites das áreas que serão utilizadas para o ecoturismo, pois, em alguns casos, o desenvolvimento desta atividade volta-se unicamente para os benefícios econômicos e não respeita verdadeiramente os limites naturais, históricos e culturais dos espaços.

Na visão de Machado e Souza (2012) toda e qualquer forma de turismo traz mudanças no ambiente em que é realizado, sendo que algumas destas transformações são positivas, como a geração de empregos, diretos ou indiretos, melhorias na qualidade de vida da população local, proporcionando um processo de desenvolvimento, especialmente, econômico.

O ecoturismo é uma proposta sustentável para o desenvolvimento socioeconômico, cultural e ambiental. Assim desenvolvendo as comunidades locais, para melhoria efetiva da qualidade de vida, nesta mesma linha de pensamento no que diz respeito aos impactos positivos do turismo sobre o meio ambiente considera-se que:

> O desenvolvimento turístico em ambientes naturais apresenta algumas vantagens que basicamente se referem a: criação de planos e programas de conservação e preservação de áreas naturais, de sítios arqueológicos, e ainda de monumentos históricos, os empreendedores turísticos passam a intervir nas medidas preservacionistas, a fim de manter a qualidade e consequentemente a atratividade dos recursos naturais e socioculturais, promove-se a descoberta e a acessibilidade de certos aspetos naturais em regiões não valorizadas, a fim de se desenvolver o seu conhecimento através de programas especiais, ecologicamente percebe-se uma utilização mais racional dos espaços e a valorização do convívio direto com a natureza (RUSCHMANN, 1997, p.56).

A questão é que o desenvolvimento de qualquer atividade econômica traz elementos positivos e negativos. Esta realidade não é diferente com o ecoturismo, em que ocorrem impactos ambientais como a redução ou extinção de recursos naturais em algumas localidades, por isso, conforme Marulo (2012) é essencial que este tipo de turismo seja desempenhado a partir de um planejamento e gerenciamento adequado dos recursos, evitando impactos negativos e estimulando os fatores positivos.

Quanto aos benefícios que podem advir com o ecoturismo para a comunidade local, se observam os seguintes fatores:

> - Econômicos: geração de emprego, renda e estímulo ao desenvolvimento econômico em vários níveis (local, regional, estadual e nacional);
> - Baixo impacto do ecoturismo: melhoria do nível sociocultural das populações locais.
> - Valor agregado às áreas protegidas: sensibilização de turistas e populações locais para a proteção do ambiente, [...].
> - O ecoturismo promove um manejo melhor das áreas protegidas: ampliação dos investimentos voltados à conservação de áreas naturais e bens culturais, [...].
> - O aumento do ecoturismo influencia, favoravelmente, as atividades educativas e científicas: Muitos ecoturistas são profissionais especializados em meio ambiente que realizam atividades econômicas potencialmente sustentáveis [...].
> - A conservação da biodiversidade é favorecida: o ecoturismo contribui para atrair a atenção sobre as espécies em perigo de extinção e facilitar sua conservação. [...].
> - Combinações vantajosas com o turismo rural e outras formas de turismo: O ecoturismo pode ser praticado combinando-se com outras formas de turismo, [...] (BUDOWSKI, 2001 *apud* DIAS, 2003, p.118).

Como aponta Lima (2004), os benefícios do ecoturismo podem sobressair aos prejuízos desde que os ecoturistas tenham a preocupação em planejar suas viagens e consciência de que devem proteger os recursos ambientais, de forma a harmonizar as atividades de ecoturismo com a preservação dos recursos, como é o caso dos hídricos.

## 2.4 LIMITES

Quando se observa a atividade turística, pode-se perceber que da mesma forma que traz benefícios, pode também causar problemas e a gerar impactos negativos. Rodrigues (2003, p.29) defende que: “[...] em todo movimento e ideia que tenha um apelo ambientalista, nem sempre há real comprometimento com a essência do conceito. Tem sido usado indiscriminadamente, com fins legítimos e não espúrios”. Percebe-se ainda, quando o turismo utiliza-se da expressão “eco” como forma de dissimular os problemas causados para o meio ambiente.

Alvarenga (2011) considera que, a prática do ecoturismo vem sendo avaliada como sustentável, mas para isso deve ser implantada a partir de planejamento e gestão de projetos, que viabilizem o uso racional dos recursos. Porém, o fato do ecoturismo ser realizado no ar, na água ou na terra e de ter a capacidade de transformar os ambientes, pode evidenciar que mesmo o ecoturismo não traz somente potencialidades, mas limites os quais devem ser observados e respeitados.

Para o autor acima, o ecoturismo deve ser desenvolvido com base em um planejamento que permita o uso adequado dos recursos e, desta forma, as gerações presentes podem utilizar-se dos recursos ambientais sem destruí-los, legando um local com memória biológica e ecológica para as próximas gerações e não apenas lembranças do que fora a exuberância da natureza.

Em diversas regiões brasileiras o ecoturismo, assegura Rotta, Luna e Weis (2006), trouxe e traz desenvolvimento, o que possibilita por vezes minimizar problemas como a miséria e falta de oportunidades nos povoados distantes, inclusive para as populações ribeirinhas. No entanto, o ecoturismo causa o temor de que possa desestruturar o ambiente, prejudicando os ecossistemas e biomas existentes.

Relatam de forma consciente Rotta, Luna e Weis (2006), que se ocorre um processo de destruição dos recursos ambientais, culturais ou sociais da localidade, o ecoturismo deixa de ser sustentável para transmutar-se em uma forma devastadora de transformação do meio ambiente.

Nesta percepção, o turismo tem gerado riqueza para as comunidades, mas, por outro lado, devido à falta de planejamento, o mesmo tem gerado as seguintes externalidades:

degradação ambiental, crescimento desordenado, especulação imobiliária e ocupação urbana desordenada.

No que diz respeito aos impactos negativos que se originam a partir do ecoturismo, Lima (2004) elenca: descaracterização do artesanato; vulgarização das manifestações tradicionais; arrogância cultural; destruição do patrimônio histórico; aumento do custo de vida; supervalorização dos bens imobiliários e consequente perda da propriedade de terras, habitações e os meios de produção por parte das populações locais; substituição de ocupações tradicionais por subempregos; esgotamento do solo e transformação negativa da paisagem pela implantação de construções e infraestrutura, além de poluição, ruído e depredação.

Conforme explica Moraes (2000), os impactos negativos que possam se originar no ecoturismo não apenas se relacionam apenas aos problemas causados ao próprio meio ambiente, mas também à comunidade local. Comenta ainda que existem diversos elementos que podem ser relacionados com o ecoturismo, por exemplo:

> - Conflitos entre usuários dos recursos e a Posse de Terra: este tipo de restrição refere-se ao conflito entre o ecoturismo e a recreação ao ar livre e as indústrias de extração de recursos, tais como as madeireiras, companhias mineradoras e a própria agricultura. [...].
> - Infra-estrutura (*sic*) Precária: A ausência de estradas, docas, aeroporto e fornecimento de água podem ser impedimentos importantes, mesmo para os ecoturistas mais aventureiros. [...].
> - Má imagem da sua Região: Algumas regiões do Brasil têm recebido uma cobertura jornalística extensa por parte dos outros países, por causa de suas práticas de desflorestamentos. Os ecoturistas estão realmente preocupados sobre as questões ambientais, mas se recusam a visitar regiões com imagem negativa. [...].
> - Acomodações de Má Qualidade: As maiorias das regiões têm serviços acomodações. Contudo, se sua região não está inserida no mercado internacional do turismo, as suas acomodações podem ser interessantes somente aos turistas nacionais. [...] (MORAES, 2000, p.70).

Neste sentido, tem-se que o ecoturismo pode trazer boas oportunidades para a comunidade local, desde que suas potencialidades sejam exploradas de forma sustentável, tendo no planejamento e gestão das atividades turísticas os principais focos de efetiva realização da sua exploração.

## 2.5 ATRATIVOS TURÍSTICOS

Os atrativos turísticos são componentes fundamentais para o produto turístico, pois é o elemento que determina a seleção do potencial das atividades turísticas, por parte do turista do local de destino de uma viagem até a localidade.

As atividades turísticas em uma determinada região a participação das comunidades locais são importantes no sentido de promover a comercialização e implantação dos equipamentos de apoio ao turista (MINISTÉRIO DO TURISMO, 2010).

A obediência às leis ambientais leva a mudança de valores e atitudes frente às questões ambientais. Bordest (1998, p. 79) observa que "[...] recursos naturais (climáticos, hídricos, de relevo, flora e fauna) a sua maior riqueza turística, além dos patrimônios históricos e culturais (danças, folclores, gastronomia, festivais, monumentos históricos, tradições)", são o que demonstra potencialidade para as atividades turísticas de uma localidade.

Os atrativos turísticos dependem de vários serviços e equipamentos de apoio para atender à demanda e as necessidades dos visitantes que possa ter maiores aproveitamentos das localidades.

> Para que o turismo seja aceito pelo visitante, toda a rede de serviços necessita funcionar e estar à disposição dele, pois alguns turistas buscam o conforto semelhante ao que possuem em seu dia a dia e outros buscam aventuras diferentes do seu cotidiano. Independentemente do tipo de atividade que o turista busca, são necessários para o maior fortalecimento das atividades turísticas os equipamentos e serviços turísticos e a infraestrutura de apoio ao turismo (DAL MEDICO, 2012, p.42).

Gloeden aponta que (2010, p. 13) os equipamentos de apoio e serviços oferecidos por uma localidade aos seus consumidores devem satisfazer suas necessidades, portanto, "infraestruturas básica e turística são compreendidas como bens e serviços ofertados e trabalhados em conjunto, para suprir de forma complexa, as expectativas dos consumidores".

O **Quadro 2** apresenta os quatro elementos dos componentes da oferta turística de uma localidade que podem interagir com as atividades ecoturísticas.

| COMPONENTES DA OFERTA TURÍSTICA | | | |
|---|---|---|---|
| FÍSICA | COMERCIAL | RECURSOS HUMANOS | RECURSOS FINANCEIROS |
| Aeroportos<br>Rodovias<br>Telecomunicações<br>Água potável e esgoto<br>Eletricidade<br>Sinalização<br>Centro de informações<br>Serviços médicos | Centros comerciais<br>Praças turísticas<br>Mercados de artesanato<br>Feira de exposições<br>Eventos comerciais<br>Praças comerciais | Universidades<br>Escolas de turismo<br>Escolas de idiomas<br>Centros de capacitação<br>Disponibilidade de recursos humanos<br>Disponibilidades de guias turísticos | Investimento público<br>Investimento privado<br>Fundos turísticos<br>Fundos de financiamento setorial<br>Programas financeiros, federais e estaduais |

**Quadro 2 - Indústrias e Setores Fornecedores**
Fonte: México (País) Secretaria de Turismo (2002, p.56)
Org: CAMPOS, M. B. N. S. (2015)

A prática do turismo está relacionada com as atrações e possibilidades, estes elementos são tratados por Bordest (2005) como fonte de recursos para produção dos tipos de turismo que atendam às necessidades e desejos humanos, desde que os lugares de turismo ofereçam os atrativos turísticos que atendam os interesses do sujeito receptor.

Nesse sentido, alguns aspectos influenciam ou mesmo permitem que o espaço seja considerado vocacionado para as atividades ecoturísticas considerando a existência de atrativos naturais como: equipamentos de apoio, serviços, infraestrutura, instituições setoriais, demanda, comunidade e outros.

O desenvolvimento desses espaços de forma sustentável, de atrativos turísticos motiva os visitantes em conhecer determinada localidade e de acordo com Pelisson:

> É necessária a existência de equipamentos e serviços de qualidade e de infraestrutura básica que permita a visitação turística em um local, por um determinado tempo. Para suprir as necessidades desta demanda real (ou da que se espera) é necessário, dispor, além da oferta original (atrativos), uma oferta agregada diversa (hotéis, restaurantes, entretenimento, transporte, dentre outros serviços) (PORTUGUEZ e PELISSON, 2014, p.7).

Nesse sentido, a orientação dessas ofertas serve para a inserção das comunidades nesses atrativos, associando a estratégia como fator principal para a sua estruturação, comercialização e roteiros ecoturísticos.

> Os produtos de ecoturismo apresentam peculiaridades que vão desde a escolha da área natural, a identificação da legislação ambiental pertinente, a seleção de atrativos naturais a serem ofertadas, as atividades contempladas, até a aplicação de um marketing responsável, associado à promoção e comercialização, observando-se o caráter ecológico – que ampliam as reflexões ambientais e a interpretação socioambiental com inserção das comunidades locais receptoras (MINISTÉRIO DE TURISMO, 2010, p.12).

Em função da rica biodiversidade é geralmente identificado o potencial mato-grossense para o ecoturismo. Mato Grosso destaca-se como um estado privilegiado por essas riquezas que se misturam entre cavernas, cachoeiras, trilhas, sítios arqueológicos e árvores retorcidas, na percepção de alguns turistas frequentadores do lugar.

### 2.5.1 Características do Cerrado

O bioma do Cerrado ocorre como área contínua no Brasil central e independente na Floresta Amazônica, Caatinga, Floresta Atlântica, Pantanal e Floresta de Pinheiros no sul do Brasil. Conforme interpretação de Marinho Filho (2010, p. 15), "o nome, a distribuição e o conceito do bioma cerrado sofreram várias modificações, desde que essa região foi designada pela primeira vez como uma unidade fitogeográfica com o nome de Oréades por Martius et al. (1840)".

Segundo explica Tesoro (1993), o Município de Rondonópolis contempla uma área coberta por cerrados e cerradões, com predominância de terras com produção agropastoril, sua vegetação é marcante e constitui um manancial turístico que pode ser explorado, desde que seja uma atividade consciente.

O cerrado é constituído de várias formações herbáceas graminosas contínuas, em geral cobertas de plantas lenhosas, cuja riqueza florística só é menor em número de espécies que a floresta tropical úmida (MORENO e HIGA, 2005). Esta condição permitiu a sua classificação nos seguintes termos: campo limpo, campo sujo, campo cerrado, cerrado e cerradão.

A partir desta concepção, Marinho Filho et al., argumenta:

> O bioma cerrado pode ser considerado uma unidade biológica que apresentou um desenvolvimento e evolução própria, evidenciado pelo grau de endemismo de animais e plantas restritos a sua área. A heterogeneidade e complexidade da distribuição das comunidades vegetais resultam da interação de diversos fatores, com o fogo, água, solo e processos históricos do passado, que atuam em diferentes escalas de tempo e espaço. Baseado na idéia de Eiten (1972) de que o cerrado seria mais bem representado pelo conceito de clímax-gradiente de Whitaker (1953), Henriques (2005) (MARINHO FILHO et al., 2010, p. 21).

Tem-se ainda que existam fatores determinantes para explicar a ocorrência de todos os tipos de vegetação existentes no bioma cerrado e os processos adaptativos associados.

Segundo a EMBRAPA (1998), a definição e uniformização de nomenclatura dos tipos fisionômicos do cerrado são muito discutidas por estudiosos e pesquisadores. Todavia, predomina o entendimento, em sua formação, desenvolvida pelo IBGE (1997), segue a classificação para as formações do cerrado de Mato Grosso:

- Cerradão (Savana florestada ou Savana densa);
- Campo Cerrado (Savana arborizada ou savana arbórea aberta);
- Parque de Cerrado ou Campo Sujo (Savana parque);
- Campo Limpo (savana gramínea lenhosa).

As características do bioma Cerrado de acordo com Moreno e Higa (2005) ocupavam 38,29% da cobertura original do Estado, recobrindo as depressões do Alto Paraguai – Guaporé, o Sul do paralelo 13º, até os limites com Mato Grosso do Sul.

Pelo avanço das fronteiras agrícolas do país, as paisagens do Cerrado foram cedendo espaços para a implantação de paisagens agrícolas, Sturza (2008, p. 1-2) afirma, “[...] este é o caso de Rondonópolis, onde as atuais paisagens agrícolas aniquilaram os antigos lugares e reduziram brutalmente a diversidade ecológica”. Portanto, reforça e configura a caracterização geográfica do Cerrado de Rondonópolis.

Quanto ao Cerrado de Rondonópolis e a sua ocupação pela agricultura, Silva (2003) argumenta que, no final de 70 do século XX, ocorreram às transformações no espaço agrário do sudeste mato-grossense sob a égide do processo de integração econômica do território brasileiro:

> No contexto do II PND, o PRODOESTE e o POLOCENTRO (Programa de Desenvolvimento dos Cerrados), criado em 1975, serviram de ponto de partida para a incorporação do cerrado ao processo de modernização conservadora da

> agricultura brasileira. Tais instrumentos de política territorial garantiram financiamentos para a criação de suporte físico e tecnológico adequado às transformações do cerrado. Com efeito, recursos financeiros foram canalizados para abertura de estradas vicinais, armazéns e silos, infra-estrutura de pesquisa, usinas de beneficiamento, frigoríficos, distritos industriais e linhas de crédito rural. Na verdade, a estratégia era viabilizar a rápida inserção de áreas de cerrado, previamente desmatadas, no complexo agroindustrial brasileiro, já concentrado no eixo Sul-Sudeste (SILVA, 2003, p. 66).

No entendimento de Silva (2003), o desenvolvimento econômico trouxe para a região mato-grossense aspectos positivos, mas também fatores que geraram ônus, especialmente no que se refere ao uso dos recursos naturais.

Nesta perspectiva, a expansão agrícola com a abertura de estradas, armazéns e silos alterou a paisagem do cerrado. A partir da década de 70, essas propostas estratégicas desenvolvimentistas começam a emergir na dinâmica estadual do centro-oeste, principalmente o cerrado mato-grossense e a região de Rondonópolis, o que ocorre também com o turismo.

### 2.5.2 Cerrado e os Impactos Ambientais do Turismo

Os impactos ambientais sobre o cerrado ocorreram em consequência do processo ocupacional, caracterizado por presença de empresas agropecuárias e forte urbanização, o que desencadeou intenso processo de desmatamento e queimadas, alterando os domínios biogeográficos, principalmente o cerrado (cf. SCHWENK, 2013).

Atividades turísticas podem acarretar esses impactos numa localidade dessa forma, é importante para as localidades receptoras conhecimento, no sentido de impedir agressões negativas, principalmente, no que diz respeito à desvalorização cultural.

Os turistas, afirma Dal Medico (2012, p.5), podem causar danos nas localidades receptoras:

> O turismo pode trazer à comunidade uma valorização de seu patrimônio, atraindo visitantes que queiram conhecer a sua localidade, explorando os seus hábitos, costumes e experiências de vida. Contudo, ainda estão em estudo às reais conseqüências que o turismo e a própria comunidade podem causar a sua cultura, pois o uso indevido dela e o turismo de massa acabam destruindo o patrimônio da localidade e, em parte, alteram a rotina cultural dos habitantes que também não estão acostumados com os hábitos turísticos.

De acordo com este autor, as atividades turísticas devem valorizar os aspectos econômicos, sociais e culturais de uma localidade, buscando a valorização dos atrativos naturais, e impedindo agressões que podem ser provocadas por essa atividade.

O **Quadro 3** exemplifica impactos socioambientais, provocados pela atividade turísticas de uma localidade:

| FATORES ASSOCIADOS COM O TURISMO | IMPACTOS POSITIVOS | IMPACTOS NEGATIVOS |
|---|---|---|
| O uso da cultura como atração turística<br><br>Contatos diretos entre turistas e moradores<br><br>Mudanças na estrutura econômica e papéis sociais<br><br>Desenvolvimento de infraestruturas<br><br>Aumento da população de Turistas | Revitalização das artes tradicionais, festivais e línguas. Acréscimo das culturas tradicionais. Ruptura dos estereótipos negativos.<br><br>Aumento das oportunidades sociais.<br><br>Maiores oportunidades econômico-sociais.<br><br>Diminuição das desigualdes sociais<br><br>Aumento das oportunidades de lazer.<br><br>Melhora das condições sanitárias, educação e melhora da qualidade de vida. | Mudança nas atividades tradicionais. Invasão da privacidade.<br><br>Aumento da comercialização. Introdução de doenças. Efeito demonstração.<br><br>Conflito e tensão na comunidade.<br><br>Perda da linguagem.<br><br>Perda de acesso às atividades de recreio e lazer.<br><br>Congestionamento, multidão, aumento da criminalidade. |

**Quadro 3 - Resumo dos Impactos Socioculturais do Turismo**
Fonte: MARULO (2012)
Org: CAMPOS, M. B. N. S. (2015)

Quanto aos impactos ambientais do turismo sobre ambientes naturais, Cruz (2001, p. 30) considera algumas premissas básicas:

- Não há atividade humana que não interfere nos ambientes, de alguma forma [...].
- Nem todo impacto sobre os ambientes naturais é, a priori, negativo. [...].
- Apesar de diversos autores ressaltarem os efeitos negativos do turismo sobre ambientes naturais, é preciso reconhecer que a degradação dos ambientes, de modo geral, não interessa ao turismo porque este tem o espaço como principal objeto de consumo. [...].

Na compreensão dos impactos ambientais, reproduzimos os paradigmas do final do século XX, a denominada questão ecológica, “degradação, conservação e preservação ambientais tornaram-se expressões comuns, apropriadas ao vocabulário cotidiano das pessoas, de modo geral, mostrando que a preservação da natureza já é uma questão de senso comum” (CRUZ, 2001, p. 26).

O processo articulado nas relações sociais e baseado na conservação dos atrativos naturais procurou compreender esse fenômeno como única prática social que consome o espaço (PORTUGUEZ et al., 2012), portanto, como a Geografia pode-se interessar pelos estudos relativos ao turismo?

A resposta para esse questionamento passa necessariamente pelo conhecimento da geografia do turismo, que envolve a dimensão socioespacial, cultural, econômica, política, ambiental e diversas áreas do conhecimento, não se referindo apenas à abordagem turística “os impactos do turismo em ambientes naturais estão associados tanto à colocação de infraestruturas nos territórios para que o turismo possa acontecer com a circulação de pessoas que a prática turística promove nos lugares” (CRUZ, 2001, p. 31).

Para uma abordagem interdisciplinar, multidisciplinar e transdisciplinar, reunimos referências sobre como os impactos ambientais contribuem com as informações sistemáticas do turismo. Para tanto, é necessário avançar em direção a um terreno transdisciplinar e pluriparadgmático, buscando conhecer o objeto de estudo – o espaço do turismo:

> No seu universo complexo e multifacetado, percorrendo os campos econômico, sociológico, antropológico, psicológico, cultural, político, jurídico, além de outros. Portando, os aportes da geografia vão até um determinado ponto a partir do qual, sente-se resvalar, a pisar em terreno movediço. Parece que é neste momento que se sente a impotência da geografia e, em conseqüência, a nossa própria impotência (RODRIGUES, 1998, p. 86).

O estado da arte tem demonstrado um tratamento do turismo por meio de abordagens neopositivistas com apoio nas informações visíveis, exatas e mensuráveis por meio de técnicas estatísticas. O tratamento do turismo por meio de abordagens críticas, considerando a atividade enquanto produto social é resultado da divisão do trabalho, da articulação dos meios de produção e, sobretudo, das condições políticas e econômicas. Tais estudos asseguram Oliveira (2008), têm denunciado impactos socioambientais e econômicos proporcionados pelo turismo enquanto produto de modo de produção capitalista.

# 3 METODOLOGIA

A metodologia conduz e estrutura a pesquisa permitindo avaliar os dados, os questionamentos, as técnicas e os resultados.

Severino (2007) explica que, a metodologia para o desenvolvimento de um estudo científico contempla caminhos ou ações que são executadas de forma a permitir o alcance dos objetivos elaborados e a resposta ao problema da pesquisa.

## 3.1 LOCALIZAÇÃO DA ÁREA DE ESTUDO

A área de estudo é Rondonópolis, município que se localiza geograficamente no sudeste do Estado de Mato Grosso, situado a aproximadamente 220 km da capital do Estado, e considerado um pólo agropecuário e econômico da região.

Neste contexto, Ross (2005, p.220) esclarece:

> [...] a produção do espaço no território de Mato Grosso, nos últimos 35 anos, é resultado da convergência de fatores naturais, técnico-científicos e político-econômico-culturais que se associam a um determinado momento da história brasileira, onde programas, projetos e planos dos governos federais estaduais voltaram-se para uma política de planejamento para o desenvolvimento econômico, sem preocupação com a equidade social e o equilíbrio ambiental.

As condições geomorfológicas, pedológicas, climáticas, bem como a introdução de tecnologias modernas de cultivo e armazenamento e financiamento levaram o estado a ampliar a sua rede urbana e a um grande crescimento demográfico.

Segundo IBGE (2014), o municipio possui uma população estimada em aproximadamente 211.718 habitantes numa área de 4.159,118 $km^2$ representando 0.48% da área do Estado com 129,2 $Km^2$ de zona urbana e 4.035,8 $km^2$ de zona rural.

De acordo Demamann (2011, p.26), a cidade está localizada em posição privilegiada, no entroncamento das rodovias federais BR-163 e BR-364. O município de Rondonópolis é servido por uma ampla rede viária, de estradas federais, estaduais e municipais, localizando-se no maior tronco rodoviário do estado, “é portão de passagem para algumas das principais cidades e estados do país como: São Paulo, Minas Gerais, Brasília, Goiânia, Campo Grande por rodovias totalmente pavimentadas [...]”.

O clima de Rondonópolis é caracterizado por uma temperatura média anual de $25^0$ C, sendo a média das máximas de $32,6^0$ C e a média das mínimas $18,6^0$ C. Setembro e Outubro são os meses mais quentes com temperaturas médias acima de $26^0$ C, e os meses de Junho com $21,9^0$ C e Julho com $22,3^0$ C são aqueles que apresentam as menores médias (SETTE, 1996).

O **Mapa1** identifica a área de estudo.

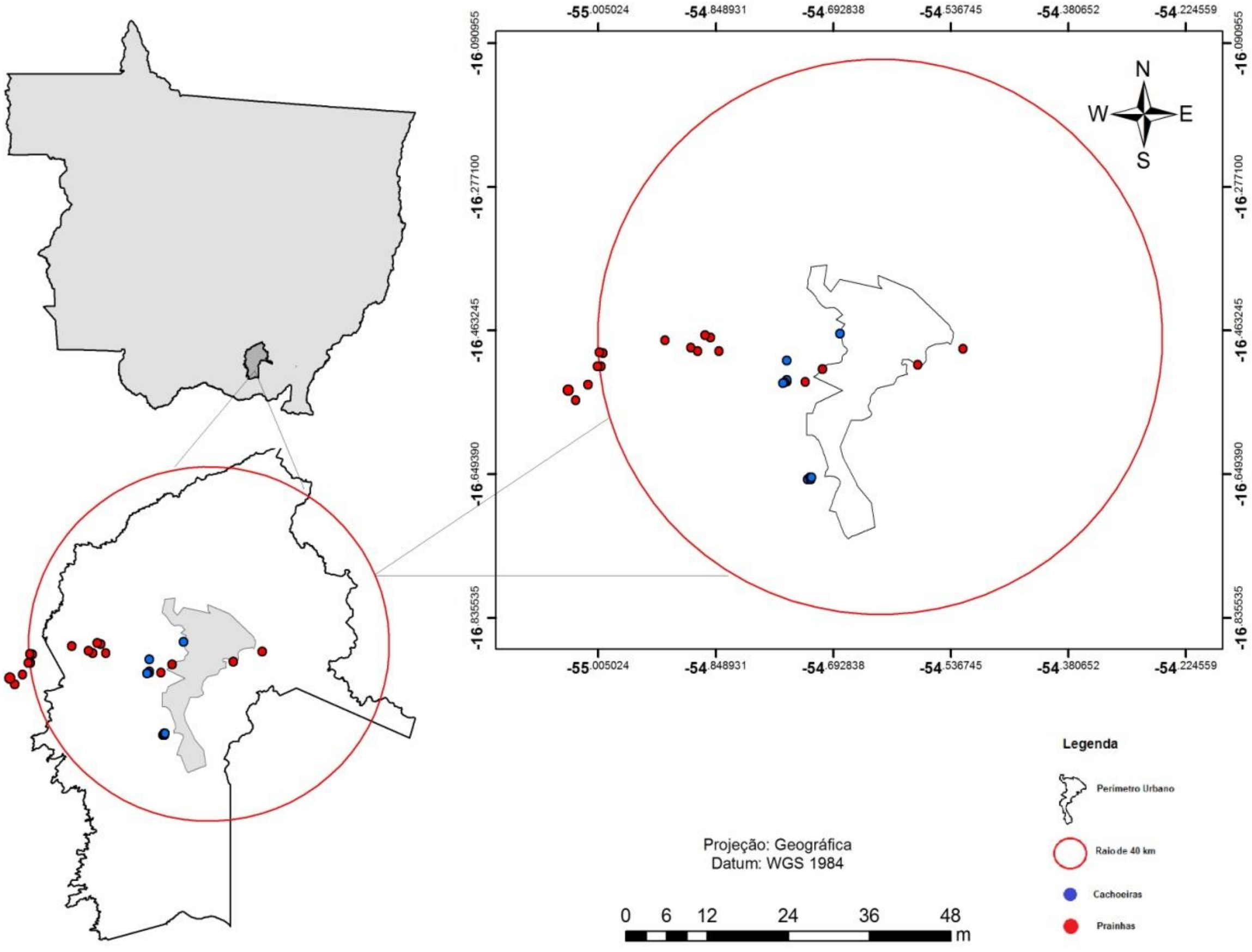

**Mapa 1 - Mato Grosso, Localização de Rondonópolis, Área de Estudo e as Potencialidades**
Fonte: IBGE. Org: CAMPOS, M. B. N. S. (2014)

No **Mapa 1**, é possível observar as potencialidades do município de Rondonópolis, localizando os lugares com intencionalidade para as práticas ecoturísticas de uso sustentável.

As áreas de potencialidade ecoturísticas disputam com a produção agropecuária espaço para realização de suas práticas recreacionais, e de acordo com Silva (2003, p. 75) “o sucesso da lavoura da soja em bases latifundiárias no sudeste mato-grossense se consolidou no início dos anos 80”.

A verticalização da produção e as potencialidades ecoturísticas, no entanto, nem sempre passam por ações planejadas que minimizam os impactos ambientais para a execução das atividades turísticas. Mas compete ao Poder Público fazer essa gestão estabelecendo instrumentos legais.

O Município de Rondonópolis como caracteriza Nardes (1997, p.31) "tem um potencial ecoturístico, ainda pouco explorado: rios com corredeiras, fontes de águas, cachoeiras e formações rochosas impressionantes. Entre as inúmeras atrações turísticas, destaca-se a cidade de Pedra".

## 3.2 ABORDAGEM METODOLÓGICA

Para atingir os objetivos propostos foram elaboradas três etapas metodológicas: Pesquisa Bibliográfica, Identificação e Seleção das Áreas de Lazer, Elaboração dos Mapas Temáticos com Potencial Ecoturístico.

A **Figura 2** demonstra as etapas metodológicas do estudo.

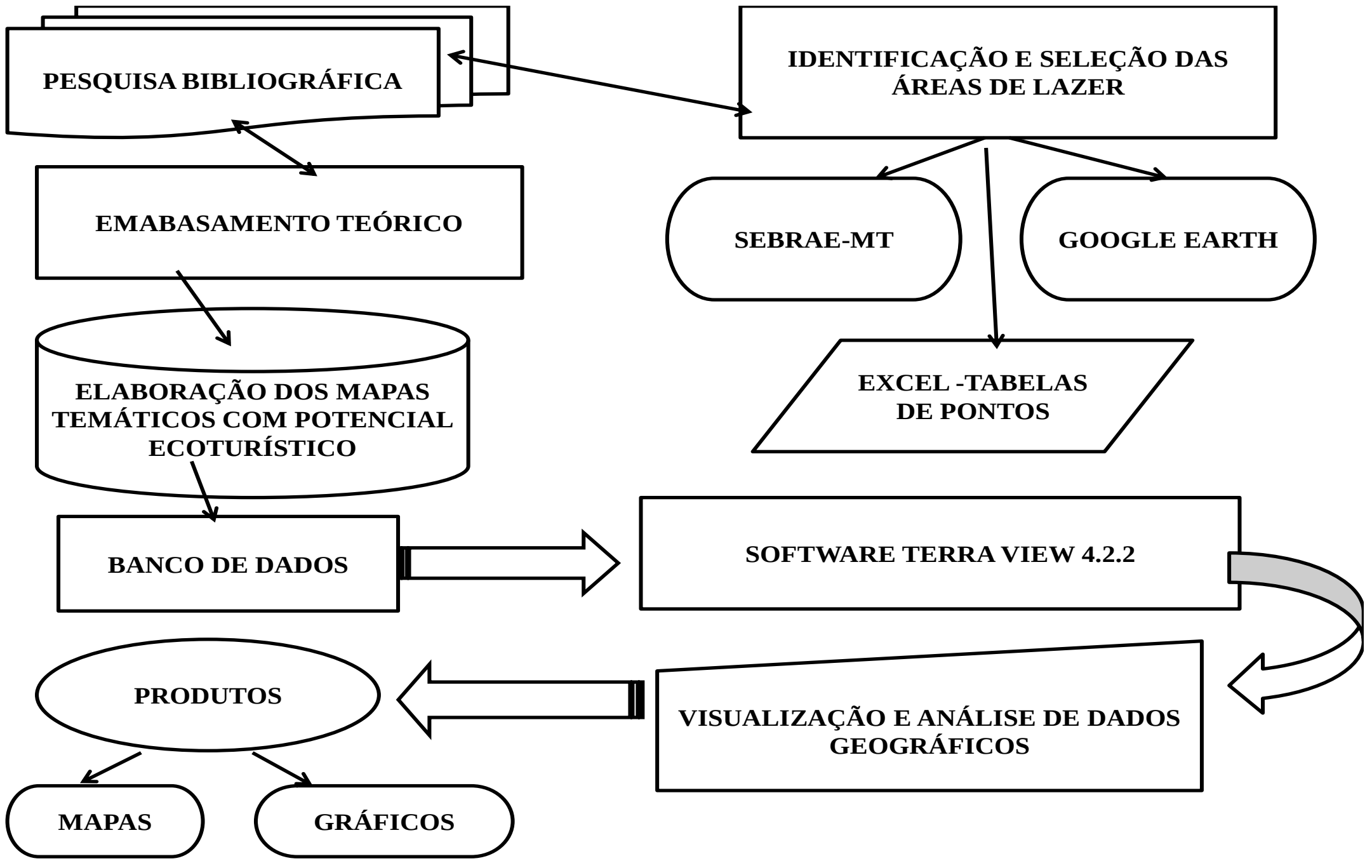

**Figura 2 - Fluxograma metodológico**
Org.: CAMPOS, Manoel B. N. S. (2015)

Etapa 1 - Pesquisa Bibliográfica

A primeira etapa consistiu-se no levantamento bibliográfico em livros, dissertações, teses, artigos, banco de dados e sites confiáveis, possibilitando interação com a temática e com o referencial teórico-metodológico.

> A pesquisa bibliográfica é desenvolvida a partir de material já elaborado, constituído principalmente de livros e artigos científicos. Embora em quase todos os estudos seja exigido algum tipo de trabalho desta natureza, há pesquisas desenvolvidas exclusivamente a partir de fontes bibliográficas. Parte dos estudos exploratórios pode ser definida como pesquisas bibliográficas, assim como certo número de pesquisas desenvolvidas a partir da técnica de análise de conteúdo (GIL, 2007, p.64).

A fundamentação teórica foi realizada por meio de diversos autores, dentre os quais citamos: Cruz (2001), Fennell (2002), Marinho e Bruhns (2003), Rodrigues (2003), Silva (2003), Veiga e Silva (2004), Moreno e Higa (2005), Nardes (2005), Teixeira (2007), Lucio (2008), Silva (2009), Alvarenga (2011), Marulo (2012) e outros.

Etapa 2- Identificação e Seleção das Áreas de Lazer

Para a identificação e seleção das áreas de lazer, foram utilizadas as coordenadas dos pontos das cachoeiras extraídas do Inventário da Oferta Turística do Serviço Brasileiro de Apoio às Micro e Pequenas Empresas de Mato Grosso (SEBRAE-MT) e as coordenadas dos pontos das prainhas extraídas de imagens do *Google Earth*.

Posteriormente, as informações foram transportadas para o programa Microsoft Excel, e foi criada uma Tabela de Pontos, a qual foi importada para o SIG Terra View 4.2.2 para elaboração dos mapas temáticos.

Etapa 3 – Elaboração dos Mapas Temáticos com Potencial Ecoturístico

A elaboração dos mapas temáticos é imprescindível para o estudo em questão. Para a criação do Banco de Dados, as informações foram obtidas com auxilio de software de Sistemas de Informações Geográficas (SIGs). As cartas temáticas: Localização da área de estudo e as potencialidades ecoturísticas de Rondonópolis; Localização das cachoeiras; Pontos ecoturísticos; Malha viária; Uso e ocupação do solo; Hidrografia e Altimétrica.

Os dados foram organizados pelo (SIGs), para análise, verificando a disposição espacial e a situação atual da vegetação. Segundo Veiga e Silva (2004, p. 190). Os SIGs ferramentas que manipulam objetos e seus atributos "é um sistema de *software* computacional com o qual a informação pode ser capturada, [...] através de uma série de planos de informação que se sobrepõe corretamente em qualquer localização". Após, foram elaboradas as representações cartográficas por meio de mapeamentos temáticos, visando à caracterização dos aspectos físicos.

## 3.3 PROCEDIMENTOS METODOLÓGICOS

Os procedimentos metodológicos se deram pela construção do marco teórico conceitual, o qual, como define Marconi e Lakatos (2001), possibilita o conhecimento do pesquisador sobre o tema analisado com o aprofundamento teórico. No caso específico desta pesquisa, foram selecionados e analisados estudos que tratavam sobre os temas potencialidades, sustentabilidade, ecoturismo, possibilidades e limites na execução de atividades ecoturísticas.

A elaboração dos mapas temáticos se deu por meio do auxílio de software de Sistemas de Informações Geográficas (SIGs). No laboratório de Geoprocessamento e Sensoriamento Remoto do Departamento de Geografia do Campus Universitário de Rondonópolis, foram identificados dez (10) pontos das cachoeiras e oito (08) das prainhas num raio de 40 km do centro da cidade usando como referência a Praça Brasil. A delimitação dos pontos ecoturísticos foi feita com base em uma imagem digital georreferenciada do *Google Earth*. Posteriormente as informações transferidas como arquivo *shap* para o SIG Terra View- 4.2.2 para confecção dos mapas.

Para a caracterização dos atrativos naturais com potencialidades ecoturísticas, criamos um banco de dados com as informações vetoriais obtidas por meio de pesquisas eletrônicas nos sites do Instituto Brasileiro de Geografia e Estatística (IBGE), Departamento Nacional de Infraestrutura de Transporte (DNIT), acervo fundiário do Instituto Nacional de Colonização e Reforma Agrária (INCRA) e Ministério do Meio Ambiente (MMA), bem como dados matriarcais, como imagens de satélite e Modelo Digital de Elevação (DEM). A espacialização das potencialidades ecoturísticas serviu como ferramenta para a manutenção e conservação dos recursos hídricos, pois revela a realidade espaço temporal das localidades. As imagens fotográficas compreendem a realidade local, bem como demonstram aspectos relevantes do processo de ocupação.

A pesquisa de campo é a articulação entre a teoria e a prática, e para Marconi e Lakatos (2006, p. 188-189) tem o objetivo de obter "informações e/ou conhecimentos acerca de um problema, para o qual se procura uma resposta, ou de uma hipótese, que se queira comprovar, ou, ainda, descobrir novos fenômenos ou as relações entre eles". O presente estudo se deu pela forma quali-quantitativa, dos resultados obtidos, o que possibilitou captar a percepção socioambiental para o desenvolvimento das atividades turísticas e o dimensionamento das potencialidades ecoturísticas do município de Rondonópolis-MT.

Os procedimentos ligados à avaliação das cachoeiras foram adaptados de Machado e Souza (2012) como segue: a) Localização; b) Distância da cachoeira até o centro da cidade e/ou limite urbano; c) Acesso mais utilizado; d) Facilidade de acesso (faixas de proximidades a rodovias pavimentadas e não pavimentadas); e) Existências de atrativos naturais (classes de uso e cobertura vegetal, áreas agrícolas, florestas, afloramento rochosos, praias); f) Atividades esportivas possíveis (natação, rapel, mergulho, banho, caiaquismo, bóia-cross, camping); g) Altura da cachoeira/ comprimento da corredeira.

Um cenário ou paisagem é considerado oferta turística quando, além do recurso natural gera benefício financeiro (cachoeiras, praias etc.), e/ou quando haja equipamentos que permitam o deslocamento e a permanência do turista no local. A atividade turística somente existirá se acompanhada de uma transferência de divisas de um centro emissor para um centro receptor (BARBOSA, 2005).

### 3.3.1 Universo da População e Tamanho da Amostra

A pesquisa contou com a participação de quatro grupos de atores sociais: os Usuários Frequentadores (UF) representados pelas pessoas que freqüentam as áreas de lazer nos atrativos turísticos do entorno da malha urbana de Rondonópolis nos finais de semanas. Nesse universo, considera-se a média de 70 pessoas por final de semana, perfazendo assim, uma população total de 280 usuários freqüentadores mensais, pois, essas localidades são abertas ao público somente aos sábados, domingos e feriados.

Ressaltando que as informações para construir o universo da pesquisa foram fornecidas pelas Associações dos Assentamentos Carimã e Gleba Rio Vermelho.

Os Moradores das Localidades (ML) são representados por 15 pessoas, que residem nas localidades e próximas às áreas de lazer. As Agências de Turismo Receptivo (ATR) são constituídas de 03 agências de turismo localizadas na área urbana de Rondonópolis. As Instituições Públicas (IP), representadas por 02 secretários municipais de Rondonópolis, o do Meio Ambiente e outro de Ciência, Tecnologia, Turismo e Desenvolvimento Econômico.

De acordo com os dados disponíveis, o número total da população (N), resulta em 300. O método utilizado para a elaboração da amostra quantitativa (A), que corresponde a 169 pessoas.

O tamanho da população foi baseado de acordo com Marulo (2012) e a amostra no método de Krejcie e Morgan (1981) qual estabelece nível de confiança de 95% e uma margem de erro (E) igual a 5% **(Anexo).** A população e a mostra foram extraídas por meio das técnicas de Amostragem Casual Simples Estratificada com Partilha Proporcional. A amostragem estratificada de acordo com Berquó (2006, p.137-138), "é aquela em que o pesquisador deseja que as subpopulações sejam representadas na amostra com a mesma proporcionalidade com que compõem a população total". O **Quadro 4** apresenta como foram extraídas a população e a amostra.

A entrevista foi executada de forma dirigida através de questionário. Foram entrevistados cento e cinqüenta e seis (156) usuários que freqüentam as áreas de lazer e os nove (09) moradores do entorno de maneira aleatória intencional, uma vez que foram selecionados aqueles moradores mais próximos à entrada das áreas de lazer, no Assentamento Carimã, e outros no Assentamento Gleba Rio Vermelhos ao longo das proximidades do Sítio do Vale Encantado, no entorno da rodovia do Peixe.

Na Secretária Municipal de Meio Ambiente, foi entrevistado o próprio Secretário de Meio Ambiente, e na Secretária Municipal de Ciência, Tecnologia, Turismo e Desenvolvimento Econômico, foi entrevistado o Secretário do Turismo.

Com relação às 02 Agências de Turismo Receptivo, foram entrevistadas as gerentes responsáveis pelas empresas. Através destes dados foi elaborado o mapeamento dos potenciais turísticos e suas demandas.

| Estrato | População | Amostra | Relação | Descrição |
|---|---|---|---|---|
| Usuários | 280 | 156 | 0,9231 | Usuários frequentadores das áreas de lazer |
| Moradores | 15 | 9 | 0,0533 | Moradores das localidades e do entorno das áreas de lazer. |
| Agências de Turismo Receptivo | 03 | 2 | 0,0118 | Empresas que ofertam roteiros turísticos em Rondonópolis. |
| Instituições Públicas | 02 | 2 | 0,0118 | - Secretário Municipal de Meio Ambiente<br>- Secretário Municipal de Ciência, Tecnologia, Turismo e Desenvolvimento Econômico. |
| Total | 300 | 169 | 0,5633 | - |

**Quadro 4 - Distribuição da Amostragem com Partilha Proporcional**
Org.: CAMPOS, M. B. N. S. (2015)

### 3.3.2 Instrumentos de Coleta de Dados

Os instrumentos de coleta de dados e a análise de um grupo específico levaram aos resultados apresentados sem qualquer forma de manipulação por parte do pesquisador, constituindo um estudo que para Gil (2007) retrata de forma fiel as informações. A coleta de dados foi desenvolvida em três momentos:

No primeiro momento, foram aplicados questionários para os moradores das localidades e do entorno **(Apêndice A)** e também para os usuários dos recursos hídricos ou praticantes de atividades ecoturísticas **(Apêndice B)**. Os questionários possibilitaram identificar as potencialidades ecoturísticas das localidades escolhidas para análise, a partir da realidade vivenciada pelos dois principais atores do ecoturismo. No questionário dos usuários também contam com a Escala de Likert de cinco pontos no seu grau de concordância ou discordância das declarações relativas à atitude sobre o uso dos recursos naturais, especialmente, os hídricos que são os mais desgastados durante as atividades de ecoturismo. As perguntas deram aos entrevistados a oportunidade de fornecer respostas claras em vez de respostas neutras e ambíguas, e sua aplicação foi mediante entrevista pessoal, de maneira aleatória intencional. Para a análise Estatística da escala foi utilizado

o coeficiente de correlação de Pearson e de Curtose. E para a comparação das médias o *Teste-t de Student* ao nível de significância de 5%.

> O consumidor constrói níveis de aceitação dos produtos e serviços, conforme suas experiências e influências sociais. Rensis Likert, em 1932, elaborou uma escala para medir esses níveis. As escalas de Likert, ou escalas somadas, requerem que os entrevistados indiquem seu grau de concordância ou discordância com declarações relativas à atitude que está sendo medida. Atribui-se valores numéricos e/ou sinais às respostas para refletir a força e a direção da reação do entrevistado à declaração. As declarações de concordância devem receber valores positivos ou altos enquanto as declarações das quais discordam devem receber valores negativos ou baixos (BRANDALISE, 2005, p. 04).

Para se obtiver o resultado final da concordância ou discordância das declarações relativas à sua atitude, foram somados todos os pontos que determinaram à média e o desvio padrão dessas declarações. Estas devem oportunizar ao entrevistado expressar respostas claras em vez de respostas neutras, ambíguas.

No segundo momento, foram realizadas visitas nas localidades utilizando um roteiro estruturado para observação, um procedimento metodológico relevante para o registro das informações e checagem dos dados secundários, contribuindo para análise e o diagnóstico do objeto de estudo. Dados que revelam e que influenciam o uso dos recursos naturais, tais como: espaço de apoio; uso e ocupação do entorno; educação ambiental; as atividades culturais e os impactos socioambientais, entre outros, conforme identificado na revisão de literatura **Apêndice E** e definido a seguir:

> A caracterização ambiental de uma determinada área da paisagem seja uma bacia hidrográfica, um município, região ou unidade ambiental, pressupõe o entendimento da dinâmica do ambiente natural com ou sem a intervenção antrópica, fundamentada na compreensão das características físico-ambientais e socioeconômicas, visando à síntese do conhecimento sobre a realidade pesquisada (NARDES, 2005, p.19).

O trabalho de campo é uma técnica fundamental no processo do planejamento turístico, tanto setorial quanto territorial, a partir do diagnóstico será possível estabelecer ações, qualificar e avaliar especialmente os atributos ecoturísticos.

No terceiro momento, foram aplicados questionários as instituições públicas **(Apêndice C)** e as agências de turismo receptivo que ofertam produtos ecoturísticos **(Apêndice D)** ambos compostos somente por perguntas fechadas. Gil (1999) define o questionário como uma técnica de investigação composta por um número mais ou menos elevado de questões apresentadas por escrito às pessoas, tendo como objetivo o conhecimento de suas opiniões, crenças, sentimentos, interesses, expectativas, situações vivenciadas.

### 3.3.3 Análise e Representação dos Dados

Análise dos dados coletados foi processada através da técnica quali-quantitativa apresentando a percepção socioambiental do desenvolvimento das atividades ecoturísticas. De acordo com Marconi e Lakatos (2006, p. 169), “uma vez manipulados os dados e obtidos os resultados, o passo seguinte é a análise e interpretação dos mesmos, constituindo-se ambas no núcleo central da pesquisa”. A representação dos dados se deu em forma de quadros, tabelas, gráficos, mapas temáticos, fotos e outros **(Capítulo 4)**.

### 3.3.4 Categorias de Análise e Variáveis da Pesquisa

No intuito de responder aos objetivos propostos no presente estudo, foi estabelecida a seguinte categorização de análise: a) a percepção dos moradores da localidade e do entorno imediato às áreas de lazer b) e a dos atores do ecoturismo.

Segundo Sturza (2006), a percepção é um processo dialético que absorve sujeito (homem) e objeto (lugar), realizando relações entre ambos, as interfaces objetivas e subjetivas, expressas ou obscurecidas, entre a globalização e a individualidade. A percepção, a vivência e a memória dos indivíduos e dos grupos sociais são elementos importantes na constituição do saber geográfico, da produção do espaço e da paisagem, que se perfaz a partir do imaginário social.

Os homens respondem ao ambiente pelos elementos por sua apreciação estética, temporária, inesperada ou percepção dos sentidos: tátil, visual e auditivo, sendo também vinculado ao sentimento de pertencimento pelos vínculos históricos da família, “a consciência do passado é um elemento importante no amor pelo lugar” (CISOTTO, 2013, p. 97).

Cada pessoa interpreta o lugar de acordo com a vivência, envolvimento e situação na temporalidade histórica. A feição ao lugar acontece, pois conforme Barbosa (2008, p. 4) destaca, os moradores estabelecem a relação entre a infância vivida e as áreas verdes do local “o apego ao lugar, por ser familiar, pela natureza, por representar o passado e pela localização, é o orgulho dos moradores. [...]. A percepção, a atitude e o valor que inferem ao meio ambiente mantêm suas características de visões de mundo muito semelhantes”.

De acordo com os aspectos caracterizados as variáveis da pesquisa são:

1- Opinião (positiva ou negativa) em relação ao uso dos recursos naturais relacionados com as atividades turísticas; a soma dos pontos nas vinte e quatros proposições da Escala Likert em relação aos benefícios socioambientais;
2- Percepção socioambiental para desenvolvimento das atividades turísticas;
3- Características (gênero, idade, grau de escolaridade, renda familiar, atividade ecoturísticas, dentre outros) dos usuários frequentadores das áreas de lazer.

Cabe ressaltar que as variáveis são aspectos, propriedades ou fatores reais ou potencialmente mensuráveis pelos valores que assumem e discerníveis em um objeto de estudo (CERVO; BERVIAN, 2002).

A perspectiva de análise introduzirá os elementos interdisciplinares na discussão do ecoturismo, enquanto processo e não apenas uma modalidade de turismo.

# 4 RESULTADOS E DISCUSSÃO

## 4.1 CARACTERIZAÇÃO DOS ATRATIVOS NATURAIS COM POTENCIALIDADES ECOTURÍSTICAS NO MUNICÍPIO

O ecoturismo ou turismo ecológico tem ganhado forças, pois com os dias agitados em centros urbanos, cada vez mais pessoas procuram áreas de lazer que possibilitem o contato com a natureza para passar os finais de semana, feriados ou mesmo feriados prolongados.

Diante de tal fato, buscou-se caracterizar por meio dos mapas temáticos as potencialidades turísticas ecológicas do município. No intuito de promover melhor aproveitando, dos recursos naturais da região, com sustentabilidade, e promovendo a melhoria da qualidade de vida das comunidades locais.

No município existem 27 cachoeiras que estão fora do perímetro urbano dentro de um raio de 40 km a partir do centro da cidade tendo como referência a Praça Brasil. São 17 cachoeiras cadastradas pelo inventário do SEBRAE/MT e localizadas nas áreas circunvizinhas próximas as Unidades de Conservação Parque Estadual Dom Osório Stoffel e Parque Ecológico João Basso localizado numa área de 3.624,57 hectares (NARDES, 2005). A presente pesquisa foi aplicada em 09 delas exceção somente a Cachoeira do Escondidinho, pois ela encontra-se no limite do perímetro urbano (**Mapa 2)**. Ao longo do Rio Vermelho dentro de um raio de 40 km quantificam-se 08 prainhas, e 10 prainhas estão fora deste raio, perfazendo um total de 18 prainhas, **Mapa 3**.

Nos **Mapas 2 e 3** é possível visualizar a Cachoeira do Escondidinho, porém ela foi retirada do estudo em função de apresentar discrepância quanto aos moradores e aos usuários, dado que essa cachoeira tem um histórico de violência e marginalidade que poderá deformar os dados desta pesquisa. Nessa localidade da Cachoeira do Escondidinho será implantado o primeiro Parque Municipal Ecológico de Rondonópolis, uma obra lançada no início de 2012, orçada em R$ 3,8 milhões, e com previsão de entrega para início de 2013, mas que está paralisada por problemas de planejamento das obras.

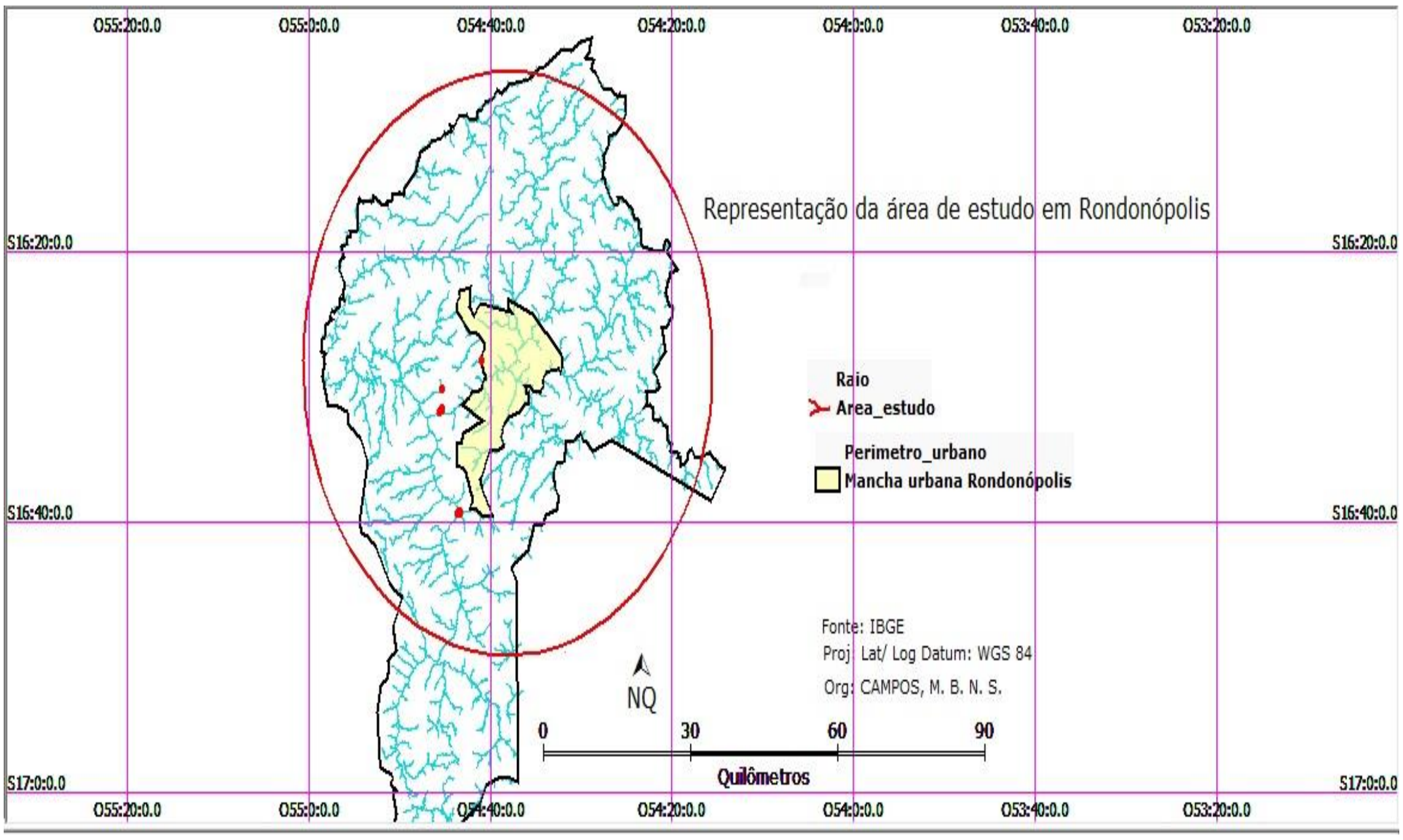

**Mapa 2 - Representação da Área de Estudo em Rondonópolis**
Fonte: IBGE
Projeção: Lat/Log Datum SAD 69
Org: CAMPOS, M. B. N. S. (2014)

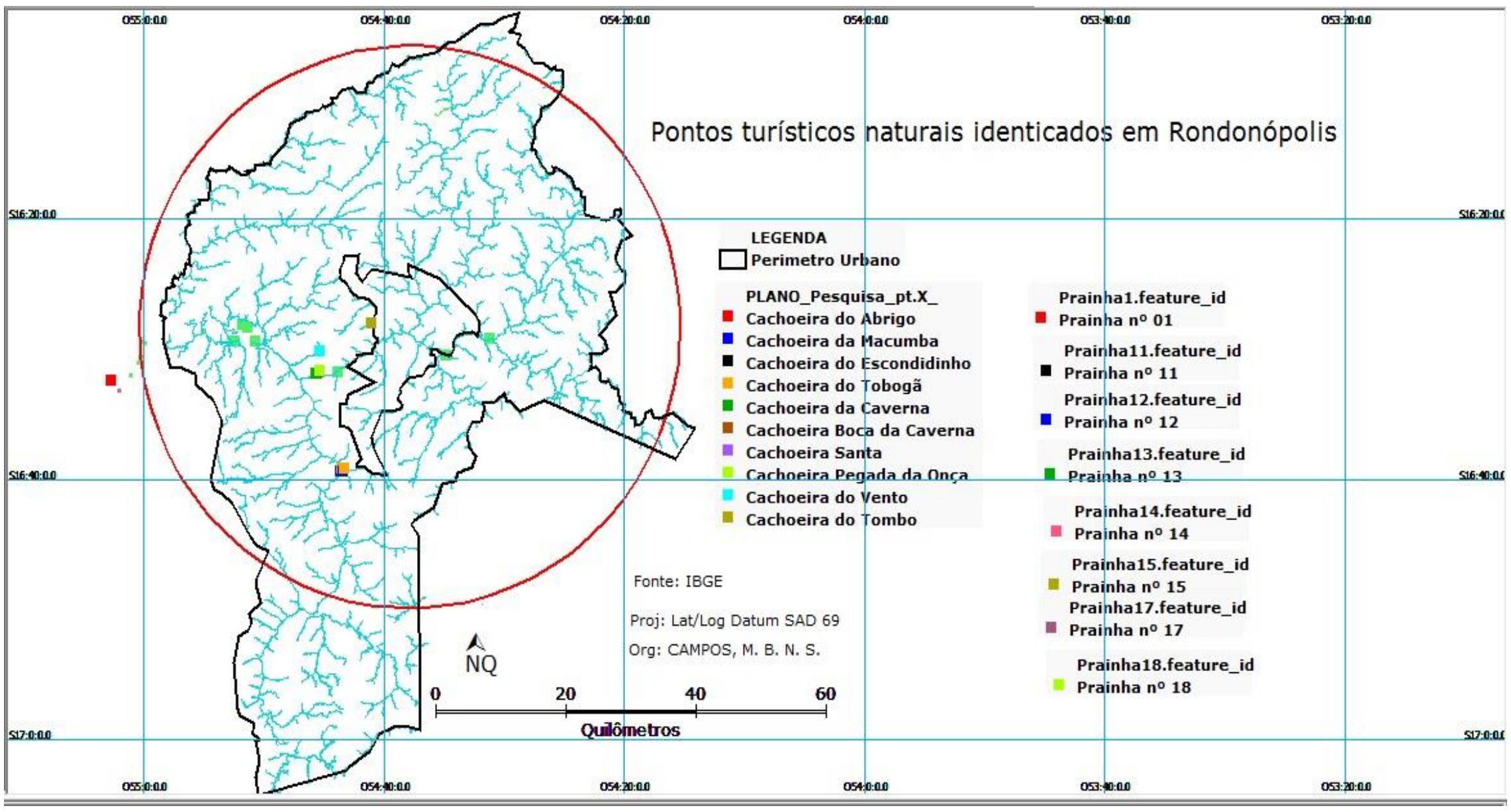

**Mapa 3 - Pontos Turísticos Identificados em Rondonópolis**
Fonte: IBGE
Projeção: Lat/Log Datum SAD 69
Org: CAMPOS, M. B. N. S.(2014)

O Mapa 4 mostra as cachoeiras e as vias de acessos representadas pela malha viária pavimentada e não pavimentada, de relevância para as possibilidades de percurso até as áreas de lazer identificadas.

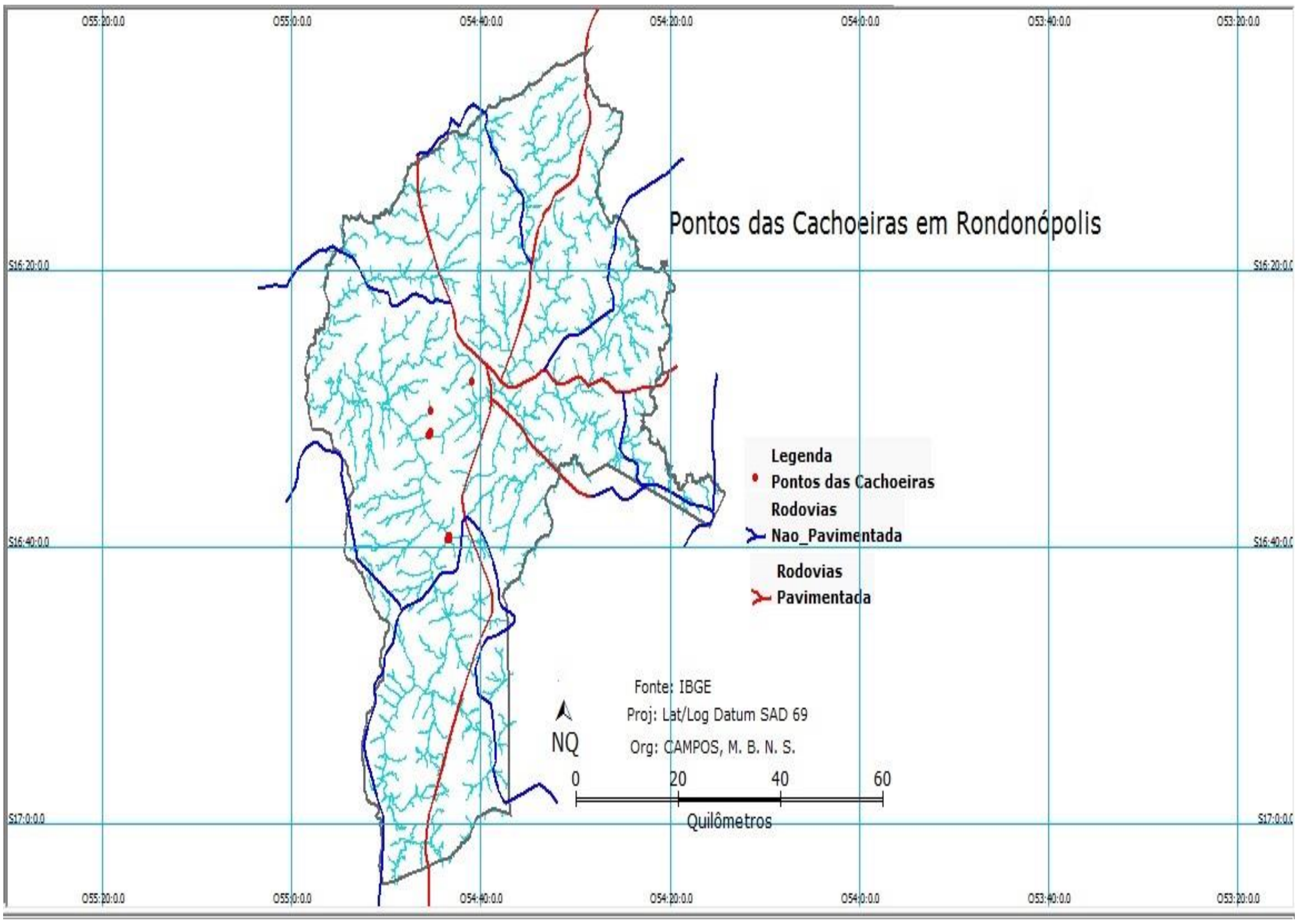

**Mapa 4 - Malha Viária incluindo as Cachoeiras, em Rondonópolis**
Fonte: IBGE
Projeção: Lat/Log Datum SAD 69
Org: CAMPOS, M. B. N. S. (2014)

Importante apontar que todas estas localidades sofreram com o passar do tempo alguma forma de degradação, mas o ecoturismo passa a ser visto como uma alternativa para a sensibilização das comunidades locais para a necessidade de conservação dos recursos naturais **Figura 3**.

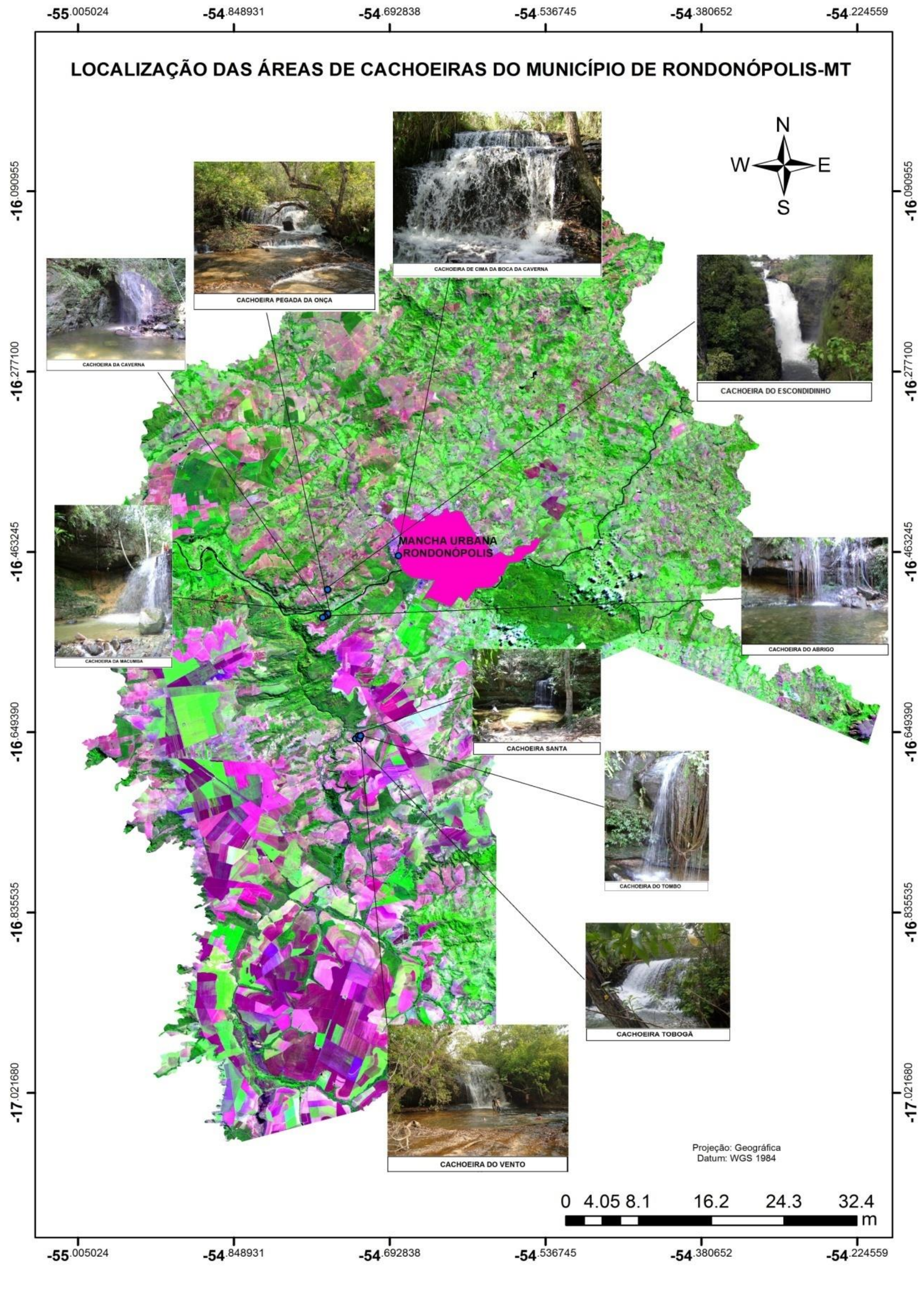

**Figura 3 - Localização das Cachoeiras nos Assentamentos Carimã e Gleba Rio Vermelho do Município de Rondonópolis – MT**
Org.: CAMPOS, M. B. N. S. (2014).

As cachoeiras destacadas na **Figura 3** podem ser identificadas pela erosão diferencial, tendo em vista a alternância das camadas de arenito com as efusivas básicas. Dos inúmeros atrativos que o município oferece, buscou-se descrever alguns que pudessem representar a diversidade de produtos turísticos que estes locais oferecem. As potencialidades destas cachoeiras estão relacionadas no **Quadro 5**:

| Nome | Potencialidades |
|---|---|
| Cachoeira Boca de Caverna | Localizada no Assentamento Gleba Rio Vermelha na zona rural. Tem como localidade mais próxima do atrativo na área urbana o Residencial Acácia. Acesso utilizado freqüentemente, Avenida K e a continuidade do km 12, até o colchete de arame que dá acesso a trilha de 600 metros, que apresenta trechos íngremes chegando à cachoeira. As atividades realizadas no atrativo natural são banho, caminhada e rapel. A cachoeira tem aproximadamente 20 metros de altura, que forma um pequeno lago com mais de 10 metros de extensão. A profundidade do lago atinge a 1,5 metros e é apropriado para o banho. A cachoeira é utilizada por praticantes de rapel. Durante o período de estiagem o fluxo de água diminui bastante. *O local não proporciona acessibilidade aos portadores de necessidades especiais.*<br>Equipamentos: sede para administração; vestiários (masculino e feminino); *playground*; cozinha comunitária; quadra esportiva de areia; recanto com mesas; pedalinhos; sanitários. |
| Cachoeira da Macumba | Localizada no Córrego do Gavião, nas proximidades do Sitio Vale Encantado. O acesso mais utilizado é pela BR - 364 saída para Cuiabá, entrada do Praia Clube, percorrendo mais de 15 km até a cachoeira. As atividades realizadas no atrativo natural são caminhadas, banho e cultos religiosos. A cachoeira tem aproximadamente 25 metros de altura, suas águas são abundantes, cristalinas e frias o ano todo. Durante o período de estiagem ocorre à diminuição do fluxo de água. *O local não proporciona acessibilidade aos portadores de necessidades especiais.*<br>Equipamentos: sede para administração; vestiários (masculino e feminino); *playground*; cozinha comunitária; quadra esportiva de areia; recanto com mesas; pedalinhos; sanitários. |
| Cachoeira Santa | Localizada na Rodovia MT - 471 no Km 5,4 tendo como referencia as Sete Placas. O acesso mais utilizado é pelo Km 22 na BR 163, e percorrendo 2,5 km até a porteira e depois mais 2,9 km até a cachoeira. As atividades realizadas no atrativo natural são caminhadas, banho e rapel. A cachoeira tem aproximadamente 35 metros de altura que deságua direto no Rio Ponte de Pedra. A ação da água no arenito esculpiu um formato de uma santa orando. A vegetação nas margens do Rio Ponte de Pedra está preservada. O local é frequentado por praticantes de rapel. A trilha para chegar onde a cachoeira deságua é emocionante atravessa mata preservada, passando por uma fenda na rocha com uma trilha íngreme. *O local não proporciona acessibilidade aos portadores de necessidades especiais.*<br>Equipamentos: sede para administração, sanitários, seis quiosques para churrascos, restaurante, salão comunitário, bar. |
| Cachoeira do Abrigo | Localizada no Assentamento Gleba Rio Vermelha na zona rural. A localidade mais próxima do atrativo na área urbana é a Residencial Acácia. O acesso utilizado frequentemente, Avenida K e a continuidade do km 12, até o colchete de arame que dá acesso a trilha de 300 metros, apresentando trechos íngremes chegando à cachoeira. As atividades realizadas no atrativo natural são banho, caminhada e rapel. A cachoeira tem aproximadamente 4 metros de altura que forma um pequeno lago. A profundidade do lago chega a 0,80 metros, apropriado para o banho com espaço para o lazer. *O local não proporciona acessibilidade aos portadores de necessidades especiais.*<br>Equipamentos: sede para administração; vestiários (masculino e feminino); |
| | *playground*; cozinha comunitária; quadra esportiva de areia; recanto com mesas; pedalinhos; sanitários. |
| Cachoeira do Escondidinho | Localizada na Avenida Arapongas, próxima da Vila Olinda. O acesso mais utilizado é por essa mesma avenida, que é a principal do bairro. A atividade realizada no atrativo natural é o banho. A Cachoeira possui aproximadamente 12 metros de altura, formando um poço. O local possui cerca de três metros de profundidade, as pessoas se divertem escalando as pedras, procurando o ponto mais alto para pular na água. O ambiente está completamente deteriorado, com resíduos sólidos e desmatamento da mata ciliar, algumas construções invadem o córrego e despejam esgoto sem nenhum tratamento. Conforme relato dos moradores o local é freqüentado por usuários de drogas e por práticas inadequadas ao ambiente de lazer. *O local não proporciona acessibilidade aos portadores de necessidades especiais.*<br>Equipamentos: nenhum |
| Cachoeira do Tobogã | Localizada na Rodovia MT - 471 no Km 5,4 tendo como referencia as Sete Placas. O acesso mais utilizado é pelo Km 22 na BR 163, e percorrendo 2,5 km até a porteira e depois mais 2,5 km até a cachoeira. As atividades realizadas no atrativo natural são caminhadas, banho e rapel. A respectiva cachoeira encontra-se no Córrego da Onça com aproximadamente 30 metros de altura em uma laje de Pedra que forma um "Tobogã'' com volume de água intenso o ano todo. A cachoeira é apropriada para a prática de rapel. Nas proximidades existem várias áreas com afloramento de água e vegetação endêmica. *O local não proporciona acessibilidade aos portadores de necessidades especiais.*<br>Equipamentos: sede para administração, sanitários, seis quiosques para churrascos, restaurante, salão comunitário, bar. |
| Cachoeira do Tombo | Localizada no Assentamento Gleba Rio Vermelha na zona rural. A localidade mais próxima do atrativo na área urbana é a Residencial Acácia. O acesso utilizado freqüentemente, Avenida K e a continuidade do km 12, até o colchete de arame que dá acesso a trilha de 100 metros, que apresenta trechos íngremes chegando à cachoeira. As atividades realizadas no atrativo natural são caminhadas, banho e rapel. A cachoeira possui aproximadamente 3 metros de altura, que deságua em uma rocha inclinada e escorregadia. Recebe essa denominação, pois as pessoas sobem na rocha para tomar banho acabam escorregando e levando um tombo. *O local não proporciona acessibilidade aos portadores de necessidades especiais.*<br>Equipamentos: sede para administração; vestiários (masculino e feminino); *playground*; cozinha comunitária; quadra esportiva de areia; recanto com mesas; pedalinhos; sanitários. |
| Cachoeira do Vento | Localizada na Rodovia MT - 471 no Km 5,4 tendo como referencia as Sete Placas. O acesso mais utilizado é pelo Km 22 na BR 163, e percorrendo 2,5 km até a porteira e depois mais 2,2 km até a cachoeira. As atividades realizadas no atrativo natural são caminhadas, banho e rapel. A cachoeira possui aproximadamente 30 metros de altura. A queda d'água produz ventos constantes, é apropriada para a prática de rapel. Nas proximidades existem varias áreas de afloramento de água e vegetação endêmica. *O local não proporciona acessibilidade aos portadores de necessidades especiais.*<br>Equipamentos: sede para administração, sanitários, seis quiosques para churrascos, restaurante, salão comunitário, bar. |
| Cachoeira Pegada da Onça | Localizada na Rodovia MT - 471 no Km 5,4 tendo como referencia as Sete Placas. O acesso mais utilizado é pelo Km 22 na BR 163, e percorrendo 2,5 km até a porteira e depois mais 1,9 km até a cachoeira. As atividades realizadas no atrativo natural são caminhadas, banho e rapel. A cachoeira possui aproximadamente 30 metros de altura, rodeada por espécies arbóreas. Na parte onde a água cai forma-se um espraiado com área para banho. Com profundidade de cerca de 0,80 metros. Devido ao desnível geomorfológico e altura dos estratos vegetacionais, a radiação solar mais intensa se dá próximo ao meio dia. A cachoeira é apropriada para a prática de rapel. Nas proximidades existem varias áreas de afloramento de água e a presença de vegetação endêmica. *O local não proporciona acessibilidade aos portadores de necessidades especiais.*<br>Equipamentos: sede para administração, sanitários, seis quiosques para churrascos, restaurante, salão comunitário, bar. |
| Cachoeira de Cima da Boca de Caverna | Localizada no Córrego do Gavião, nas proximidades do Sitio Vale Encantado. O acesso mais utilizado é pelo Km 8,5 na BR - 364 saída para Cuiabá, entrada do Praia Clube, percorrendo mais de 15 km até a cachoeira. As atividades realizadas no atrativo natural são caminhadas, banho e cultos religiosos. A cachoeira possui aproximadamente 6 metros de altura e deságua formando um pequeno poço natural, que é utilizado para banho. Existem muitas espécies vegetais arbustivas e arbóreas no percurso, o que dificulta a entrada de luz solar na cachoeira. Próximo às margens do local existem várias áreas de afloramento de água, que estão sendo degradadas, diminuindo o fluxo de água da nascente. A trilha de acesso à cachoeira atravessa áreas de pequenas várzeas provocando desmatamento e compactação do solo. *O local não proporciona acessibilidade aos portadores de necessidades especiais.*<br>Equipamentos: sede para administração; vestiários (masculino e feminino); *playground*; cozinha comunitária; quadra esportiva de areia; recanto com mesas; pedalinhos; sanitários. |

**Quadro 5 - As Potencialidades Ecoturísticas no Município de Rondonópolis**
Org: CAMPOS, M. B. N. S. (2014)

O Assentamento Carimã faz limite com o Parque Estadual Dom Osório Stoffel e fica nas proximidades da Reserva Particular do Patrimônio Natural Parque Ecológico João Basso, unidades de conservação, com significativo patrimônio ecológico e diverso atrativo turístico. Não é possível realizar observações de fragmentos isolados dos espaços, porém devem-se considerar os fenômenos na sua totalidade, avaliando o entorno e a montante dos cursos hídricos e considerando os atrativos.

Na **Figura 4** percebe-se a localização das cachoeiras no Assentamento Carimã, existente no Córrego Água Fria, curso d’água que é afluente do Ribeirão Ponte de Pedra, juntamente no local onde este forma um Cânion. Já na **Figura 5** são visualizadas as cachoeiras do Assentamento da Gleba Rio Vermelho todas pertencentes ao Córrego do Gavião, tributário do Rio Vermelho.

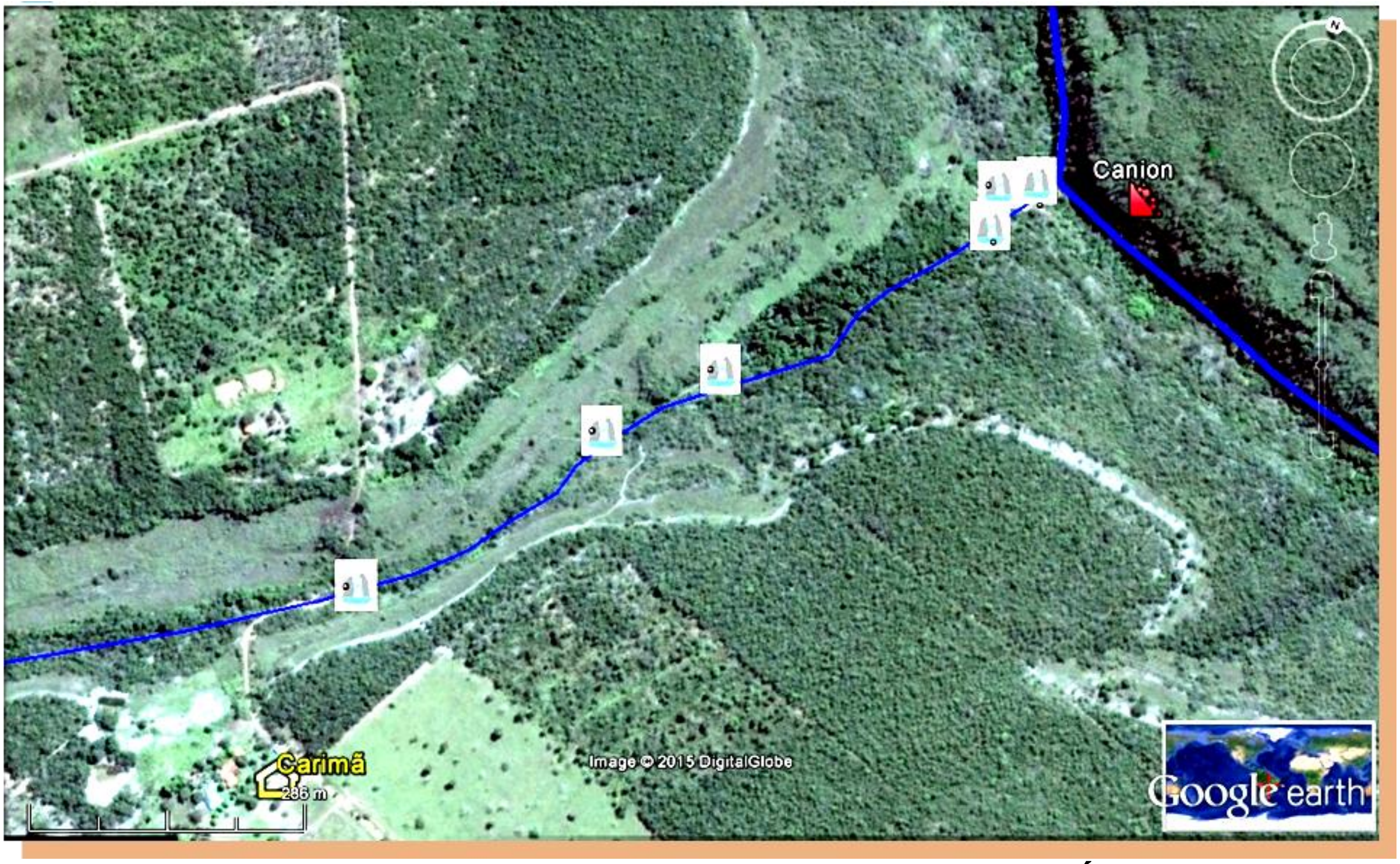

**Figura 4 - Assentamento Carimã destacando as cachoeiras do córrego Água Fria até desaguar no Ponte de Pedra onde forma um cânion**
Fonte: *Google Earth* (2015)
Org.: CAMPOS, M. B. N. S. (2015)

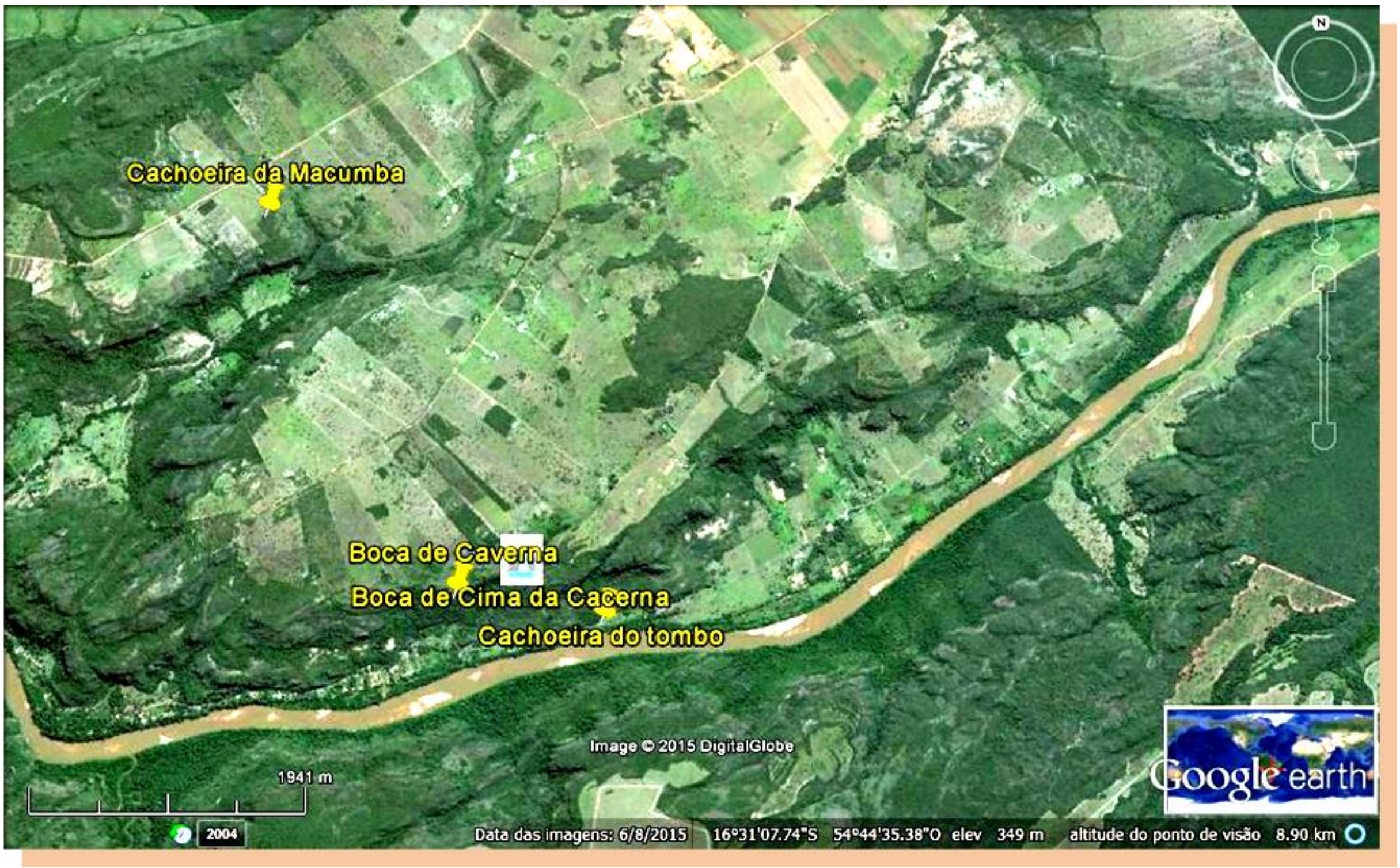

**Figura 5 - Assentamento Gleba Rio Vermelho destacando as cachoeiras até desaguar no Rio Vermelho**
Fonte: *Google Earth* (2015)
Org.: CAMPOS, M. B. N. S. (2015)

Na **Figura 6** é possível visualizar a localização das cachoeiras nos Assentamentos Carimã e Gleba Rio Vermelho município de Rondonópolis.

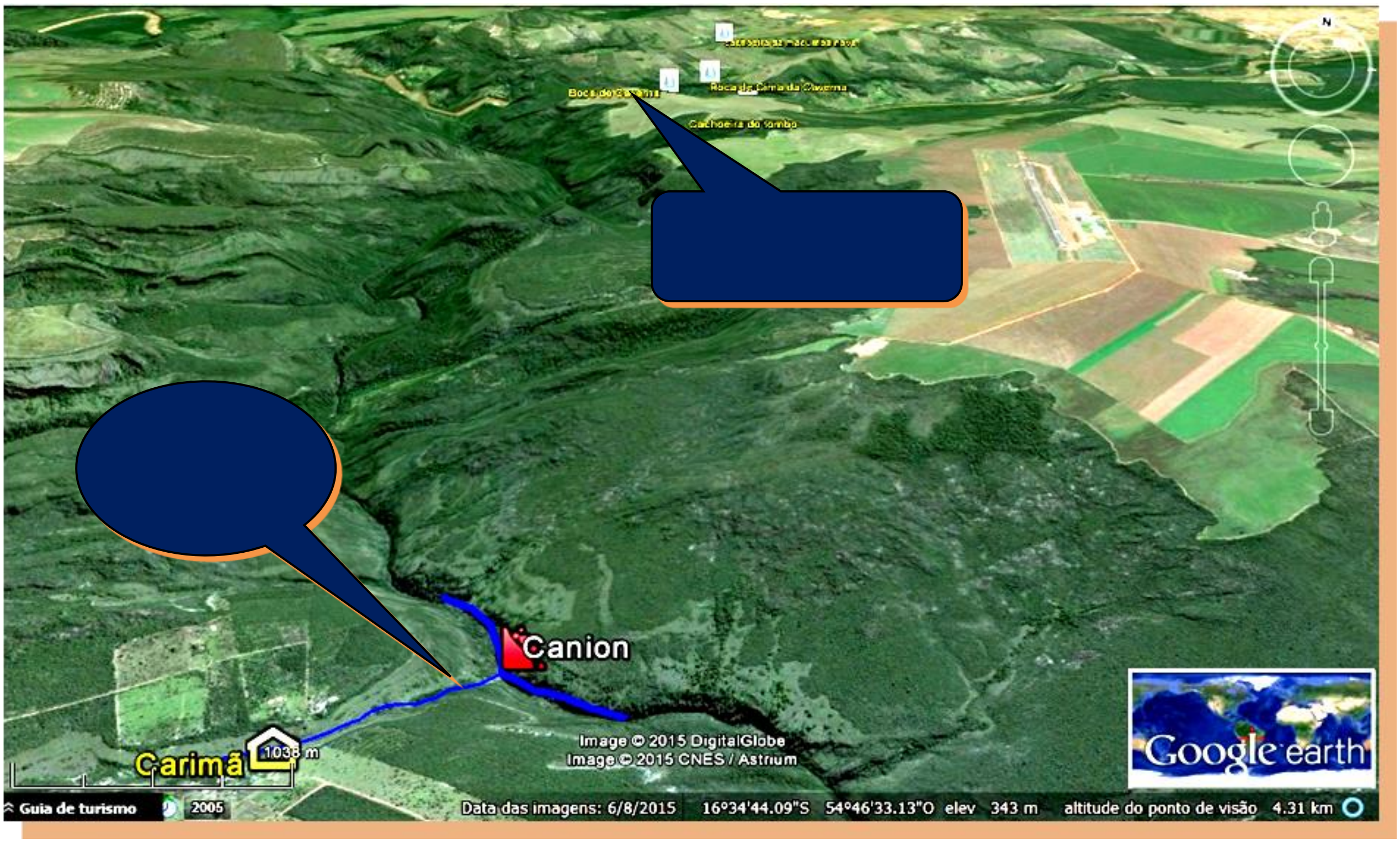

**Figura 6 - Localização dos Assentamentos Carimã e Gleba Rio Vermelho em Rondonópolis - MT**
Fonte: *Google Earth* (2015)
Org.: CAMPOS, M. B. N. S. (2015)

Como observado na pesquisa e relatado por moradores, no local não existe sinalização turística adequada e várias trilhas alternativas são criadas, geralmente sobre áreas de manancial aflorante as quais são impactadas com a supressão vegetacional e a compactação pelo pisoteio **Foto 1**. Além disso, lixo é lançado nos cursos hídricos e nas trilhas; ocorrem também pichações em árvores e rochas.

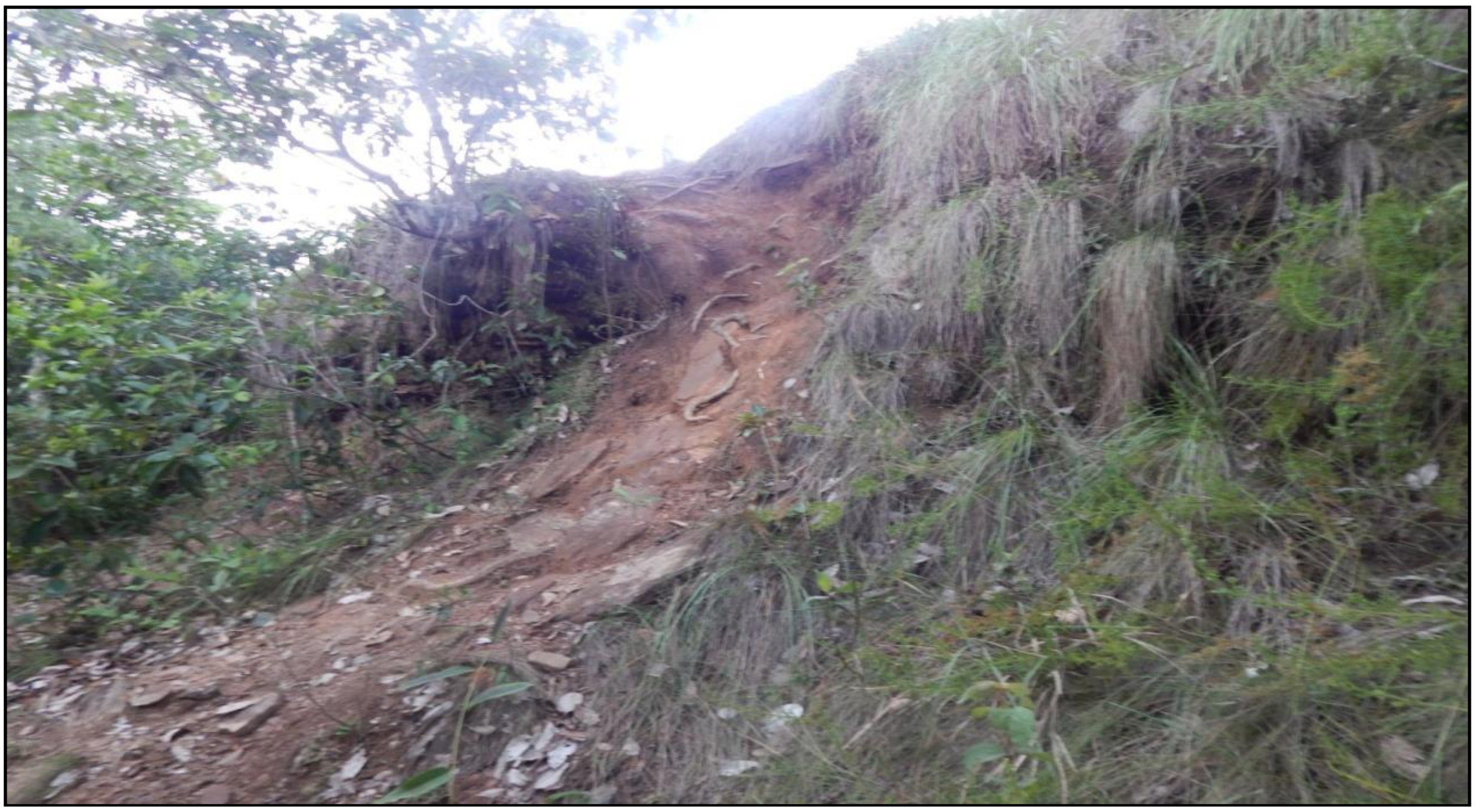

**Foto 1 - Trecho degradado em processo de erosão em uma das trilhas abertas para dar acesso às cachoeiras do Assentamento Carimã**
Org.: CAMPOS, M. B. N. S. (2014)

Através do que foi exposto, verifica-se que é de suma importância ocupação do espaço geográfico de forma ordenada, seguindo critérios conservacionistas baseados na legislação vigente e em critérios técnicos, principalmente nas áreas mais frágeis, uma vez que alterações como a extração da vegetação causam grandes prejuízos a este ambiente **Foto 2.**

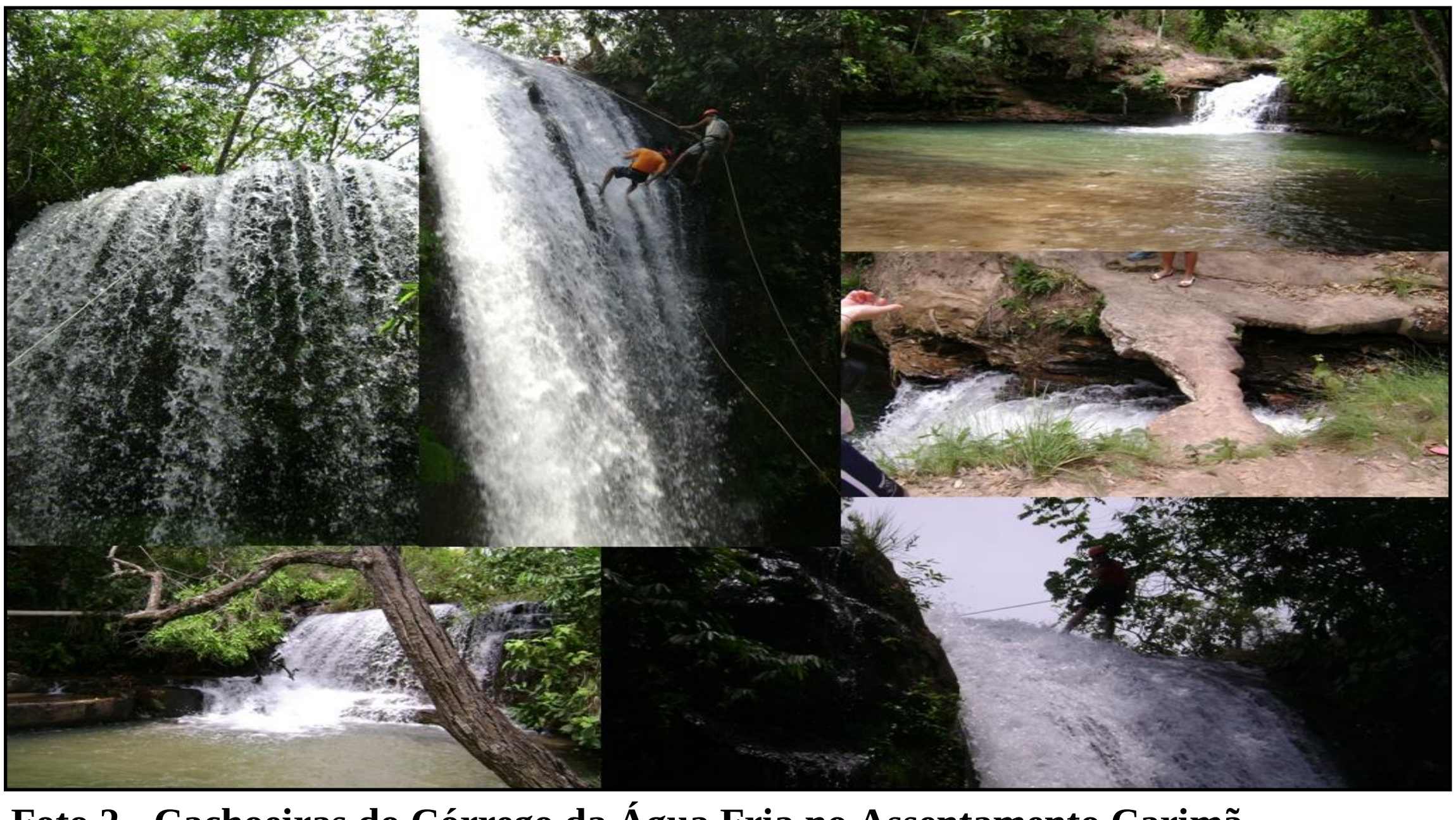

**Foto 2 - Cachoeiras do Córrego da Água Fria no Assentamento Carimã**
Org.: CAMPOS, M. B. N. S. (2014)

A partir dos dados, percebe-se a crescente importância do planejamento ambiental e de medidas conservadoras para essas áreas, tais como: o planejamento e construção de estruturas suspensas para os turistas, evitando o pisoteio e o desmatamento; restabelecimento dos espaços degradados; limitação do número de visitantes e a elaboração de estudos de capacidade de carga do lugar; sinalização adequada, entre outras. Medidas que podem suavizar os impactos nas áreas mais frágeis e intensamente utilizadas por visitantes do assentamento além de fornecer melhoria na qualidade de vida dos frequentadores, proporcionada pelo contato com áreas rurais e com vegetação natural conservada.

Como se observa na **Figura 7** ao longo do Rio Vermelho encontram-se cachoeiras e as prainhas, cabe aos moradores desses espaços, porém, lutar por sua conservação e sustentabilidade, e para isso é primordial um planejamento que envolva todos os atores sociais: frequentadores, moradores, instituições públicas e agências de turismos.

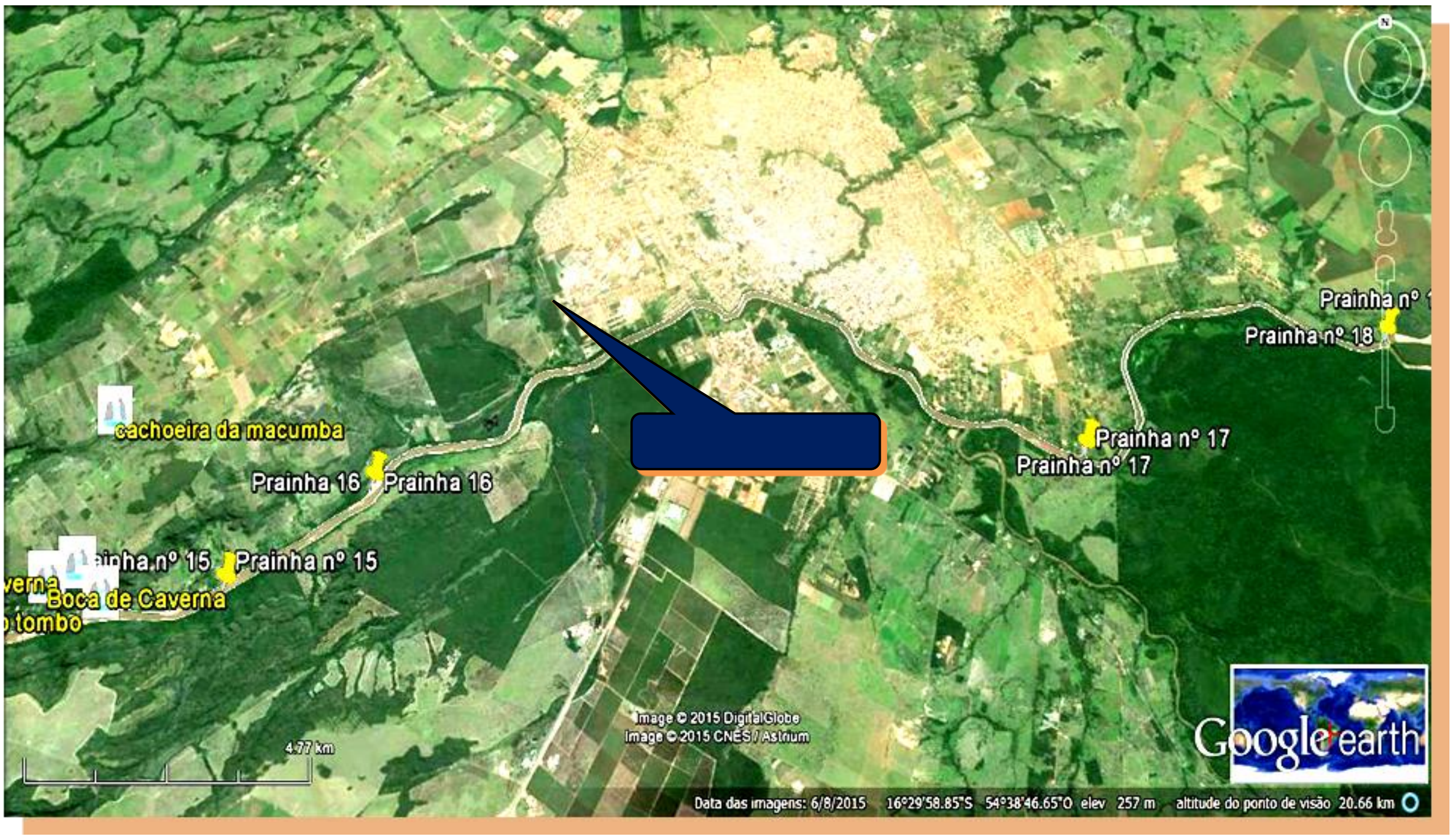

**Figura 7 – As Cachoeiras e as Prainhas ao longo do Rio Vermelho**
Fonte: *Google Earth* (2015)
Org.: CAMPOS, M. B. N. S. (2015)

## 4.2 ATORES SOCIAIS, GESTORES PRIVADOS E PÚBLICOS

Esta pesquisa contou com a participação de quatro grupos de atores sociais: os usuários frequentadores representados pelas pessoas que visitam as áreas de lazer nos finais de semana e/ou feriados, os moradores, pessoas que residem nas localidades e proximidades das áreas de lazer, o turismo receptivo, por agências de turismo localizadas na área urbana de Rondonópolis e as Instituições Públicas representadas pelo Secretário do Meio Ambiente e o Secretário de Ciência, Tecnologia, Turismo e Desenvolvimento Econômico.

Para Moreno (2005, p.195) o turismo ecológico e rural são formas potenciais viáveis em muitos municípios do Estado de Mato Grosso. Mas, por ele a valorização da cultura da localidade deve estar articulada a conservação do ambiente de forma harmônica, para a prática de atividades ecoturísticas. Ele afirma ainda, que o sucesso do ecoturismo encontra-se condicionado à boa vontade e envolvimento dos moradores das localidades.

A economia de Mato Grosso é apoiada pela atividade agropecuária que vem crescendo nas últimas décadas, em contrapartida a população rural vem reduzindo em relação ao crescimento da população urbana, e os moradores residentes nas áreas de lazer também são afetados por essa dinâmica.

De acordo com os resultados da pesquisa realizada nos locais das áreas de lazer o número de homens é ligeiramente superior ao sexo feminino **(Gráfico 1).**

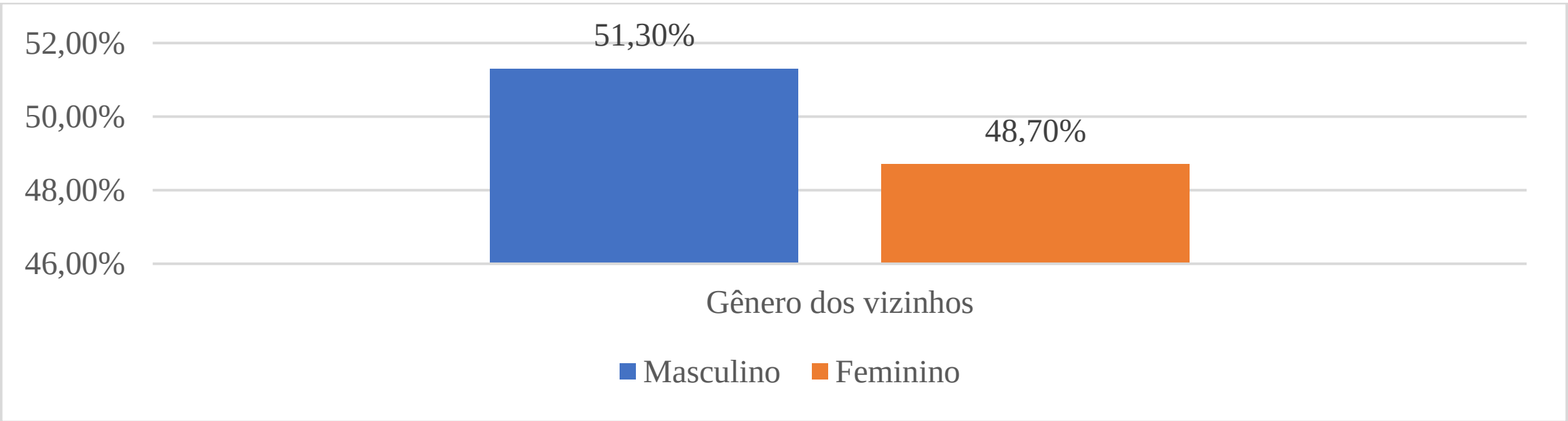

**Gráfico 1 – Gênero dos vizinhos das áreas de lazer.**
Fonte: Pesquisa (2015).

Como mostra o **Gráfico 2** a maior frequência de faixa etária dos moradores destas áreas de lazer está entre 36 e 45 anos.

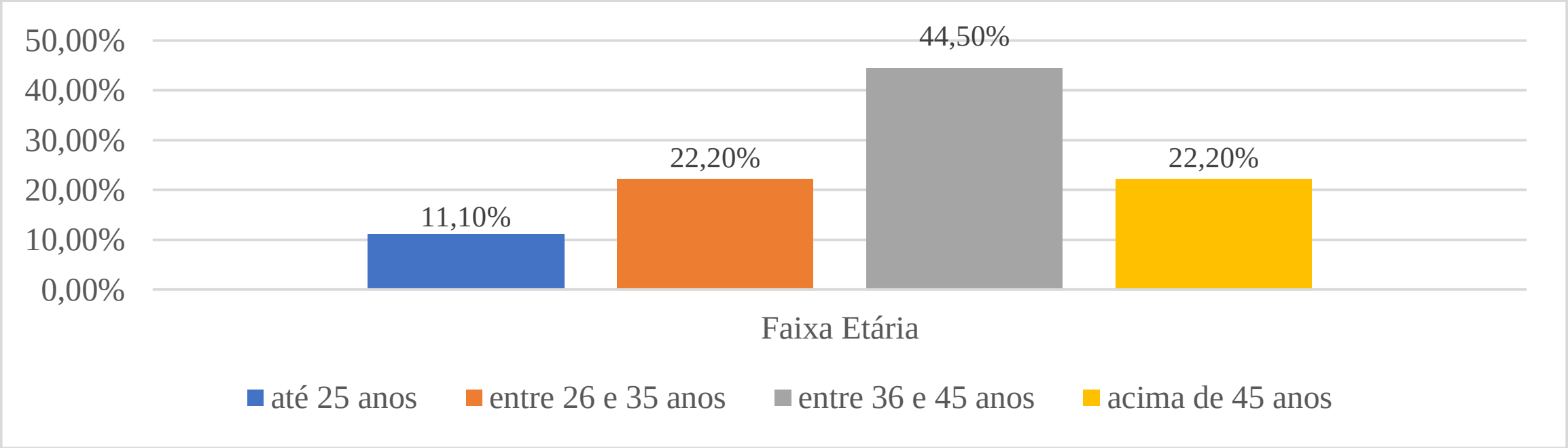

**Gráfico 2 – Faixa etária dos vizinhos das áreas de lazer.**
Fonte: Pesquisa (2015).

No que se refere ao grau de escolaridade, verifica-se um equilíbrio entre aqueles que possuem o ensino médio incompleto e o ensino superior incompleto, existindo ainda um percentual de 11,1% que não responderam por falta de conhecimento educacional, conforme aponta o **Gráfico 3**.

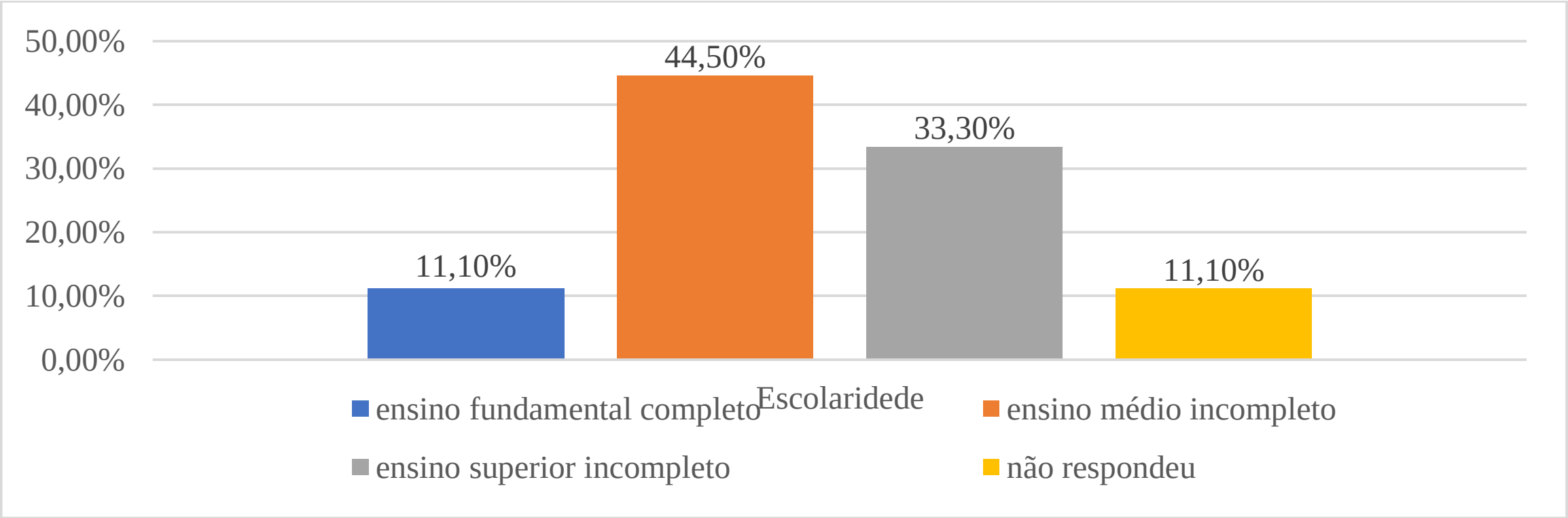

**Gráfico 3 – Grau de escolaridade dos vizinos das áreas de lazer.**
Fonte: Pesquisa (2015).

A maioria das residências acomoda entre 5 e 7 pessoas, como mostra o **Gráfico 4**, salientado assim, a necessidade das famílias desenvolverem atividades ecoturísticas para garantir retorno financeiro, evitando o êxodo rural e a fragilização econômica como consequência do desemprego ou subempregos nos centros urbanos, que muitas vezes gera a hipertrofia.

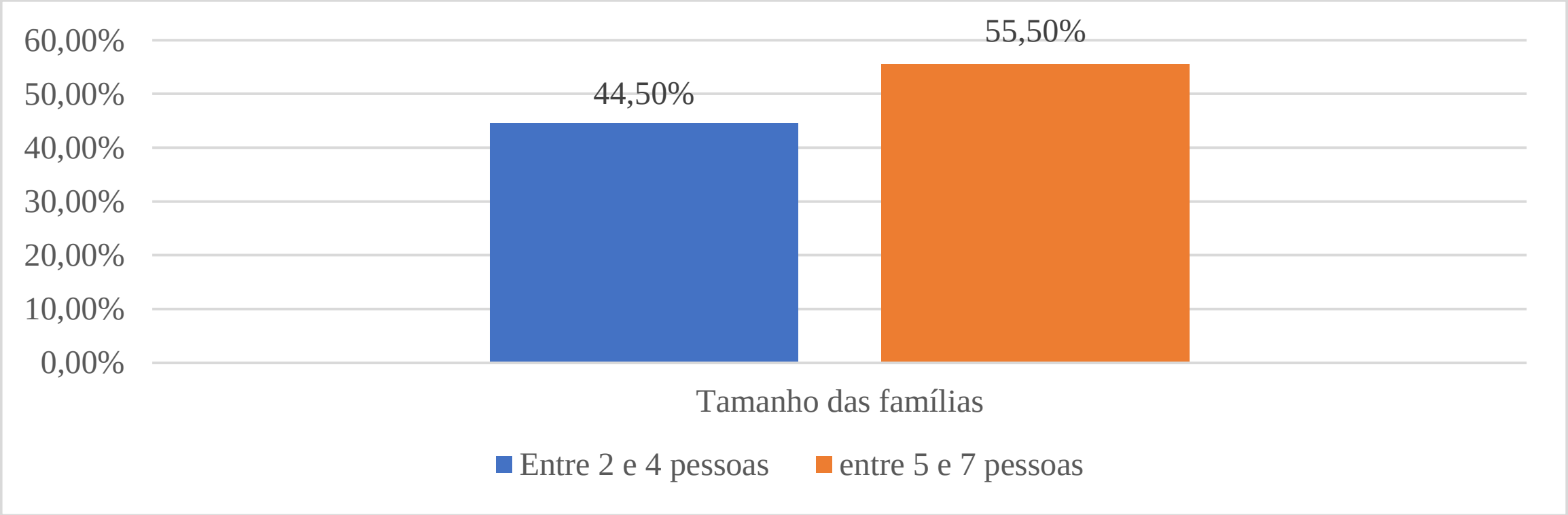

**Gráfico 4 – Tamanho das famílias vizinhas às áreas de lazer.**
Fonte: Pesquisa (2015)

O **Gráfico 5** indica que a renda familiar apresenta-se bastante discrepante, a grande maioria possui renda entre R$ 1.001,00 e R$ 3.000,00, enquanto uma minoria possui renda acima de R$ 3.001,00. Sem dúvida, este é dado demonstra uma inversão quanto à quantidade de pessoas residentes nestas localidades **(Gráfico 4)**.

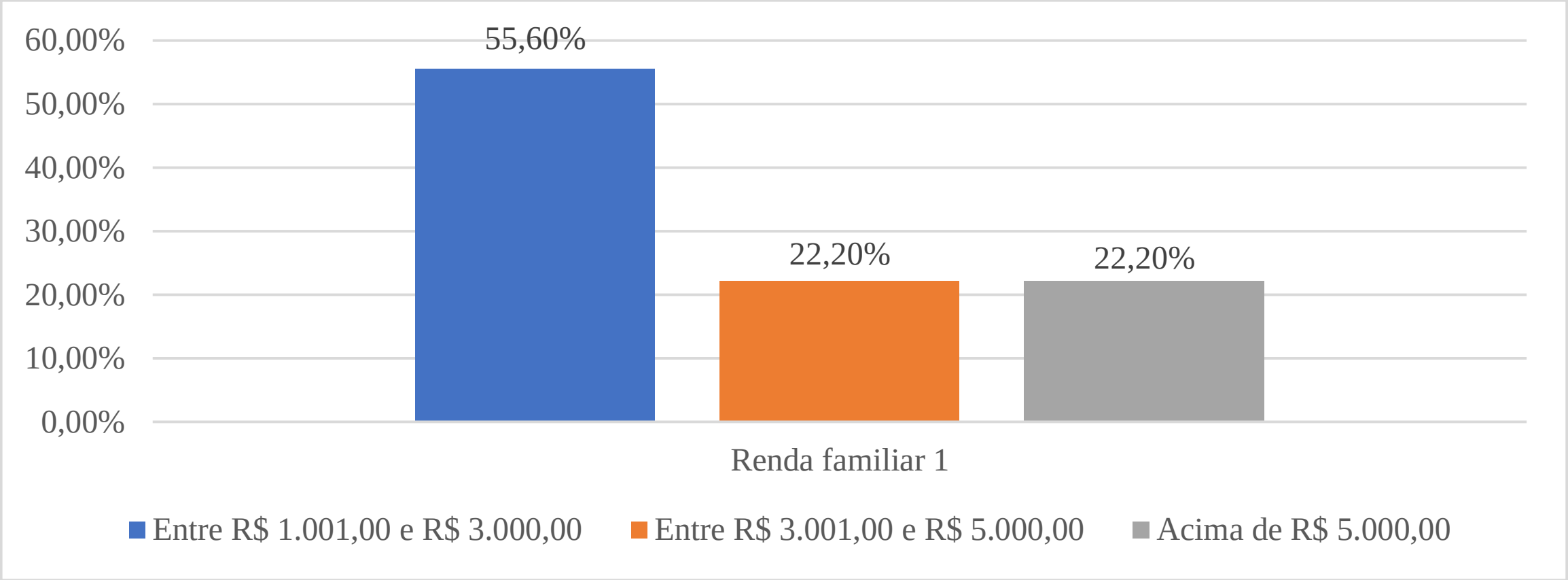

**Gráfico 5 – Renda familiar dos vizinhos das áreas de lazer.**
Fonte: Pesquisa (2015).

O **Gráfico 6** ilustra o tempo que os mesmos residem nas localidades. A maioria respondeu entre 2 e 5 anos, o que chama a atenção neste dados é o fato de existir um grande contingente de novos moradores, principalmente jovens, pois 33,3% responderam ter até 25 anos e 22,2% entre 26 e 35 anos segundo dados do **(Gráfico 2)**. O tempo de permanência nas localidades está relacionado com diversos fatores, dentre os quais a renda familiar e a atividade laboral.

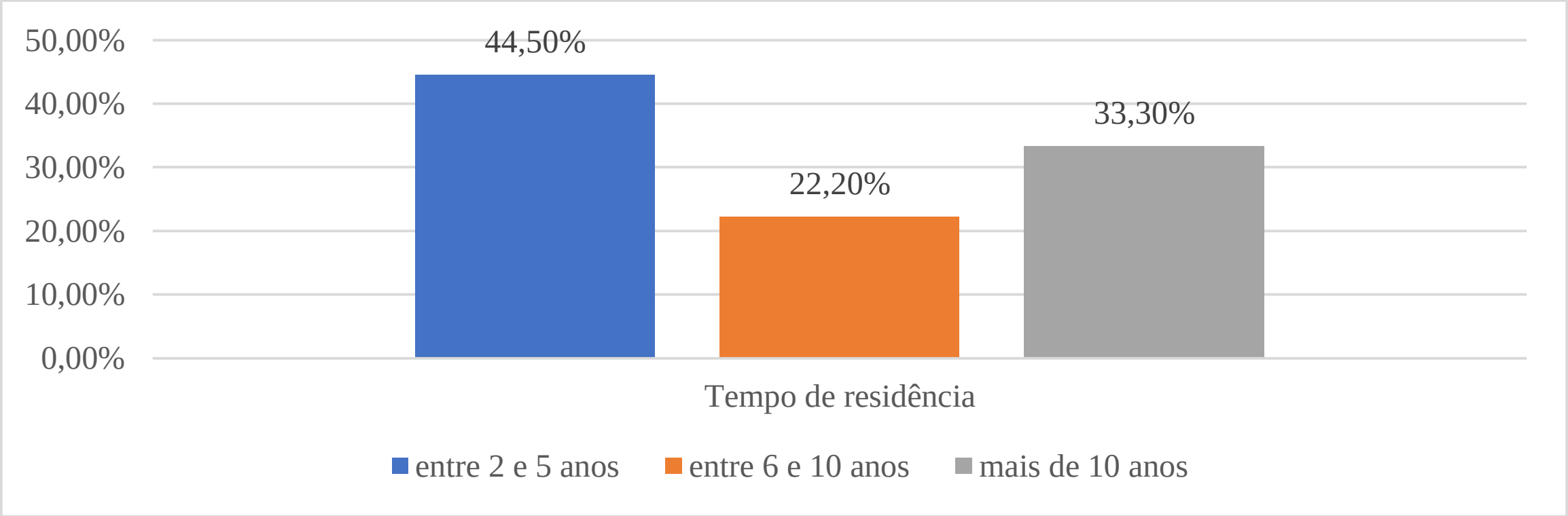

**Gráfico 6 – Tempo que os vizinhos residem nas localidades**
Fonte: Pesquisa (2015)

No **Gráfico 7**, apresenta expressiva parcela de pessoas que não conhecem a diferença entre turismo e ecoturismo, ou seja, desconhecem a importância de desenvolver o turismo com a conservação dos recursos naturais, especialmente, os hídricos que se apresentam em processo de escassez em todo o território nacional.

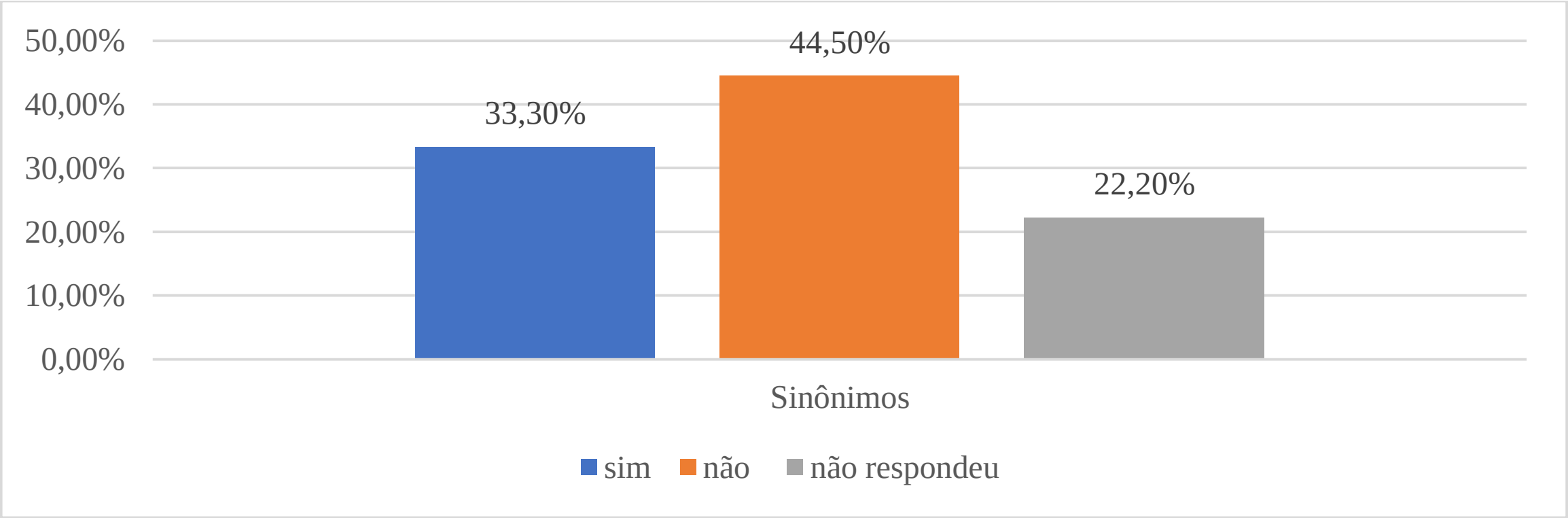

**Gráfico 7 – Turismo e ecoturismo são sinônimos**
Fonte: Pesquisa (2015)

Como esclarece Schneider (2006), há que se considerar que ambos estão fortemente interligados e fazem parte de um mesmo contexto de aproveitamento de recursos naturais ou artificiais para o desenvolvimento de lazer e atividades turísticas.

O **Gráfico 8** refere-se ao benefício econômico do ecoturismo para os moradores das localidades, que de forma mais ampla, acentua-se positivamente, tendo em vista que 77,8% o avaliaram entre bom e excelente, demonstrando inclusive aprovação com relação a essa atividade.

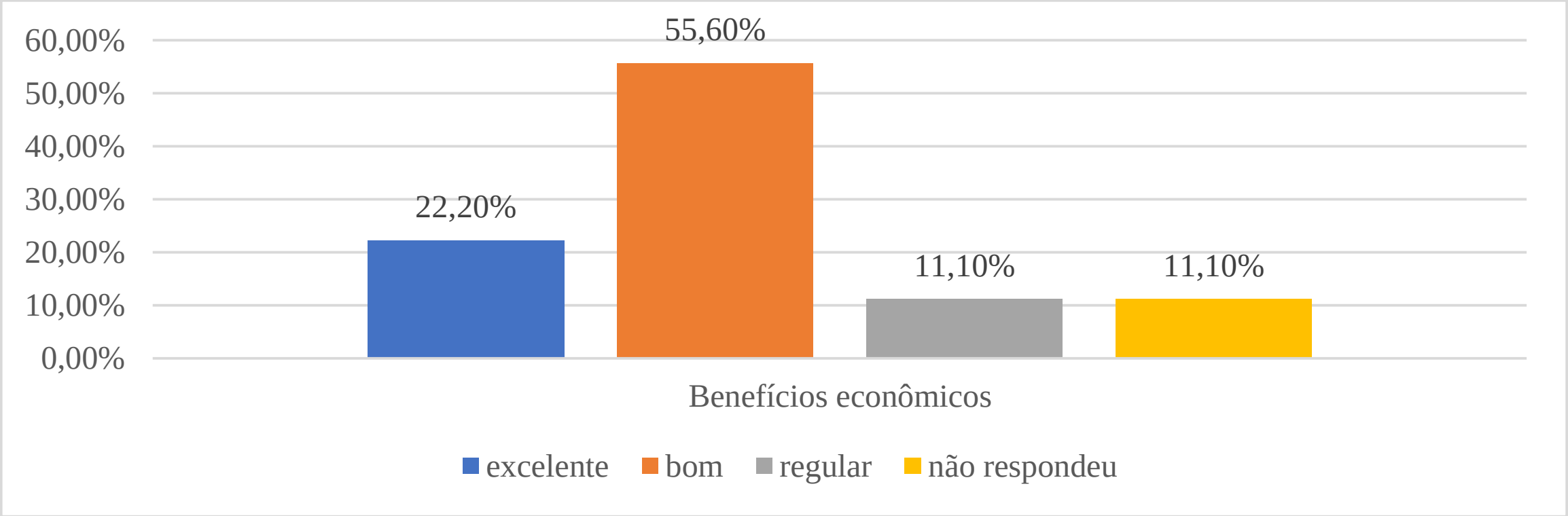

**Gráfico 8 – Benefício econômico dos frequentadores para os moradores**
Fonte: Pesquisa (2015)

Deve-se ressaltar que a relação moradores/frequentadores não é de todo positiva, já que 44,5% destacaram-na como regular, conforme mostra **Gráfico 9.**

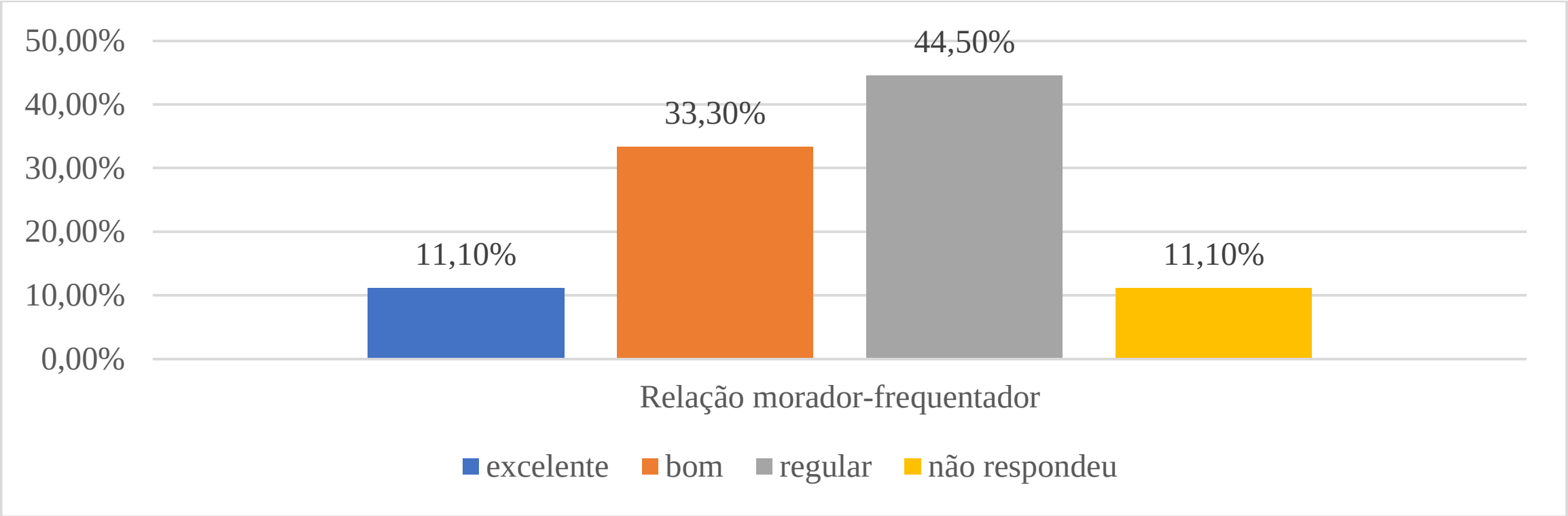

**Gráfico 9 – Avaliação da relação morador-frequentador**
Fonte: Pesquisa (2015)

Conforme o entendimento de Brasil (2008) somente existe ecoturismo quando há no processo preocupação com a conservação dos recursos naturais do local em que é desenvolvida a atividade ecoturística.

Em uma análise quantitativa do **Gráfico 10**, entretanto, percebe-se que 89% dos moradores possuem pouca preocupação com a conservação dos recursos naturais de forma sustentável.

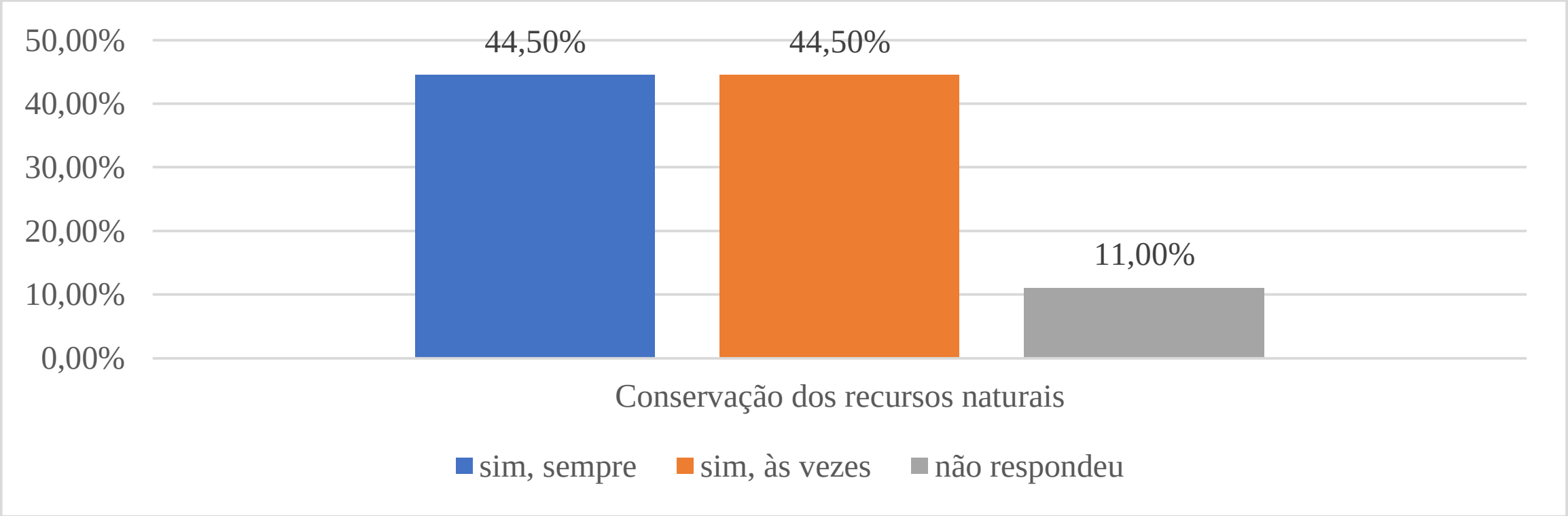

**Gráfico 10 – Preocupação dos moradores com a conservação dos recursos naturais**
Fonte: Pesquisa (2015)

A participação dos moradores em atividades que causam impactos ambientais é um dado relevante. Como se observa no **Gráfico 11**, para os pesquisados, ocorre reduzidos impactos ambientais nas localidades, alguns consideram inexistentes, e alguns nem mesmo responderam. Todavia, nas observações *in loco* identificamos a compactação, erosão e perda do solo; assoreamento dos rios; lixo lançado nos cursos hídricos e nas trilhas; pichações nas árvores e rochas; afastamento da fauna; desequilíbrio ecológico causado pelo desmatamento e queimadas e redução da biodiversidade.

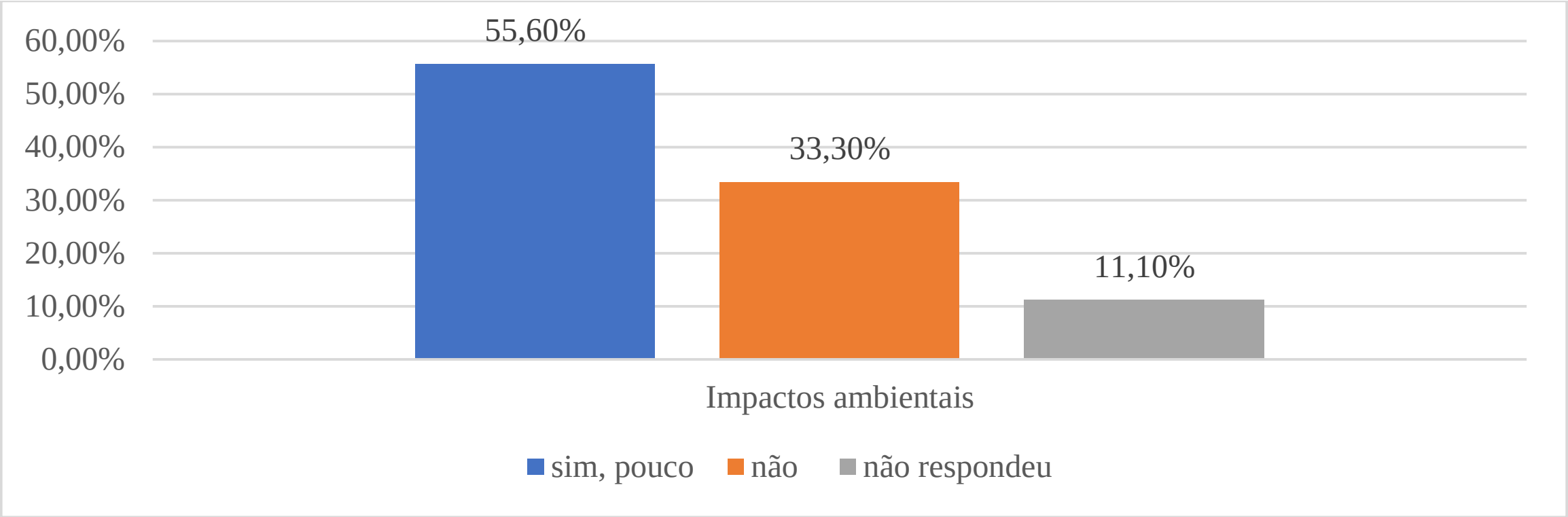

**Gráfico 11 – Impactos ambientais causados pelas atividades segundo os moradores das localidades**
Fonte: Pesquisa (2015)

De acordo com dados do **Gráfico 12**, o principal benefício do ecoturismo para a comunidade é o econômico. É necessário salientar que o ecoturismo tem relevância no campo econômico, apesar disso, o seu desenvolvimento não deve se embasar somente nestes benefícios.

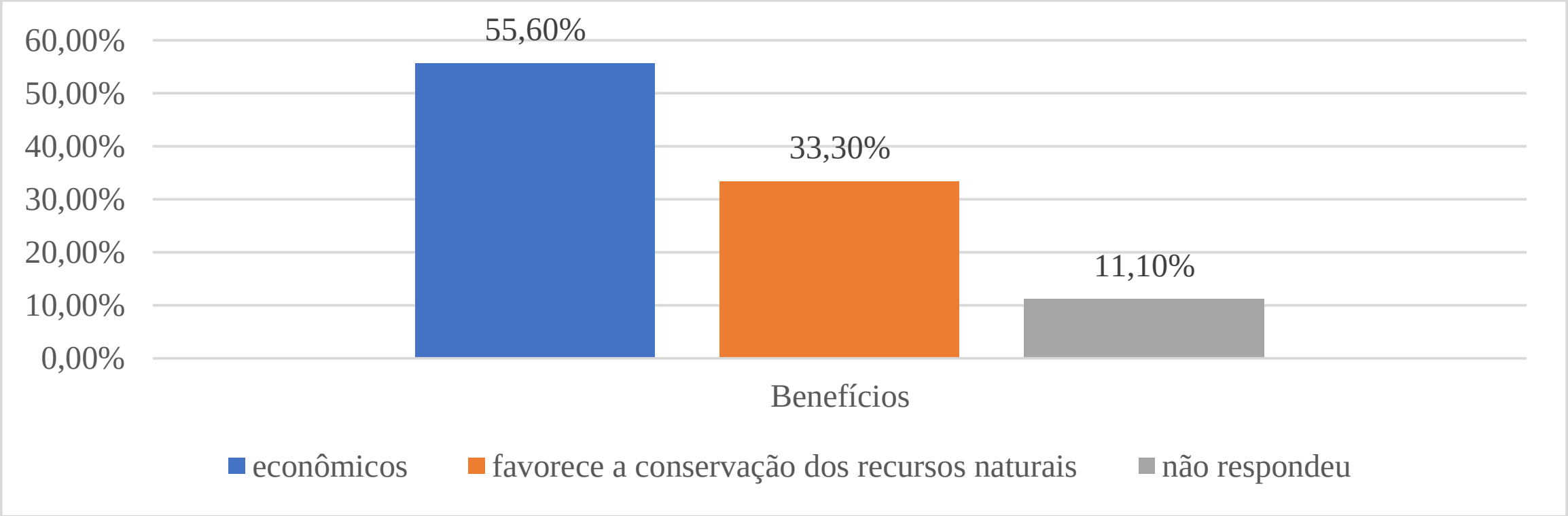

**Gráfico 12 – Benefícios trazidos pelos usuários frequentadores para as localidades**
Fonte: Pesquisa (2015)

As características das atividades ecoturísticas exigem atitudes e comportamentos, que remetem à mudanças dos padrões de produção e consumo com postura socioambiental responsável no uso dos recursos naturais, que conforme aponta o **Gráfico 13**, pois a maioria considera a destruição dos recursos hídricos como a principal problemática trazida pelos ecoturistas para a comunidade.

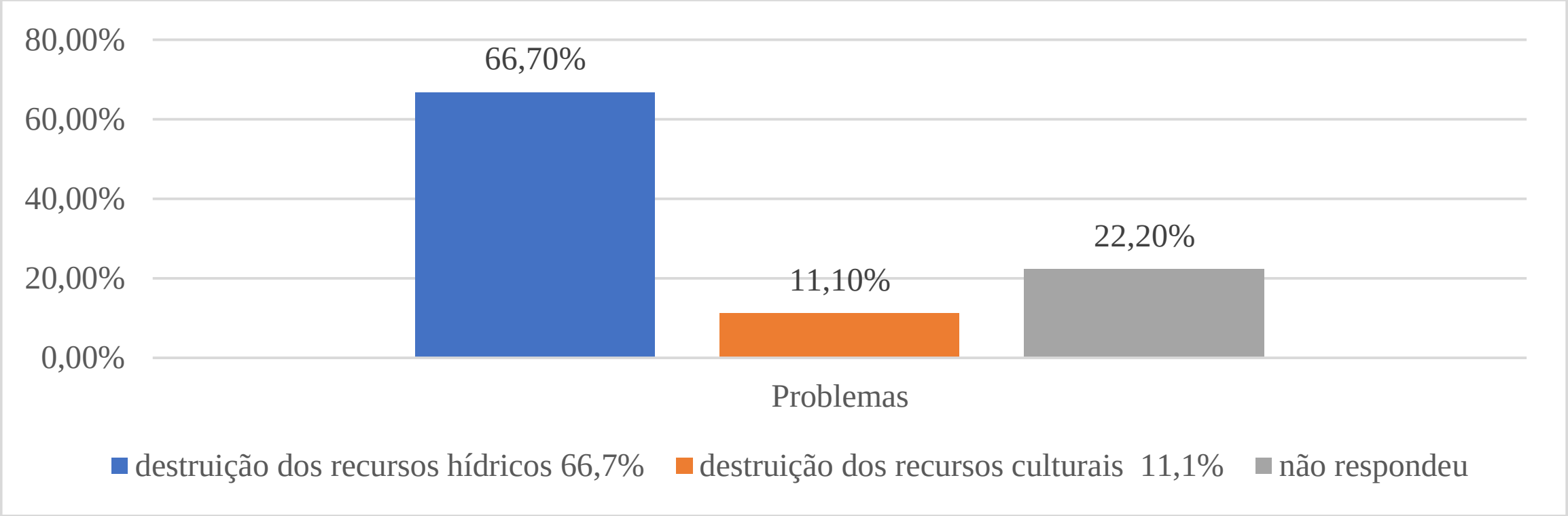

**Gráfico 13 – Problemas trazidos pelos frequentadores para as localidades**
Fonte: Pesquisa (2015)

Nestas localidades existe um número significativo de atividades, o que aponta para a necessidade de promoção do que as áreas oferecem com divulgação e roteiros ecoturísticos. Só assim, será possível potencializar os atrativos turísticos nas localidades receptoras.

### 4.2.1 Frequentadores das Áreas de Lazer

O uso sustentável das potencialidades ecoturísticas é relevante para que mudanças de uso dos recursos hídricos ocorram, pois a sensibilização dos usuários envolve interligação entre a parte socioambiental e aspectos econômicos. Os gestores levam em conta apenas aspectos econômicos sem considerar a percepção dos usuários frequentadores, fator importante para o desenvolvimento das atividades ecoturísticas de uma localidade.

As atividades ecoturísticas atraem pessoas de diferentes sexos, seja pelo convívio com a natureza ou a qualidade de vida que esta atividade oferece. Muito embora, conforme o **Gráfico 14** haja a predominância do gênero masculino, normalmente as mulheres possuem uma maior capacidade de distinção de ambiente e em geral são mais detalhistas que os homens, o que leva suas opiniões a serem mais criteriosas na avaliação dos mínimos detalhes existentes, principalmente nos atrativos hídricos das localidades.

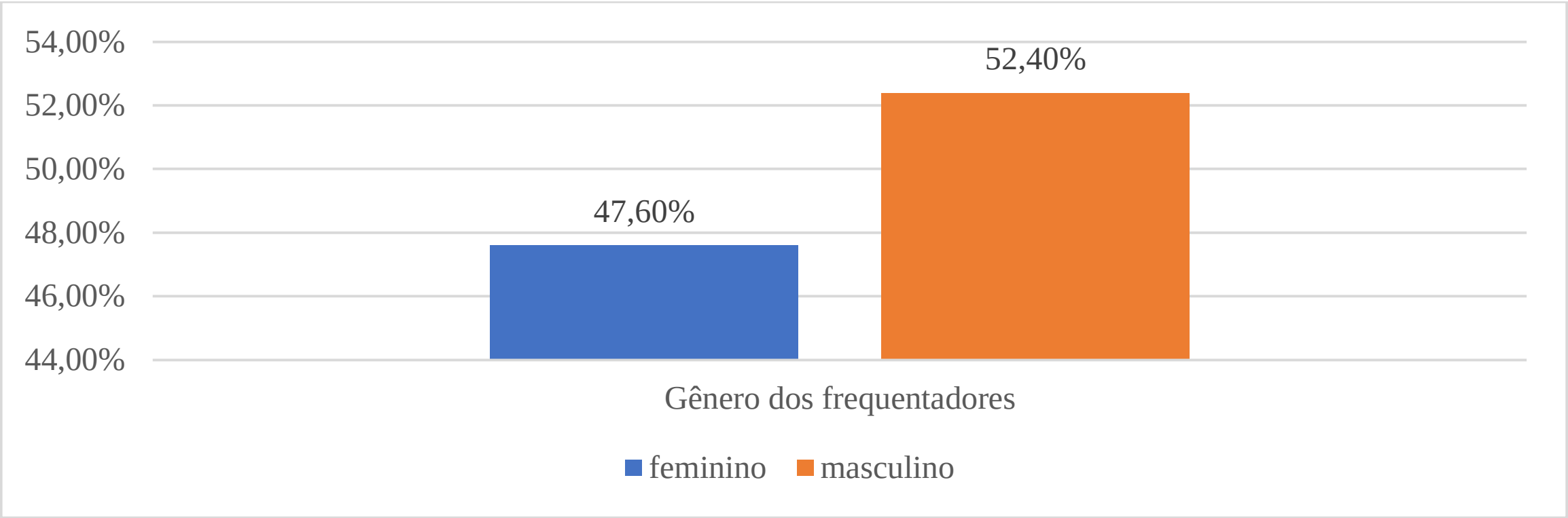

**Gráfico 14 – Sexo dos frequentadores**
Fonte: Pesquisa (2015)

A faixa etária dos frequentadores figura uma pequena predominância de pessoas entre 36 e acima de 45 anos **(Gráfico 15)**, o que revela que certa maturidade dos frequentadores, uma vez que o turismo ecológico surge na possibilidade do contato com a natureza, relaxar a mente e exercitar o corpo, o que evidencia que as áreas de lazer são excelente refúgio para descansar e libertar o estresse da cidade.

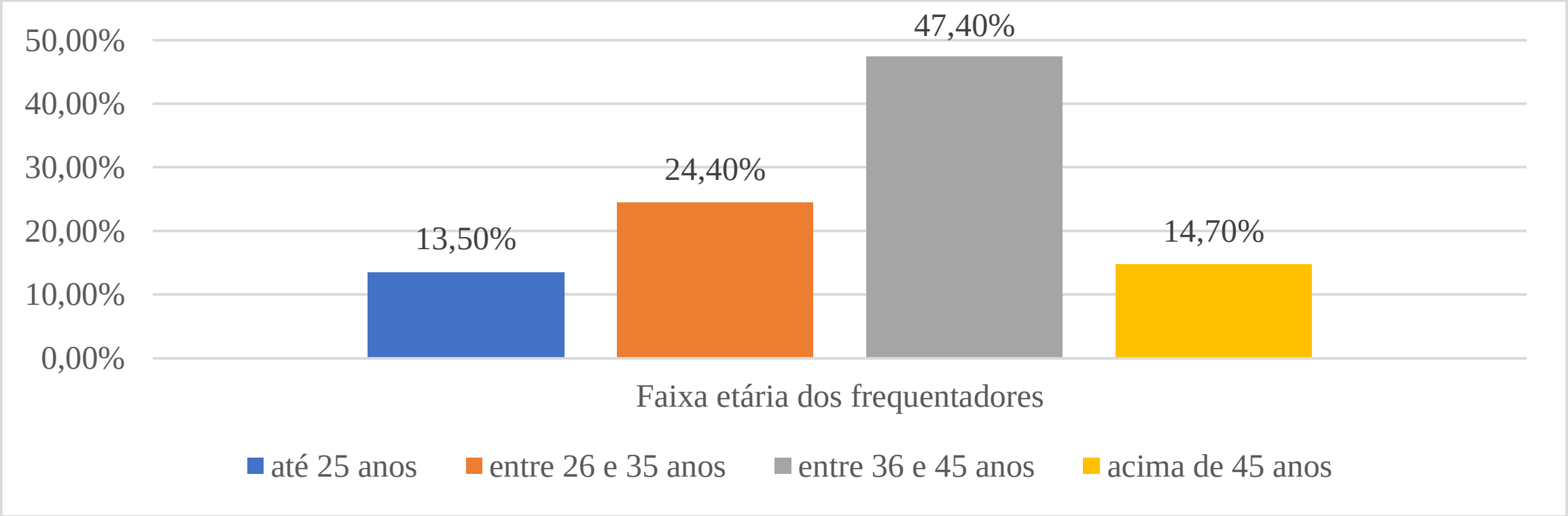

**Gráfico 15 – Faixa etária dos frequentadores**
Fonte: Pesquisa (2015)

Interessante salientar que Sousa (2011) menciona que o turismo, especialmente o ecoturismo, é uma atividade que proporciona qualidade de vida, o que pode justificar um fluxo maior de pessoas nas faixas etárias mais elevadas, inclusive na terceira idade.

Quanto ao índice do grau de escolaridade dos frequentadores, observou-se a predominância do ensino superior completo, o que pode revelar que as atividades de ecoturismo são desenvolvidas por pessoas com mais elevado nível de conhecimento acadêmico, pois o lazer insere-se como alternativa para utilização de tempo livre, proporcionando descontração e descanso das atividades diárias **(Gráfico 16).**

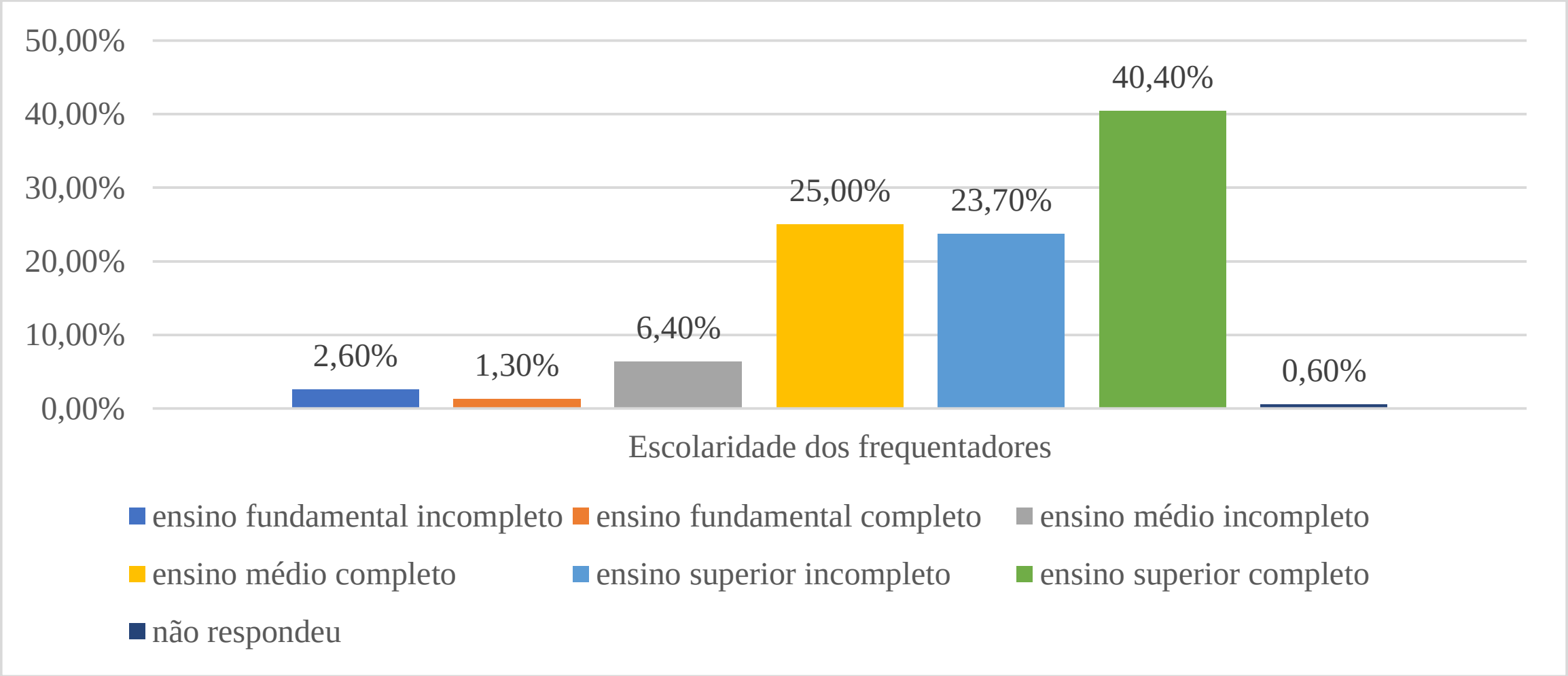

**Gráfico 16 – Grau de escolaridade dos frequentadores**
Fonte: Pesquisa (2015)

A renda familiar dos frequentadores apontou, em sua maioria entre, renda entre R$ 3.001,00 e mais de R$ 5.000,00. Importante salientar que a renda familiar é fator preponderante para a prática de lazer, o que revela um contingente de pessoas com renda significativa a se engajar em passeios ecoturísticos **(Gráfico 17).**

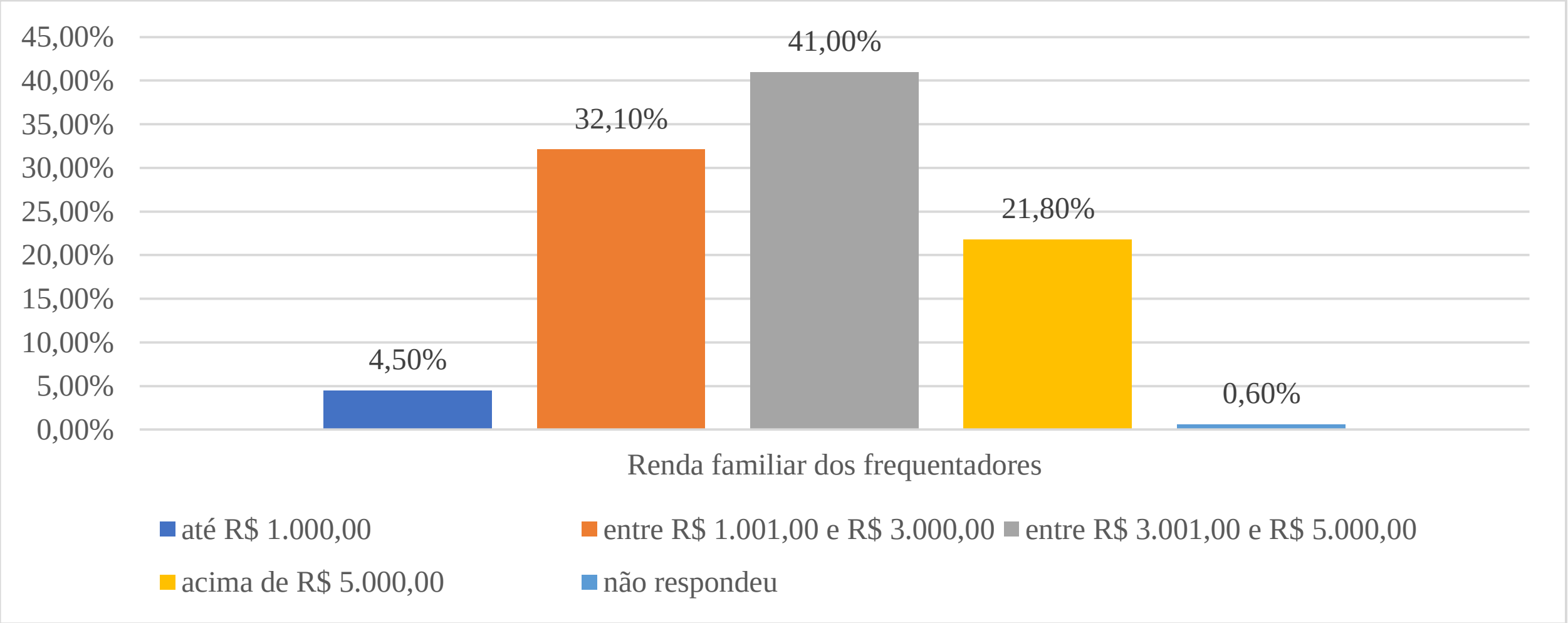

**Gráfico 17 – Renda familiar dos frequentadores**
Fonte: Pesquisa (2015)

A partir da observação do **Gráfico 18**, percebe-se que, o ecoturismo vem ganhando maior espaço nas discussões acadêmicas e sociais nos últimos cinco anos. A verdade é que no município essa atividade está em franco crescimento há mais de uma década, e muito embora essa prática venha crescendo continuamente, ainda existem limitações na relação sustentabilidade e ecoturismo, ressaltando a importância de fortalecer a sua sensibilização e responsabilidade dos ecoturistas com o passar dos anos e a continuidade deste lazer.

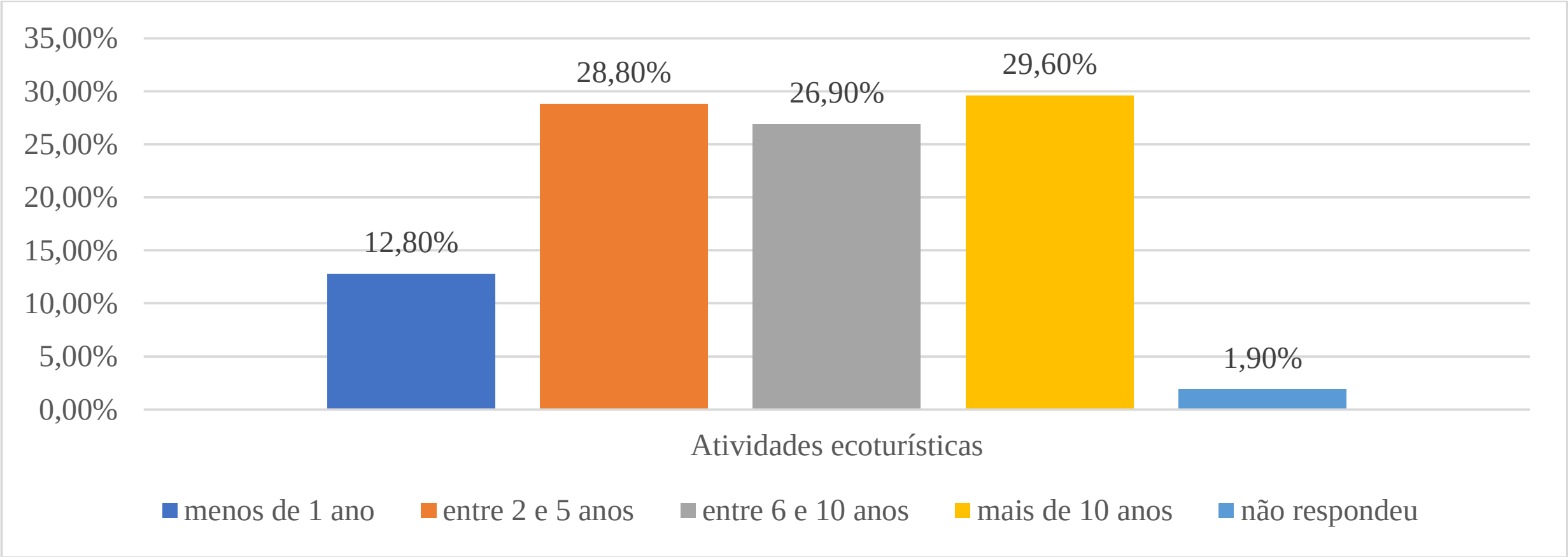

**Gráfico 18 – Tempo que desenvolve atividades ecoturísticas**
Fonte: Pesquisa (2015)

Mesmo que existam interesses econômicos, sociais, ambientais e culturais para desenvolver as atividades turísticas, Silva, Santos e Benevides (2007) evidenciam problema do uso inadequado dos recursos naturais, fato este que ocorre de forma especial nos países em desenvolvimento como é o do Brasil.

O **Gráfico 19** evidencia que, na opinião de 51,9% dos ecoturistas entrevistados o ecoturismo e o turismo são atividades diferentes, ao passo que 45,5% consideraram que é a mesma atividade.

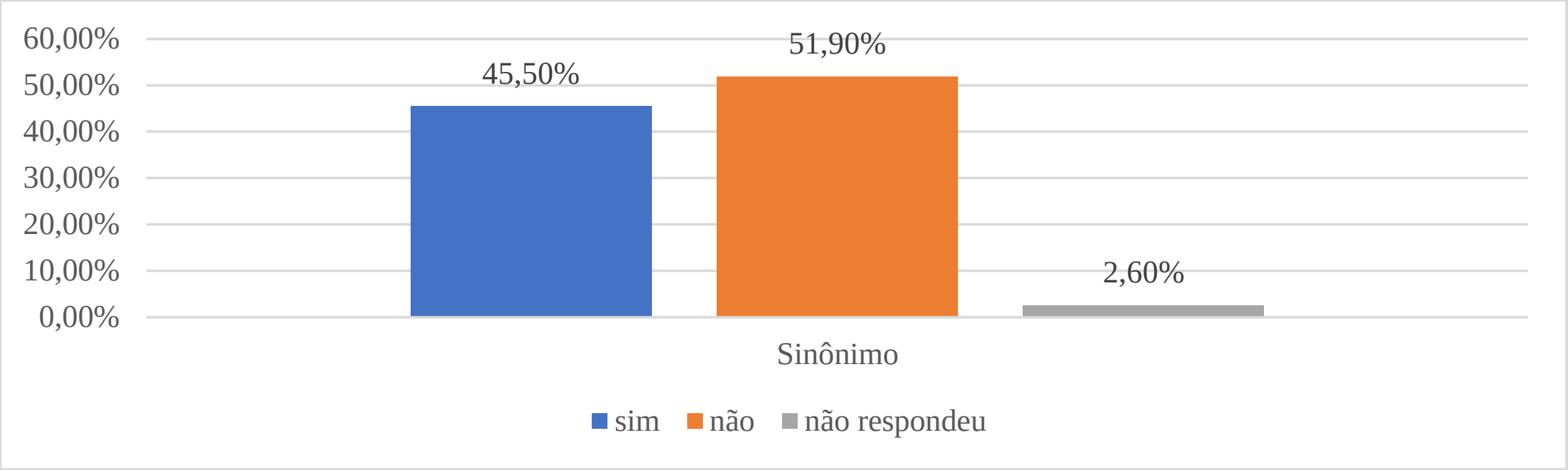

**Gráfico 19 – Turismo e ecoturismo é a mesma atividade**
Fonte: Pesquisa (2015)

O ecoturismo abrange os recursos naturais e as atividades desenvolvidas em meio à natureza. Silva, Santos e Benevides (2007) frisam que ele possibilita o aporte econômico e social para pessoas que vivem em locais mais distantes, beneficiando qualidade de vida para os moradores das áreas em que é desenvolvida essa atividade, pois o turismo é uma atividade que pode também trazer estes aportes, bem como desenvolver-se na área urbana ou na área rural.

Quanto à interferência do turismo na oferta de recreação no município, a maioria 74,4% dos atores pesquisados têm a percepção de que o turismo contribuiu para o aumento das opções de recreação. Enquanto que a minoria de 14,8% declarou que não percebe que o turismo não interfere ou não contribua para a ampliação das opções de recreação nas localidades **(Gráfico 20).** Cruz (2001) defende que não existe atividade humana que não altere de alguma forma o ambiente, neste sentido, a prática de turismo ou ecoturismo evidencia-se como um dos elementos transformadores da realidade vivenciada nas localidades mencionadas.

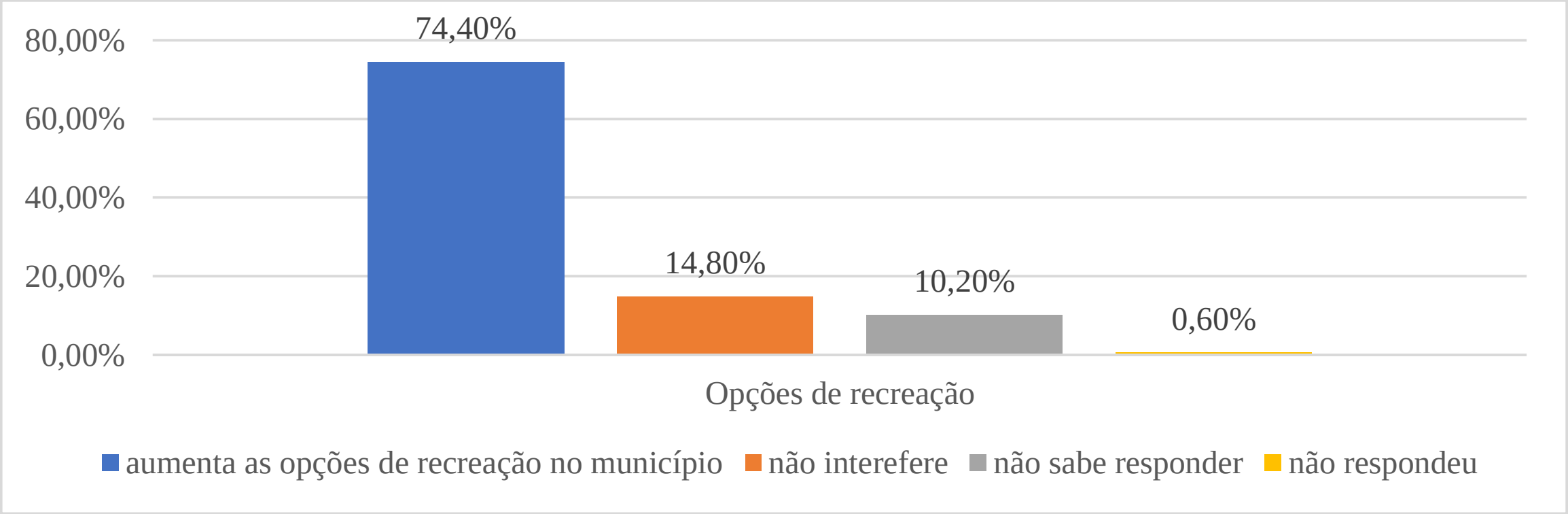

**Gráfico 20 – Em sua opinião o turismo as opções de recreação no município?**
Fonte: Pesquisa (2015)

As atividades ecoturísticas, por vezes, geram efeitos negativos, pois o uso dos recursos naturais, especificamente os hídricos, em alguns casos pode gerar problemas para o meio ambiente e/ou as localidades receptoras.

Quanto ao conhecimento das áreas de lazer e outros atrativos, 67,9% dos atores acreditam que o turismo sustentável ajuda na utilização dos recursos hídricos, bem como, 32,1% não tem conhecimento a respeito **(Gráfico 21)**. Cruz (2001) argumenta que quanto aos impactos do turismo sobre ambientes naturais, a conservação dos recursos somente ocorre com um turismo ambiental e socialmente correto, em que exista o comprometimento de todos para a conservação dos recursos naturais.

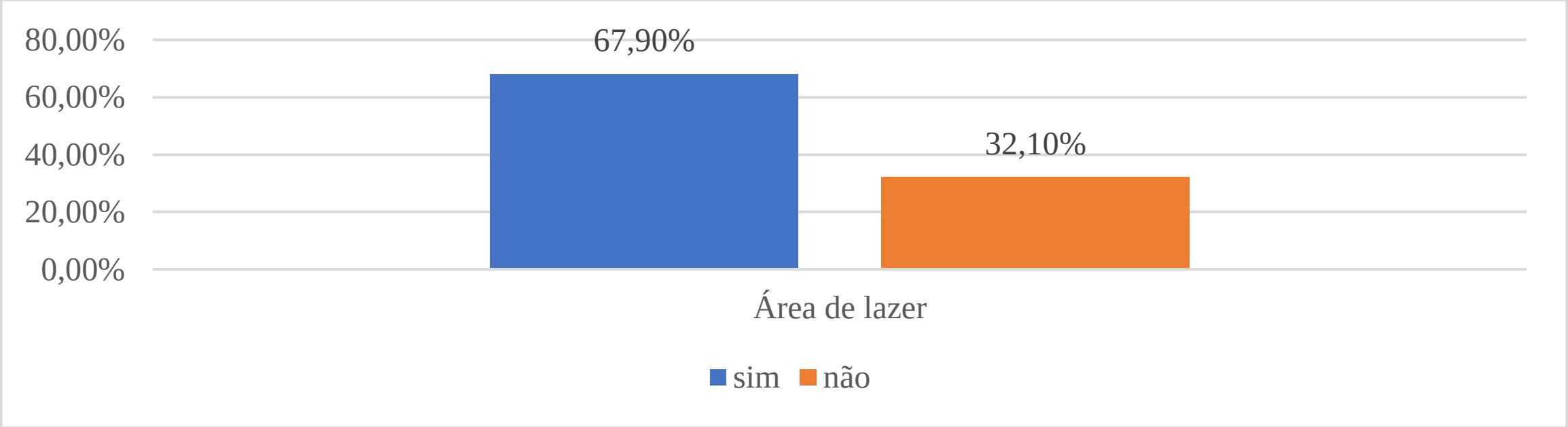

**Gráfico 21 – Tem conhecimento da área de lazer**
Fonte: Pesquisa (2015)

Os aspectos positivos relacionam-se com a situação das áreas de lazer, principalmente o contato com a natureza, não percebem sinais de depredação à flora existente. Quanto à fauna, nota-se uma convivência pacífica entre os usuários e as espécies encontradas, a beleza cênica da paisagem, os quais se configuram em atrativos naturais aproveitados no desenvolvimento do turismo ecológico local (SANTOS et al, 2007).

De acordo com o **Gráfico 22**, no quesito avaliação das áreas de lazer, a maioria as considera boas, para o desenvolvimento das atividades turísticas. Como explicam Wallace e Pierce (1996 *apud* FENNELL, 2002) um dos objetivos do ecoturismo é a contribuição para a conservação da área em que é desenvolvida a atividade ecoturística com a participação dos moradores das localidades.

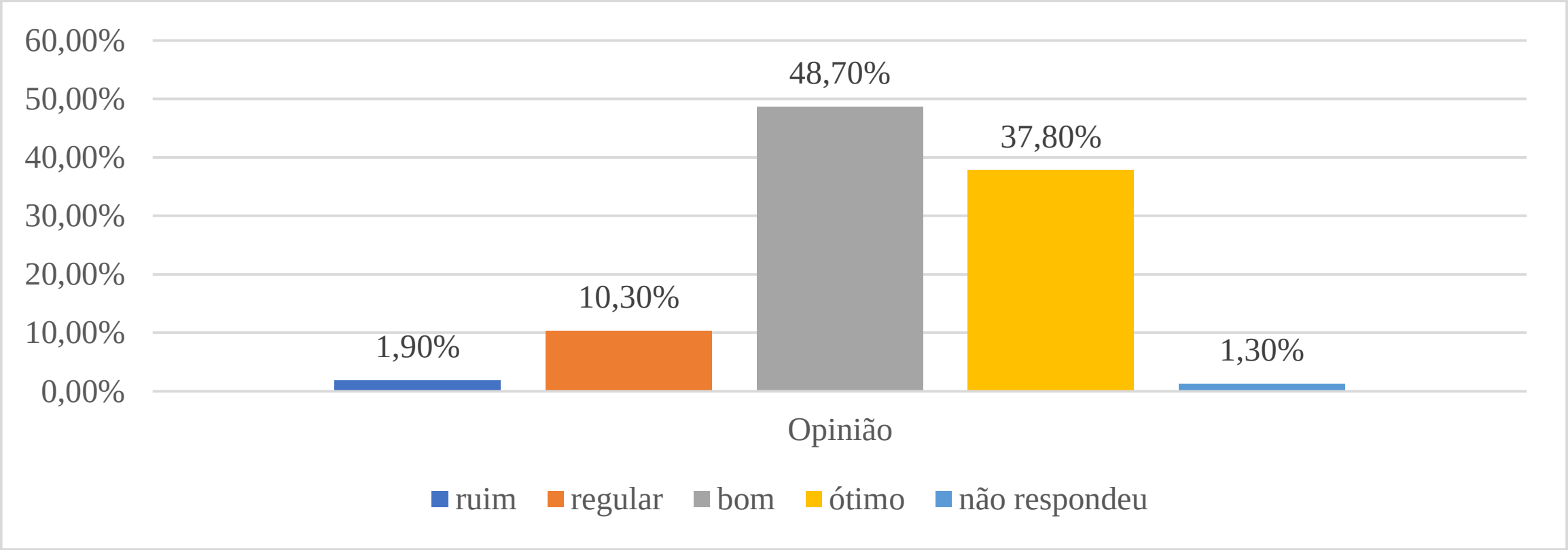

**Gráfico 22 – Opinião sobre a respectiva área**
Fonte: Pesquisa (2015)

Não se pode avaliar uma área sem ter a sensibilização de que toda e qualquer atividade gera mudança, por isso todos são responsáveis por conservá-la, especialmente os usuários que têm no ambiente o seu lazer.

Como forma de se deslocar até a área de lazer ecoturística 73,7% usa o automóvel. Mas, apenas 16,7% de motocicleta **Gráfico 23.**

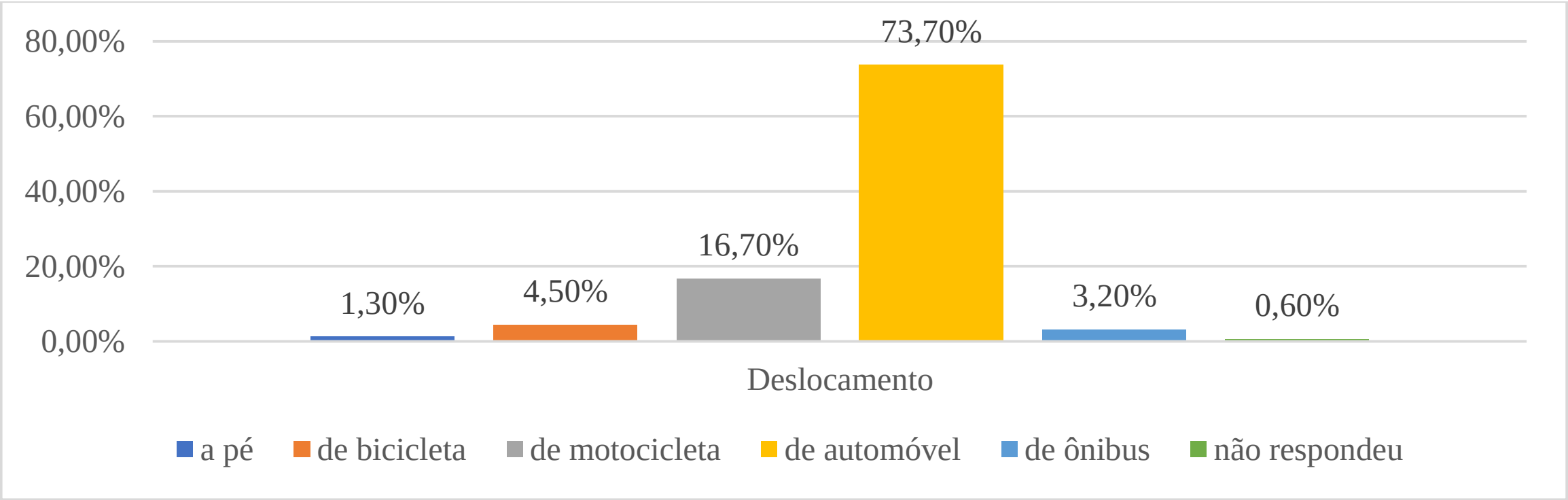

**Gráfico 23 – Forma de deslocamento para a área de lazer**
Fonte: Pesquisa (2015)

Os dados revelam ainda que o ecoturismo é praticado por pessoas que moram distantes das áreas de lazer, visto que 42,3% percorrem em média 35 km para chegar à localidade. Enquanto 28,2% fazem mais que 40 km **(Gráfico 24).**

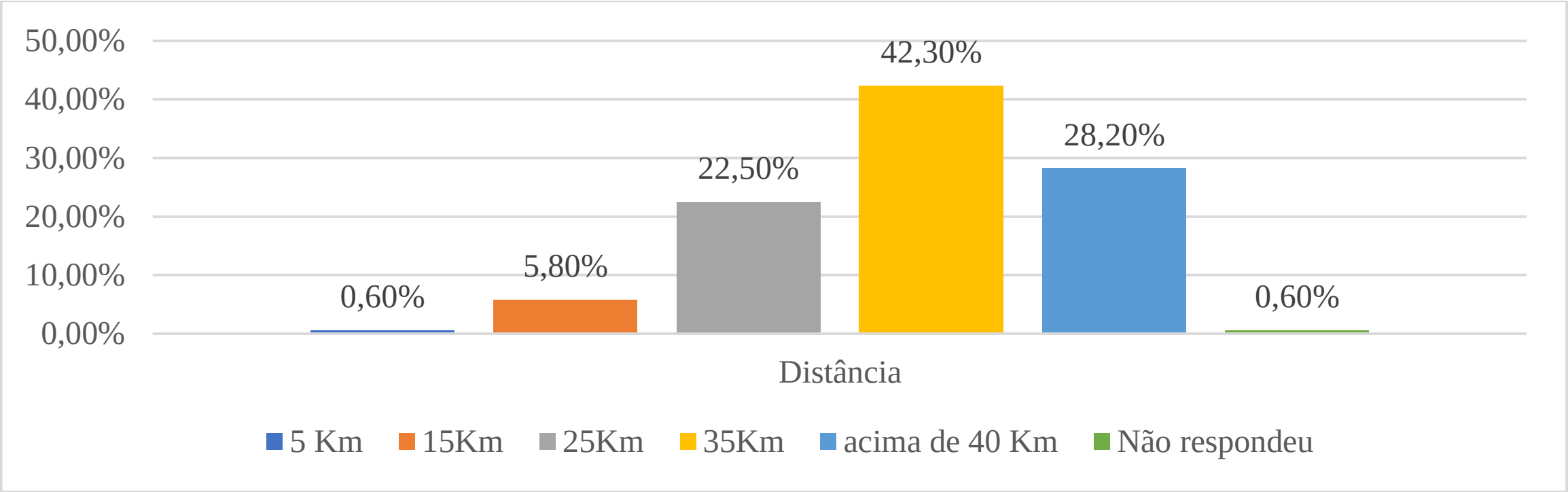

**Gráfico 24 – Distância para chegar à área de lazer**
Fonte: Pesquisa (2015)

O meio de transporte mais utilizado para se chegar às áreas de lazer é o automóvel, dado confirmado pelo **Gráfico 23**. Evidenciando que as oportunidades diferentes de lazer que as localidades oferecem como caminhada nas trilhas, banho e rapel, atraem pessoas que desejam descansar e, ao mesmo tempo, buscam uma alternativa de lazer em meio à natureza (MINISTÉRIO DO TURISMO 2010).

Os usuários se preocupam com a conservação dos recursos naturais somam 75%, sendo dados do **Gráfico 25**. O que evidencia, sobretudo, que os mesmos executam o ecoturismo de forma consciente, primando pela conservação, principalmente dos recursos hídricos. Dados do Ministério do Turismo (2010) explicam que a gestão socioambiental de uma área de lazer, principalmente dos recursos naturais e sua conservação minimizam os impactos negativos das atividades ecoturísticas em uma localidade.

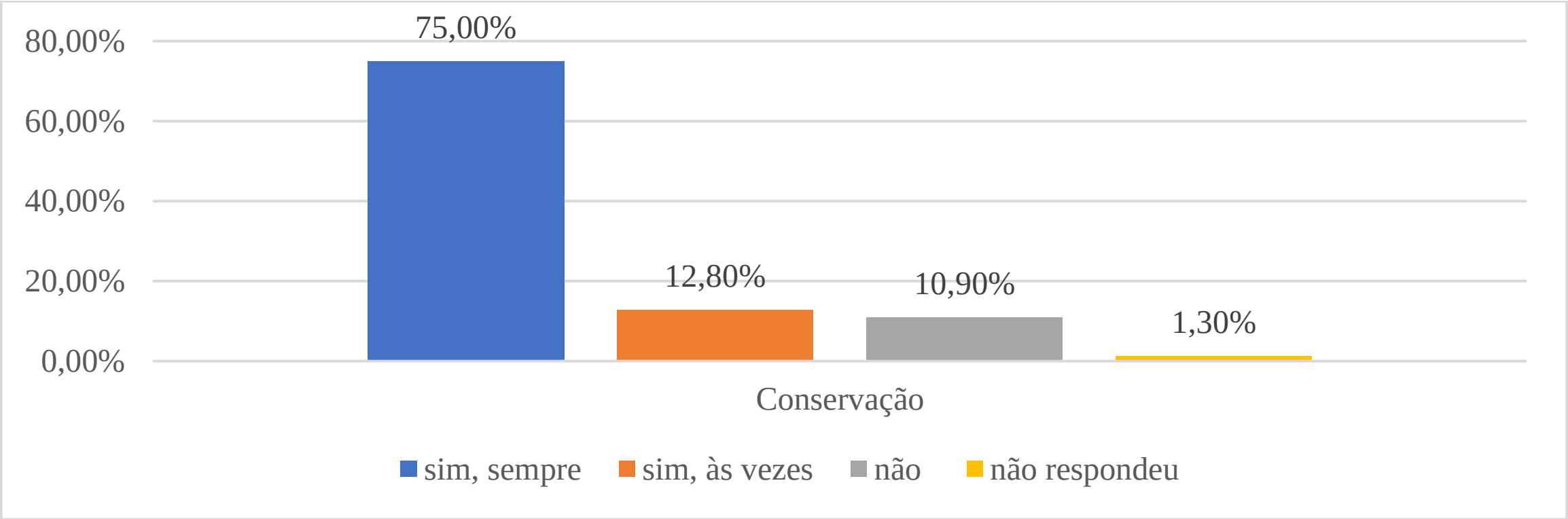

**Gráfico 25 – Preocupa-se com a conservação dos recursos naturais**
Fonte: Pesquisa (2015)

Ressalta-se que um dos principais aspectos a ser observado pelos adeptos a essa modalidade figura em maximizar os pontos positivos e minimizar os negativos, ações complexas e que nem sempre são alcançadas pelos usuários.

A maioria dos usuários pesquisados 53,8% acredita que os frequentadores não causam impactos a natureza das localidades **(Gráfico 26)**. Em concordância com Ruschmann (1997) e dados do Ministério do Turismo (2010), o ecoturismo pode causar impactos positivos e negativos, cabe, assim, o controle dos prejuízos e o aprimoramento dos benefícios com atividades ecoturísticas, o que deve ser uma ação realizada por usuários, moradores, empresários que atuam nas áreas, poder público e a comunidade em geral.

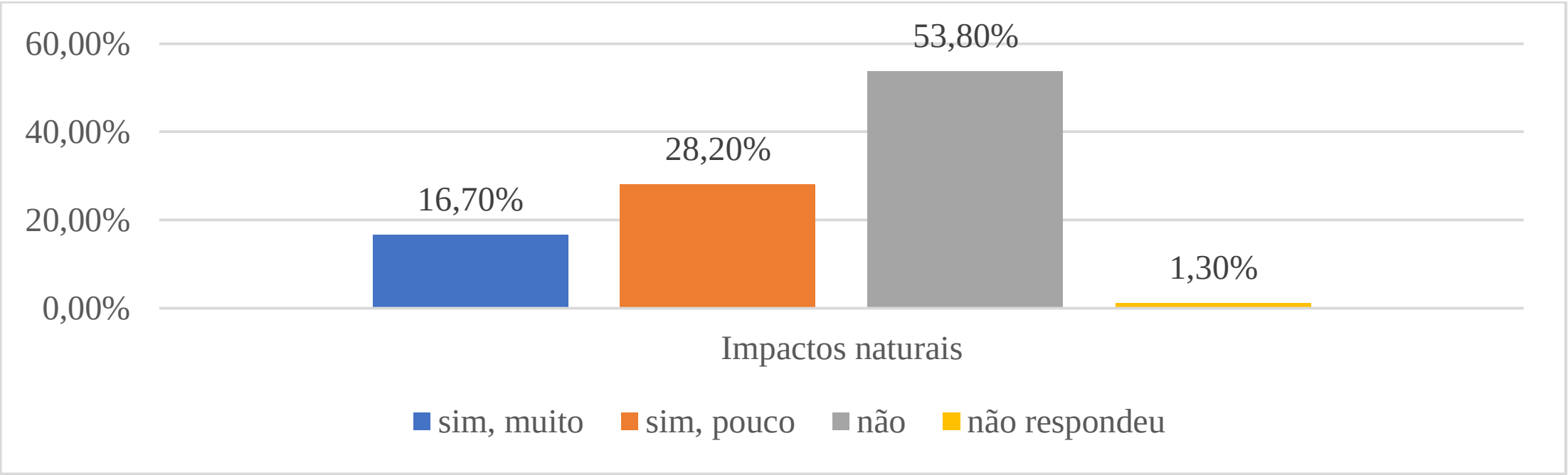

**Gráfico 26 – O ecoturismo vem causando impactos naturais nas localidades**
Fonte: Pesquisa (2015)

Apresenta-se, por meio do **Gráfico 27** a percepção dos usuários com relação à conservação da área de lazer, configurando um problema existente no desenvolvimento das atividades ecoturísticas em Rondonópolis, pois 46,2% definem a conservação da localidade como regular, enquanto, 36,5% avaliaram-na como boa, o que levanta uma questão que deve ser amplamente discutida e trabalhada, para benefício não apenas os usuários, mas também as comunidades onde estas estão inseridas.

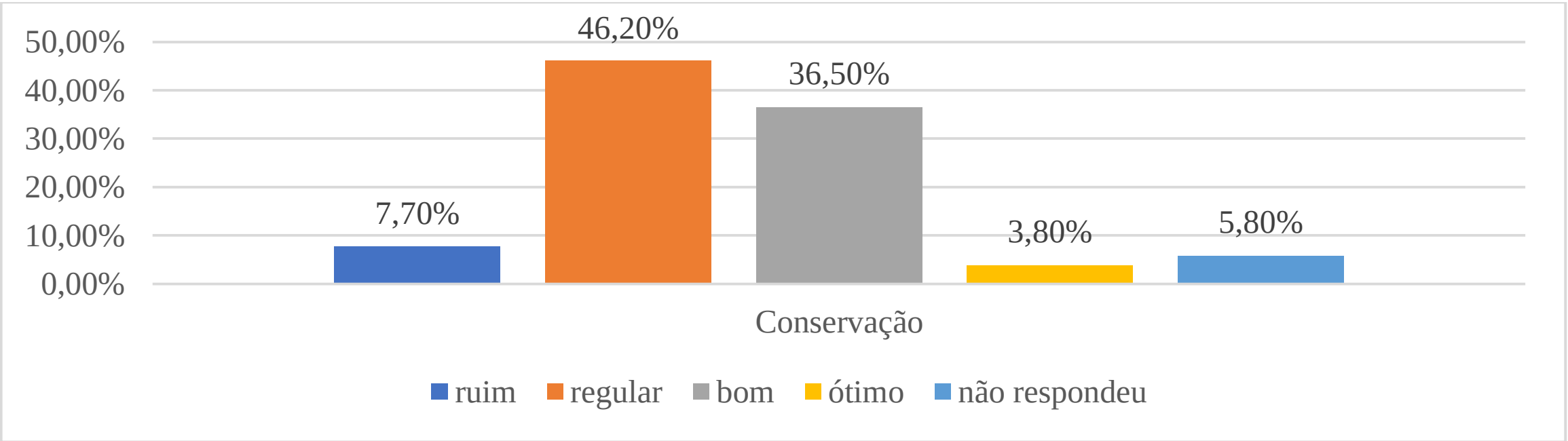

**Gráfico 27 – Avaliação da conservação da área de lazer pelos frequentadores**
Fonte: Pesquisa (2015)

Impactos negativos, ou seja, perniciosos para os recursos naturais, devem ser controlados de forma a minimizar suas consequências e auxiliar na conservação dos recursos naturais, para Rodrigues (1998) a população é quem deve ter maior controle em relação a essa conservação.As agências de turismo são responsáveis por grande parte da expansão vivenciada pelo ecoturismo no Brasil, pois atuam diretamente na exploração desta atividade, porém necessitam desempenhar participação mais efetiva na conservação das áreas de lazer. No contexto desta pesquisa, 37,9% dos usuários avaliam como regular a atuação das agências de turismo e 37,2% consideram-na ruim, dados do **Gráfico 28.**

Em concordância com Lacerda e Cândido (2013), destaca-se ainda, que a falta de sensibilização na conservação dos recursos hídricos pode constituir fragilidade no posicionamento de sustentabilidade destas empresas.

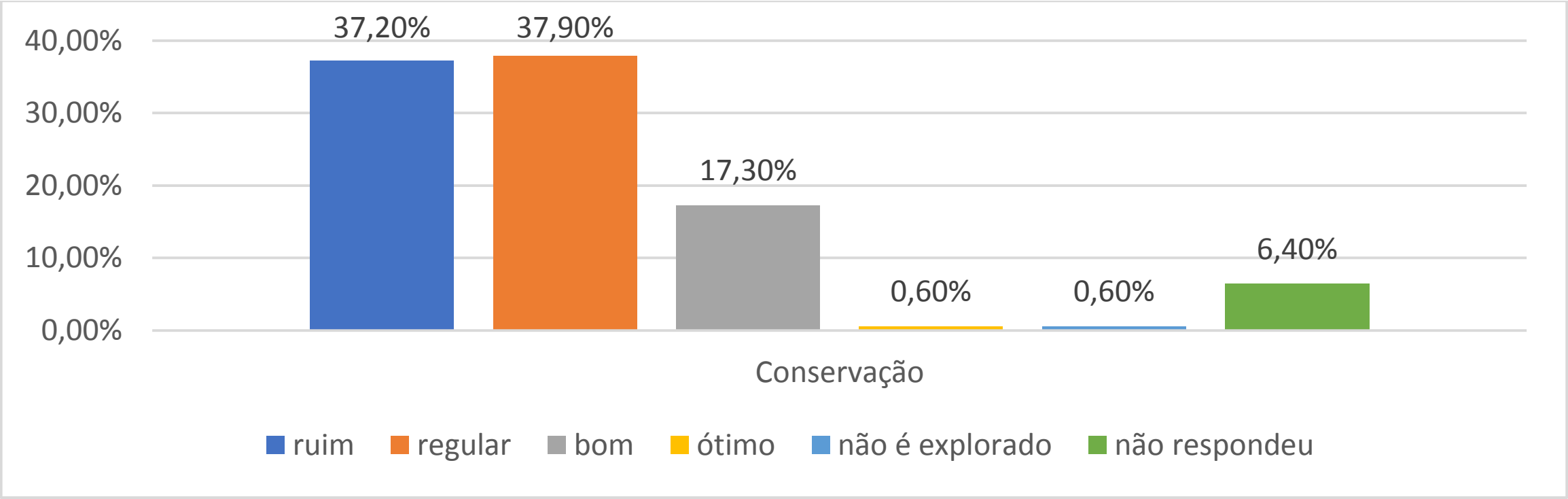

**Gráfico 28– Avaliação da conservação da área de lazer pelas agências de turismo segundo os usuários**
Fonte: Pesquisa (2015)

Conforme Sousa (2011) somente pode ocorrer um processo de sustentabilidade quando existir o respeito ao patrimônio natural e cultural em toda e qualquer localidade turística em que o ecoturismo vem sendo executado como forma de exploração econômica.

Dos usuários pesquisados 42,9% consideram como bom a conservação dos recursos naturais pelos moradores, como mostra o **Gráfico 29.** Entretanto, tem-se a compreensão de que os moradores deveriam ser os mais preocupados em proteger os recursos naturais existentes em sua área, pois as atividades ecoturísticas oferecem recursos financeiros que podem viabilizar uma vida mais confortável, existem, no entanto, casos em que os usuários não têm uma relação harmoniosa com os moradores.

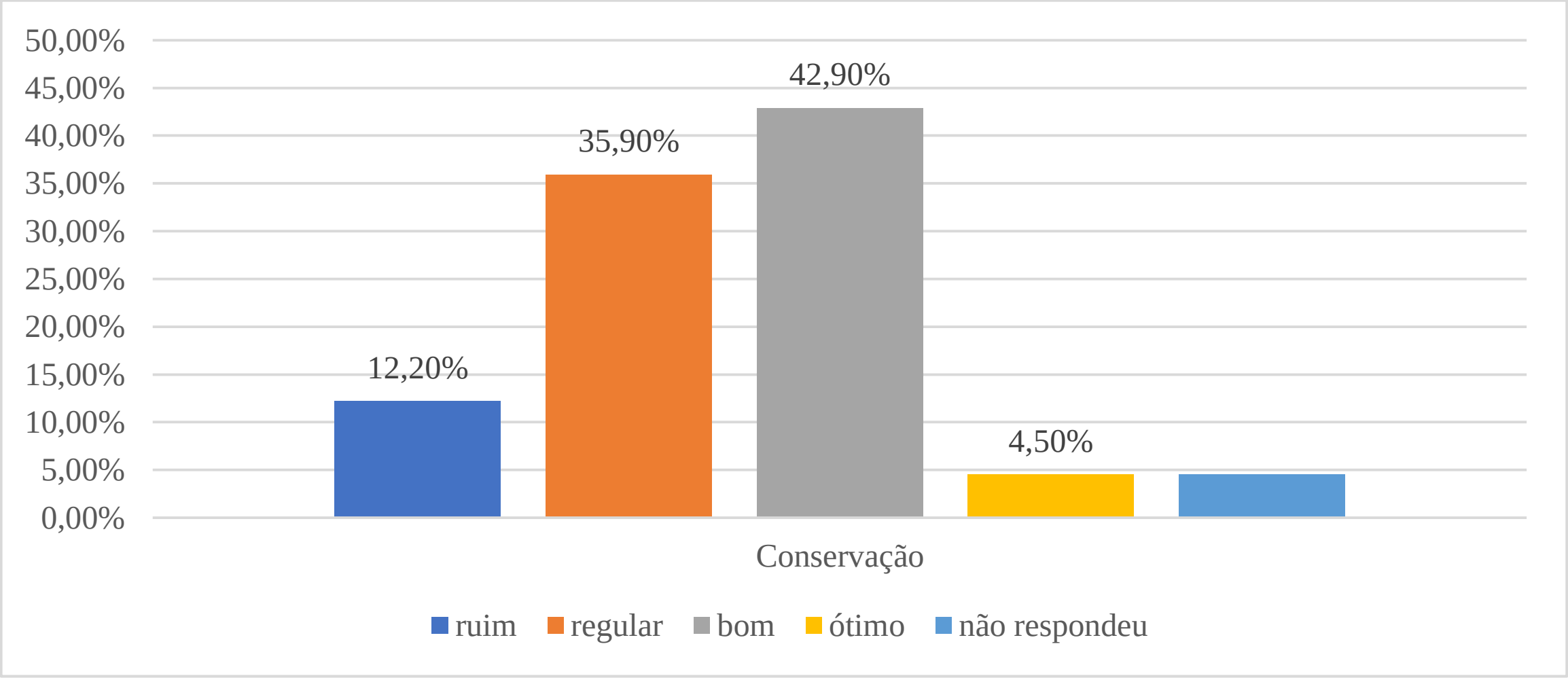

**Gráfico 29 – Avaliação da conservação da área de lazer pelos moradores**
Fonte: Pesquisa (2015)

As belezas naturais de uma localidade não podem perder sua originalidade em prol do conforto dos usuários e do crescimento econômico dos moradores, esquecendo-se da conservação dos recursos naturais e culturais. Hetzer (1965 *apud* FENNELL, 2002) afirma que os recursos naturais e a cultura da comunidade anfitriã não podem ser afetados de forma negativa pelo ecoturismo, o que demanda sustentabilidade e sensibilização dos agentes receptores.

Em relação ao uso dos recursos naturais para o desenvolvimento das atividades de serviços turísticos é entendido como percepção de satisfação e, auxilia juntamente com outras variáveis na criação da imagem do atrativo ou destino turístico (MONDO e FIATES, 2015).

Com o intuito de identificar quais categorias são mais relevantes para o uso sustentável dos recursos hídricos do município de Rondonópolis, procedeu-se a realização do *Teste*-t de *Student* para comparação de média (amostras independentes). Os índices de relevância dos indicadores foram aglutinados por categorias, gerando um índice de cada categoria. A **Tabela 1** apresenta os testes de comparação das três categorias do estudo.

Os indicadores sobre o uso dos recursos hídricos foram avaliados a partir de Escala Likerts de 1 a 5 pontos para cada resposta dada pelos usuários frequentadores, sendo distribuídas de acordo com a alternativa escolhida entre as categorias: benefícios econômicos; benefícios socioambientais e impactos socioambientais. Portanto, foi possível apurar os seguintes parâmetros: os que concordam parcialmente, ou seja, os menos favoráveis ou que avaliaram como poucos benefícios obtiveram a menor média (3,41) (desvio padrão de 1,53), obtendo a maior média, aqueles que concordam totalmente ou os mais favoráveis 4,04 (desvio padrão 1,26), indicando que o turismo traz benefícios. Assim, confirma-se que a utilização dos recursos hídricos, relacionada com atividades ecoturisticas para usuários frequentadores nas localidades pesquisadas é significativamente positiva, demonstrando coeficientes excelentes de confiabilidade das categorias em todos os casos.

| BENEFÍCIOS ECONÔMICOS | *Média* | *Desvio Padrão* |
|---|---|---|
| 1.Estradas pavimentadas<br>2.Estradas não pavimentadas<br>3.Melhoria das vias de acesso<br>4.Fonte de renda<br>5.Gera emprego | 4,04 | 1,26 |
| BENEFÍCOS SOCIOAMBIENTAIS | | |
| 1.Utilização dos recursos naturais com sustentabilidades<br>2.Conservação do patrimônio histórico e cultural<br>3.Divulgação das potencialidades ecoturísticas<br>4.Fiscalização<br>5.Propostas e sensibilização de educação ambiental | 3,95 | 0,96 |
| IMPACTOS SOCIOAMBIENTAIS | | |
| 1.Assoreamento dos rios<br>2.Presença de lixo<br>3.Pichação e marcas em árvores e rochedos<br>4.Retirada de espécie vegetais com fins ornamentais<br>5.Ampla acessibilidade a população<br>6.Recebe apoio de instituições e entidades voltadas ao meio ambiente<br>7.Limite de capacidade de carga<br>8.Contaminação das águas pelo uso intensivo de agrotóxicos<br>9.Desmatamento<br>10.Queimadas<br>11.Descaracterização da vegetação natural<br>12.Afastamento e extinção da fauna<br>13.Falta de manifestações culturais<br>14.Gastronomia | 3,41 | 1,53 |

**Tabela 1 - Percepção sobre o uso dos recursos hídricos relacionados com as atividades ecoturísticas**
Org: CAMPOS, M. B. N. SILVA (2015)

Por sua natureza, as atividades ecoturísticas se desenvolvem num contexto que implica o uso dos recursos naturais, o que resulta em dados de proporções alarmantes de danos ambientais decorrentes dessas práticas defende Marullo (2012).

A **Tabela 2** apresenta a correlação entre os índices das três categorias analisadas: benefícios econômicos, benefícios socioambientais e impactos socioambientais, comprovando a distribuição de normalidade das variáveis por meio de coeficientes de assimetria de *Pearson* e de *Curtose*, cujos valores variam aproximadamente de 3, -3 e +3 (BERQUÓ, 2006).

| CATEGORIAS | Médias | Desvio-padrão | Erro amostral | Pearson | Curtose |
|---|---|---|---|---|---|
| Benefícios econômicos | 4,04 | 1,26 | 0,09 | -1,09 | 1,73 |
| Benefícios socioambientais | 3,95 | 0,96 | 0,07 | -1,72 | 1,78 |
| Impactos socioambientais | 3,41 | 1,53 | 0,07 | -2,14 | 0,74 |

**Tabela 2 - Correlação entre os índices da escala Likerts de acordo com as categorias**
Org: CAMPOS, M. B. N. SILVA (2015)

A pesquisa sugere que o uso dos recursos hídricos relacionados com as atividades ecoturísticas esteja voltado para uma prática sustentável, por meio do envolvimento dos moradores e usuários frequentadores das localidades objeto de estudo, em consonância com o pensamento de (PORTUGUES et al., 2012).

## 4.3 AS AGÊNCIAS DE TURISMO QUE OFERTAM ROTEIROS TURÍSTICOS EM RONDONÓPOLIS-MT

As agências de viagem e turismo locais são mais voltadas ao turismo emissivo, isto é, promovem mais saídas de pessoas para fora do município do que recebem turistas, dado que não há uma definição de marketing de produtos, roteiros e potencialidades turísticas da região.

Neste estudo, participaram duas agências de turismo de Rondonópolis, que oferecem roteiros turísticos no município uma localizada à Rua Otávio Pitaluga, denominada Empresa A, e outra na Avenida Júlio Campos, denominada Empresa B, ambas atuando em roteiros principalmente urbanos.

Com relação à oferta de produtos ecoturísticos, a Empresa A declarou que trabalha somente com ecoturismo nacional, e a Empresa B que trabalha com ecoturismo nacional e internacional, o que aponta esta atividade como uma tendência mundial. No entanto, ambas as agências enfatizaram a não existência de roteiros para o ecoturismo no município. Uma das agências a Empresa A destacou inclusive que a Secretaria de Turismo em âmbito do Estado não trabalha com ecoturismo, enquanto a outra agência a Empresa B frisou que existem roteiros das potencialidades, mas que as empresas não exploram a relação entre a atividade com cunho educacional e conservacionista, premissa do ecoturismo.

O Município de Rondonópolis tem potencial turístico associado aos recursos naturais, assunto que deve estar presente em discussões acerca do desenvolvimento das atividades turísticas das localidades.

As principais sugestões para a consolidação do ecoturismo em Rondonópolis vieram da participação da representante da Empresa A, que mencionou investimentos como: “acesso, pavimentação asfáltica, divulgação e infraestrutura para os usuários, restaurantes, banheiros e lanchonetes”, considerando somente fatores estruturais, relacionados com o dispêndio de recursos financeiros para sua efetivação, suprimindo a sensibilização dos moradores, usuários, gestores públicos e privados que atuam na área do ecoturismo quanto a necessidade de conservação dos recursos naturais, culturais e sociais, ou a sustentabilidade das atividades.

A exploração do ecoturismo, somente envolvendo aspectos econômicos ou financeiros, não atende aos paradigmas e valores da sustentabilidade, pois os recursos hídricos existentes nas localidades devem ser conservados e protegidos para as gerações futuras.

## 4.4 AS INSTUIÇÕES PÚBLICAS

O desenvolvimento de atividades turísticas no município, para os gestores públicos é positivo, eles consideram necessária a criação de um pólo de educação ambiental, no sentido de contribuir para a prática de atividades ecoturísticas, principalmente nas localidades envolvidas neste estudo.

As duas instituições públicas que fizeram parte deste estudo foram a Prefeitura Municipal de Rondonópolis e a Secretaria Municipal de Meio Ambiente (SEMMA), ambas responsáveis diretamente pelo ecoturismo.

Ambas as instituições entendem que o ecoturismo está articulado à conservação do meio ambiente o que aparentemente aponta para uma visão do discurso de totalidade.

Os gestores consideram que o ecoturismo traz benefícios econômicos para o município. E destacaram que Educação Ambiental contribui para esta atividade ressaltando a importância do trabalho realizado nas escolas, especialmente, nas escolas municipais, que devem primar por uma educação consciente e voltada para o resgate de valores, que se fundem na visão cidadã de responsabilidade socioambiental, tendo em sua essência a sustentabilidade.

A Secretária Municipal de Ciência, Tecnologia, Turismo e Desenvolvimento Econômico relatou que desenvolve ações voltadas para o ecoturismo, mas não identificou as modalidades executadas. E o representante da SEMMA afirmou não desenvolver ações de ecoturismo, justificando que a maioria das áreas com tais potencialidades se localiza em propriedades particulares.

A existência de contradição na opinião das instituições públicas demonstra que ainda há um longo caminho a ser percorrido para que exista efetivamente uma política de valorização do ecoturismo e um real processo de sustentabilidade em Rondonópolis.

## 5 CONSIDERAÇÕES FINAIS

Nas últimas décadas ocorreu uma busca pelo turismo ecológico e consequente fuga dos centros urbanos, esta atividade surge como possibilidade de contato com a natureza, para relaxar a mente e exercitar o corpo, fazendo de áreas de lazer, como as abordadas nesta pesquisa, um excelente refúgio para descansar e libertar o estresse da cidade buscando a qualidade de vida em atividades turísticas.

Rondonópolis localiza-se numa região de natureza exuberante e com forte potencial para desenvolvimento de atividades turísticas, devido a traços geográficos como a presença de afloramentos rochosos, região de encosta e topos de morros, forma-se um ambiente de grande riqueza e constituindo um cenário propício ao ecoturismo. Há que se ressaltar, porém, a fragilidade desses ecossistemas, já que algumas áreas de lazer do municípios analisados nesta pesquisa, como as cachoeiras e as prainhas estão localizadas nas áreas circunvizinhas e próximas às Unidades de Conservação Dom Osório Stoffel e Parque Ecológico João Basso (NARDES, 2005), e atividades turísticas nessas localidades atraem nos finais de semana e/ou feriados um fluxo de visitantes, o que vem gerando impactos negativos ao ambiente.

Em razão do exposto, verifica-se que é de suma importância a ocupação do espaço geográfico de forma ordenada, seguindo critérios conservacionistas baseados na legislação vigente e critérios técnicos, principalmente nas áreas mais frágeis. Pois, alterações nas Áreas de Preservação Permanente (APPs) podem ocasionar à deterioração da função ambiental, comprometer a qualidade dos recursos hídricos, a biodiversidade, a flora, a fauna, desproteger o solo e afetar o bem estar da população.

Com este estudo foi possível compreender que o ecoturismo pode contribuir para a conservação dos recursos hídricos no município de Rondonópolis. A sensibilização quanto ao uso racional dos recursos naturais especialmente dos hídricos, deve estar presente em todas as atividades ecoturísticas, envolvendo todos os atores deste processo: sejam os moradores das localidades em que são realizadas as atividades ecoturísticas; os usuários frequentadores; as empresas que exploram as atividades turísticas e o poder público.

As atividades ecoturísticas no município estão ainda incipientes, não há um roteiro sistematizado e consistente com os atributos ambientais, principalmente em relação aos recursos hídricos. O ecoturismo se faz presente nas áreas de lazer analisadas, no entanto, para ser realmente "ecológico" deve favorecer a conservação desses ambientes, harmonizando o benefício socioambiental como o mais significativo parâmetro em relação à exploração econômica, buscando a diminuição dos impactos negativos nesses locais e a conservação dos recursos naturais existentes.

As técnicas de sensoriamento remoto e geoprocessamento utilizadas demonstraram a contribuição das geotecnologias nos estudos ambientais, especificamente no que diz respeito aos recursos hídricos, que assim como os demais procedimentos metodológicos usados foram eficazes para a compreensão deste objeto de estudo.

O trabalho de campo evidenciou a existência de nove cachoeiras com potencial para desenvolver as práticas ecoturísticas, nas áreas de lazer. Algumas, no entanto, se encontram em estágio de degradação socioambiental, provavelmente causado pela falta de ações e normas para visitação aos locais, capacidade de carga e ordenamento.

As políticas públicas municipais voltadas para o desenvolvimento das atividades ecoturísticas devem contemplar as áreas de lazer como as cachoeiras e as prainhas, podendo contribuir para a ampliação de ações e projetos envolvendo os atores sociais. Pois, tanto os moradores das localidades, os usuários frequentadores, as empresas de turismo quanto o poder público devem praticar e incorporar em suas atuações os princípios da conservação e sustentabilidade.

As sugestões aqui apresentadas podem representar significativa contribuição na sustentabilidade da biodiversidade, pois minimizam os impactos negativos gerados pelo desenvolvimento das atividades ecoturísticas nas localidades e no entorno das áreas de lazer. Nesse sentido, considera-se que o uso dos recursos hídricos do município com sustentabilidade garante proteção à biodiversidade em longo prazo e por essa razão devem ser integradas em um contexto que envolve os moradores, os usuários, agências de turismo e poder público.

A partir da análise do contexto, sugere-se que essas áreas de lazer contemplem: estabelecimento de roteiro com seus atributos ambientais e socioculturais, calendário definindo periodicidade de visitação das cachoeiras, quedas d'água, das prainhas e das trilhas ecológicas, pavimentação das rodovias de acesso às áreas de lazer, sinalização para melhoria de acessibilidade, informações quanto às práticas executadas para a sustentabilidade dos locais.

Para a realização das práticas turísticas propostas nessas áreas de lazer com sustentabilidade dos recursos hídricos no município, necessita-se do poder público para auxiliar na implementação de políticas apropriadas de ecoturismo que respeitem as culturas locais e favoreçam a conservação da biodiversidade nas localidades e áreas circunvizinhas.

## REFERÊNCIAS BIBLIOGRÁFICAS

# APÊNDICES

## APÊNDICE A – QUESTIONÁRIO PARA OS MORADORES DAS LOCALIDADES PRÓXIMOS DAS ÁREAS DE LAZER

O presente questionário tem como objetivo a coleta de dados relacionados às potencialidades turísticas para o desenvolvimento do ecoturismo em Rondonópolis-MT.

Ao responder as questões estarão auxiliando na construção de um estudo científico para a apresentação de resultados, de forma a demonstrar a necessidade do uso com racionalidade dos recursos naturais, especialmente, os hídricos que são os mais prejudicados para a realização de atividades ecoturísticas.

Obrigado pela colaboração!
Manoel Benedito Nirdo da Silva Campos

### I. CARACTERÍSTICAS SOCIOECONÔMICAS

1. Sexo:
( ) feminino
( ) masculino

2. Faixa etária:
( ) até 25 anos
( ) entre 26 e 35 anos
( ) entre 36 e 45 anos
( ) acima de 45 anos

3. Grau de escolaridade:
( ) ensino fundamental incompleto
( ) ensino fundamental completo
( ) ensino médio incompleto
( ) ensino médio completo
( ) ensino superior incompleto
( ) ensino superior completo

4. Quantas pessoas moram em sua casa?
( ) somente 1 pessoa
( ) entre 2 e 4 pessoas
( ) entre 5 e 7 pessoas
( ) acima de 7 pessoas
5. Qual a renda familiar?
( ) até R$ 1.000,00

( ) entre R$ 1.001,00 e R$ 3.000,00
( ) entre R$ 3.001,00 e R$ 5.000,00
( ) acima de R$ 5.000,00

6. Há quanto tempo residem nesta localidade?
( ) menos de 1 ano
( ) entre 2 e 5 anos
( ) entre 6 e 10 anos
( ) mais de 10 anos

7. Em sua opinião turismo e ecoturismo são sinônimos?
( ) sim
( ) não

8. Benefícios econômicos do ecoturismo para as localidades?
( ) excelente
( ) bom
( ) regular
( ) ruim
( ) péssimo

9. Como avalia a relação moradores/ecoturistas?
( ) excelente
( ) bom
( ) regular
( ) ruim
( ) péssimo

## II. PERCEPÇÃO SOCIOAMBIENTAL PARA DESENVOLVIMENTO DE ATIVIDADES TURÍSTICAS

10. Há preocupação dos ecoturistas com a conservação dos recursos naturais?
( ) em sua maioria evitam a destruição dos recursos naturais
( ) somente alguns evitam a destruição dos recursos naturais
( ) não existe ecoturista preocupado com a conservação dos recursos naturais

11. Impactos ambientais causados pelas atividades em sua localidade?
( ) sim, muito
( ) sim, pouco
( ) não

12. Quais são os benefícios trazidos pelo ecoturista para as localidades?
( ) econômicos
( ) favorece a conservação dos recurso naturais
( ) baixo impacto dos recursos naturais

( ) Não traz benefícios nenhum

13. Quais são os problemas trazidos usuários frequentadores?
( ) destruição dos recursos culturais
(  ) destruição dos recursos sociais
(  ) destruição dos recursos hídricos

## APÊNDICE B – QUESTIONÁRIO PARA OS USUÁRIOS FREQUENTADORES DAS ÁREAS DE LAZER

O presente questionário tem como objetivo a coleta de dados relacionados às potencialidades turísticas para o desenvolvimento do ecoturismo em Rondonópolis-MT.

Ao responder as questões estarão auxiliando na construção de um estudo científico para a apresentação de resultados, de forma a demonstrar a necessidade do uso com racionalidade dos recursos naturais, especialmente, os hídricos que são os mais prejudicados para a realização de atividades ecoturísticas.

Obrigado pela colaboração!
Manoel Benedito Nirdo da Silva Campos

| I. CARACTERÍSTICAS SOCIOECONÔMICAS |
|---|

1. Sexo:
( ) feminino
( ) masculino

2. Faixa etária:
( ) até 25 anos
( ) entre 26 e 35 anos
( ) entre 36 e 45 anos
( ) acima de 45 anos

3. Grau de escolaridade:
( ) ensino fundamental incompleto
( ) ensino fundamental completo
( ) ensino médio incompleto
( ) ensino médio completo
( ) ensino superior incompleto
( ) ensino superior completo

4. Qual a renda familiar?
( ) até R$ 1.000,00
( ) entre R$ 1.001,00 e R$ 3.000,00
( ) entre R$ 3.001,00 e R$ 5.000,00
( ) acima de R$ 5.000,00

5. Há quanto tempo desenvolve atividades ecoturísticas?
( ) menos de 1 ano
( ) entre 2 e 5 anos
( ) entre 6 e 10 anos
( ) mais de 10 anos

6. Em sua opinião turismo e ecoturismo são a mesma atividade?
( ) sim
( ) não

7. Em sua opinião o turismo...

( ) Aumenta as opções de recreação no município
( ) Não interfere
( ) Não sabe responder

## II. PERCEPÇÃO SOCIOAMBIENTAL PARA DESENVOLVIMENTO DE ATIVIDADES TURÍSTICAS

8. Você conhece esta área de lazer?
( ) Sim ( ) Não
Qual a opinião sobre a respectiva área?
( ) Ruim ( ) Regular ( ) Bom ( ) Ótimo. Justifique: _______________________

9. Como você chega à área de lazer?

( ) a pé ( ) de bicicleta ( ) de moto ( ) de carro ( ) de ônibus ( ) outros _________

10. Qual a distância que você percorre para chegar à área de lazer?

( ) 5 Km ( ) 15 Km ( ) 25 Km ( ) 35 Km ( ) acima de 40 Km

11. Você se preocupa com a conservação dos recursos naturais?
( ) sim, sempre
( ) sim, às vezes
( ) não

12. Acredita que o ecoturismo praticado nessa localidade vem causando impactos naturais?
( ) sim, muito
( ) sim, pouco
( ) não

13. Como avalia a conservação dessa área de lazer?
- Pelo usuário? ( ) Ruim ( ) Regular ( ) Bom ( ) Ótimo
- Pelas agências de turismo? ( ) Ruim ( ) Regular ( ) Bom ( ) Ótimo
- E pelos moradores? ( ) Ruim ( ) Regular ( ) Bom ( ) Ótimo

III - ESCALA DE LIKERT - PERCEPÇÃO SOBRE O USO DOS RECURSOS NATURAIS RELACIONADOS COM AS ATIVIDADES TURÍSTICAS.

14. Instrução: Compare os itens abaixo assinalando de acordo com sua percepção aquele que expresse sua opinião.

| | | | | | |
|---|---|---|---|---|---|
| Pode-se afirmar se não há diferença significativa entre os benefícios econômicos, sociais e ambientais no uso dos recursos naturais nas localidades receptoras das áreas de lazer?<br>Legenda: CT= Concordo Totalmente; C= Concordo; I= Indiferente; D= Discordo; DT= Discordo Totalmente. | | | | | |
| BENEFÍCIOS ECONÔMICOS | | | | | |
| 1.Estradas pavimentadas | DT | D | I | C | CT |
| 2.Estradas não pavimentadas | DT | D | I | C | CT |
| 3. Melhoria das vias de acesso | DT | D | I | C | CT |
| 4. Fonte de renda | DT | D | I | C | CT |
| 5. Gera emprego | DT | D | I | C | CT |
| BENEFÍCOS SOCIOAMBIENTAIS | | | | | |
| 1.Utilização dos recursos naturais com sustentabilidades | DT | D | I | C | CT |
| 2. Conservação do patrimônio histórico e cultural | DT | D | I | C | CT |
| 3.Divulgação das potencialidades ecoturísticas | DT | D | I | C | CT |
| 4. Fiscalização | DT | D | I | C | CT |
| 5. Propostas e sensibilização de educação ambiental | DT | D | I | C | CT |
| IMPACTOS SOCIOAMBIENTAIS | | | | | |
| 1. Assoreamento dos rios | DT | D | I | C | CT |
| 2.Presença de lixo | DT | D | I | C | CT |
| 3.Pichação e marcas em árvores e rochedos | DT | D | I | C | CT |
| 4.Retirada de espécie vegetais com fins ornamentais | DT | D | I | C | CT |
| 5.Ampla acessibilidade a população | DT | D | I | C | CT |
| 6.Recebe apoio de instituições e entidades voltadas | DT | D | I | C | CT |
| ao meio ambiente | | | | | |
| 7.Limite de capacidade de carga | DT | D | I | C | CT |
| 8. Contaminação das águas pelo uso intensivo de agrotóxicos | DT | D | I | C | CT |
| 9. Desmatamento | DT | D | I | C | CT |
| 10. Queimadas | DT | D | I | C | CT |
| 11.Descaracterização da vegetação natural | DT | D | I | C | CT |
| 12. Afastamento e extinção da fauna | DT | D | I | C | CT |
| 13.Falta de manifestações culturais | DT | D | I | C | CT |
| 14.Gastronomia | DT | D | I | C | CT |

## APÊNDICE C – QUESTIONÁRIO DAS INSTITUIÇÕES PÚBLICAS

### I. DADOS E CARACTERÍSTICAS DA INSTITUIÇÃO

1 – Nome da Instituição_________________________________________

2 – Representante legal _________________________________________

3 – Endereço ________________________________________________

4 – Telefone: ________________________________________________

5 – E-mail; __________________________________________________

6 – Origem: _________________________________________________

7 – Objetivos; _______________________________________________

8 – Área de atuação: Zona Urbana (  ) Zona Rural

### II. PERCEPÇÃO SOCIOAMBIENTAL EM RELAÇÃO AO DESENVOLVIMENTO DE ATIVIDADES TURÍSTICAS

9) O ecoturismo e o meio ambiente estão associados?

Sim (  ) Não (  ) Não sei responder (  )

10) Existem propostas para a prática do desenvolvimento do ecoturismo?

Sim (  ) Não (  ) Não sei responder (  )

11) O ecoturismo influencia no desenvolvimento do município?

Sim (  ) Não (  ) Não sei responder (  )

12) Em sua opinião, quem tem responsabilidade para o desenvolvimento das atividades ecoturísticas?

(  ) Órgão ambiental (  ) Órgão de turismo (  ) Turistas (  ) Moradores (  ) Outros

13) A Secretaria Municipal desenvolve ações de ecoturismo?

Sim (  ) Não (  ) Não sei responder (  )

**APÊNDICE D – QUESTIONÁRIO PARA AS AGÊNCIAS DE TURISMO RECEPTIVO**

I. DADOS E CARACTERÍSTICAS DA EMPRESA

1 – Nome da Empresa_______________________________________________

2 – Representante legal ___________________________________________

3 – Endereço ___________________________________________________

4 – Telefone: ___________________________________________________

5 – E-mail; _____________________________________________________

6 – Origem: ____________________________________________________

7 – Objetivos; __________________________________________________

8 – Área de atuação: Zona Urbana ( ) Zona Rural

II. PERCEPÇÃO SOCIOAMBIENTAL EM RELAÇÃO AO DESENVOLVIMENTO DE ATIVIDADES TURÍSTICAS

9) Existem roteiros ecoturísticos que contemplam o município?

Sim ( ) Não ( ) Não sei responder ( )

10) Em que âmbito a sua empresa oferta o produto ecoturístico?

Municipal ( ) Regional ( ) Nacional ( ) Internacional ( )

11) Quais são as sugestões para a consolidação do ecoturismo em Rondonópolis?

## APÊNDICE E – ROTEIRO DE OBSERVAÇÃO

| ASPECTOS A CONSIDERAR | ELEMENTO A OBSERVAR |
|---|---|
| Espaço de apoio | Ambiente possui espaço adequado para recepção dos turistas. |
| Ocupação | Sinais de apropriação e demarcação das localidades – placa de identificação; uso de nome em objetos e lugares. |
| Recursos naturais | Potenciais existentes: Flora; Fauna e outros. |
| Atividades de educação ambiental | Fiscalização; Conservação; Desmatamento; Degradação e outros. |
| Atividades culturais | Manifestações culturais: Dança; Gastronomia e outros eventos. |
| Impacto socioambiental | -Impactos sociais positivos e negativos.<br>-Impactos ambientais positivos e negativos. |

# ANEXO

## ANEXO – TABELA PARA CÁLCULO DO TAMANHO DA AMOSTRA

Determinação do tamanho da amostra a partir do tamanho da população

| N* | A* | N | A | N | A |
|---|---|---|---|---|---|
| 10 | 10 | 220 | 140 | 1200 | 291 |
| 15 | 14 | 230 | 144 | 1300 | 297 |
| 20 | 19 | 240 | 148 | 1400 | 302 |
| 25 | 24 | 250 | 152 | 1500 | 306 |
| 30 | 28 | 260 | 155 | 1600 | 310 |
| 35 | 32 | 270 | 159 | 1700 | 313 |
| 40 | 36 | 280 | 162 | 1800 | 317 |
| 45 | 40 | 290 | 165 | 1900 | 320 |
| 50 | 44 | 300* | 169* | 2000 | 322 |
| 55 | 48 | 320 | 175 | 2200 | 327 |
| 60 | 52 | 340 | 181 | 2400 | 331 |
| 65 | 56 | 360 | 186 | 2600 | 335 |
| 70 | 59 | 380 | 191 | 2800 | 338 |
| 75 | 63 | 400 | 196 | 3000 | 341 |
| 80 | 66 | 420 | 201 | 3500 | 346 |
| 85 | 70 | 440 | 205 | 4000 | 351 |
| 90 | 73 | 460 | 210 | 4500 | 354 |
| 95 | 76 | 480 | 214 | 5000 | 357 |
| 100 | 80 | 500 | 217 | 6000 | 361 |
| 110 | 86 | 550 | 226 | 7000 | 364 |
| 120 | 92 | 600 | 234 | 8000 | 367 |
| 130 | 97 | 650 | 242 | 9000 | 368 |
| 140 | 103 | 700 | 248 | 10000 | 370 |
| 150 | 108 | 750 | 254 | 15000 | 375 |
| 160 | 113 | 800 | 260 | 20000 | 377 |
| 170 | 118 | 850 | 265 | 30000 | 379 |
| 180 | 123 | 900 | 269 | 40000 | 380 |
| 190 | 127 | 950 | 274 | 50000 | 381 |
| 200 | 132 | 1000 | 278 | 75000 | 382 |
| 210 | 136 | 1100 | 285 | 100000 | 384 |

*N = Tamanho da População

*A = Tamanho da Amostra

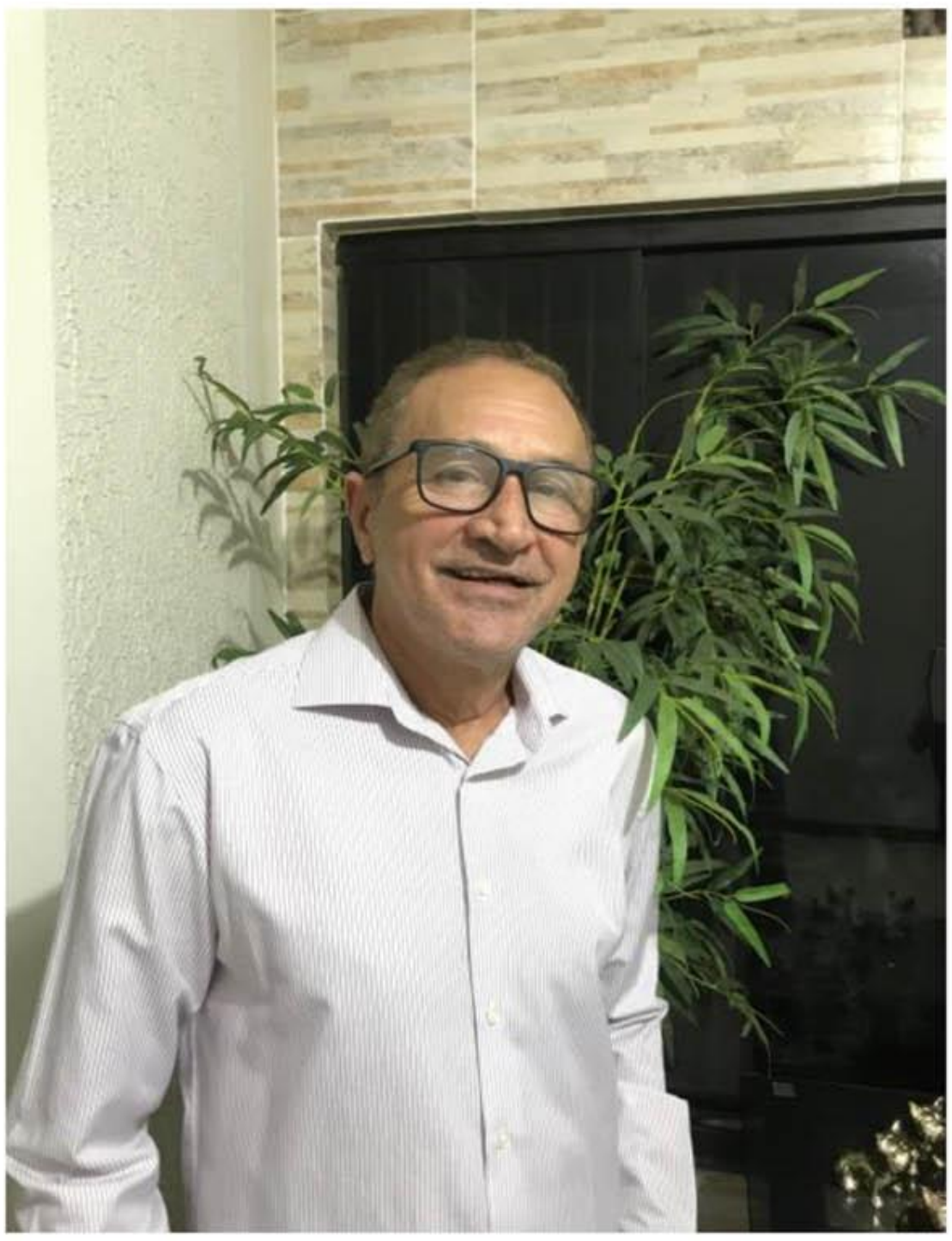

# MINHA HISTÓRIA

A minha primeira experiência com o ensino ocorreu em 1988, numa turma de Secretariado, na Escola Estadual de 1º e 2º Graus Raimundo Pinheiro em Cuiabá-MT. Fiz a graduação em Licenciatura Plena Em Matemática pela Universidade Federal de Mato Grosso. Em seguida concluí a Especialização em Formação Docente à Nível Superior com carga horária de 375/horas pela Universidade Federal de Mato Grosso, bem como, a Especialização em Estatística com carga horária de 360/horas pela Universidade de Marília, UNIMAR. Concluí o Mestrado em Geografia pelo Campus Universitário de Rondonópolis da Universidade Federal de Mato Grosso. O Doutorado em Ciências da Educação foi concluído na Universidad Tecnologica Intercontinental, UTIC, Paraguai e Convalidação pela Universidade Federal do Rio de Janeiro.

Atualmente sou professor Associado I do Departamento de Matemática do Instituto de Ciências Exatas e Naturais da Universidade Federal de Rondonópolis/MT. Tenho experiência na área de Matemática e Estatística, com ênfase em Ciências Exatas e da Terra, atuando principalmente com os seguintes temas: modelagem matemática, inclusão digital, processo de ensino-aprendizagem, software educativo, software livre e informática. No Departamento de Matemática, decidiu-se que todo docente vinculado a ele pode ministrar aula em qualquer curso daquelas disciplinas que está sob sua responsabilidade. Assim, além de responsáveis pela maior parte do curso de Matemática, nas modalidades Bacharelado e Licenciatura. Atuamos também nos cursos que têm em suas estruturas curriculares as disciplinas de Matemática e Estatística, a saber: Ciências Contábeis, Biologia, Engenharia Agrícola e Ambiental, Engenharia Mecânica, Pedagogia, Biologia, Sistema de Informação, Biblioteconomia, Psicologia e Enfermagem.

Ocupei os cargos de Pró-Reitor do Campus Universitário de Rondonópolis da Universidade Federal de Mato Grosso; Direção do Instituto de Ciências Exatas e Naturais do Campus Universitário de Rondonópolis da Universidade Federal de Mato Grosso, assumi também a Chefia do Departamento de Matemática e a Coordenações do Curso de Matemática e outras atividades administrativas pertinentes a Instituição. Destacamos que sempre procurei transformar minhas aulas num ambiente agradável e propício para a aprendizagem. Mesmo assim, tenho clareza do quanto ainda devo acrescentar à minha caminhada como Professor.